\documentclass[article]{aa}
\usepackage{graphicx}
\usepackage{eufrak}
\usepackage{txfonts}
\usepackage{eufrak}
\usepackage{natbib}
\bibpunct{(}{)}{;}{a}{}{,}

\begin{document}

\title{Dust driven mass loss from carbon stars\\ as a function of stellar parameters}
\subtitle{I. A grid of Solar-metallicity wind models}
\titlerunning{Dust driven mass loss from carbon stars as a function of stellar parameters}

\author{Lars Mattsson\thanks{\email{mattsson@astro.uu.se}}\and Rurik Wahlin \and Susanne H\"ofner}
\institute{Dept. of Physics and Astronomy, Div. of Astronomy and Space Physics, Uppsala University, Box 515, SE-751 20 Uppsala, Sweden}
         
\offprints{Lars Mattsson}

\date{Received date; accepted date}

\abstract
{Knowing how the mass loss of carbon-rich AGB stars depends on stellar parameters is crucial for stellar evolution modelling, as well as for the 
understanding of when and how circumstellar structures emerge around these stars, e.g., dust shells and so-called detached shells of expelled gas.}
{The purpose of this paper is to explore the stellar parameter space using a numerical radiation hydrodynamic (RHD) model of carbon-star atmospheres,
including a detailed description of dust formation and frequency-dependent radiative transfer, in order to determine how the mass loss of carbon 
stars changes with stellar parameters.}
{We have computed a grid of 900 numeric dynamic model atmospheres (DMAs) using a well-tested computer code. This grid of models covers most of the 
expected combinations of stellar parameters, which are made up of the stellar temperature, the stellar luminosity, the stellar mass, the 
abundance of condensible carbon, and the velocity amplitude of the pulsation.}
{The resultant mass-loss rates and wind speeds are clearly affected by the choice of stellar temperature, mass, luminosity and the abundance of 
available carbon. In certain parts of the parameter space there is also an inevitable mass-loss threshold, below which a dust-driven wind is not 
possible. Contrary to some previous studies, we find a strong dependence on the abundance of free carbon, which turns out to be a critical 
parameter. Furthermore, we have found that the dust grains that form in the atmosphere may grow too large for the commonly used small particle 
approximation of the dust opacity to be strictly valid. This may have some bearing on the wind properties, although further study of this problem 
is needed before quantitative conclusions can be drawn.}
{The wind properties show relatively simple dependences on stellar parameters above the mass-loss threshold, while the threshold itself is of a more
complicated nature. Hence, we chose not to derive any simplistic mass-loss formula, but rather provide a mass-loss prescription in the form of an
easy-to-use FORTRAN routine (available at http://coolstars.astro.uu.se). 
Since this mass-loss routine is based on data coming from an essentially self-consistent model of mass loss, it may 
therefore serve as a better mass-loss prescription for stellar evolution calculations than empirical formulae. Furthermore, we conclude that there 
are still some issues that need to be investigated, such as the r\^ole of grain-sizes.}
\keywords{Stars: AGB and post-AGB -- Stars: atmospheres -- Stars: carbon -- Stars: circumstellar matter -- 
          Stars: evolution -- Stars: mass loss -- Hydrodynamics -- Radiative transfer}

\maketitle

\section{Introduction}
How the mass loss of carbon-rich AGB stars (carbon stars) depends on stellar parameters is not very well known. 
It is crucial, however, to have that kind of information in many contexts, such as for stellar evolution modelling. A variety of empirical 
\cite[e.g.,][]{Vassiliadis93,Groenewegen93,Groenewegen95} as well as theoretical \cite[e.g.,][]{Blocker95, Arndt97, Wachter02, Wachter08} formulae 
with different parameterisations are available in the literature and the agreement between them is rather poor. These simple mass-loss 
prescriptions are nevertheless used in many models of stellar evolution, sometimes without enough consideration of their applicability in specific 
cases.  

Determining the mass-loss rates observationally is complicated and riddled with many possible sources of uncertainty. Also, the fundamental stellar 
parameters, such as effective temperature and mass, can be hard to determine observationally with high precision, which makes it very difficult to 
empirically determine the mass-loss rate as a function of stellar parameters with sufficient accuracy. Empirical studies as such can only provide a 
limited amount of information, since one cannot easily disentangle the effects of individual stellar parameters. The reason for this is of course that 
stellar evolution and observed stellar parameters are connected, and we can only observe each star at one evolutionary stage. Observations can 
tell us about the mass loss for a given set of stellar parameters, but the exact dependences on these parameters can unfortunately not be determined
easily due to selection effects. Existing empirical relations suffer from selection effects, with a very steep dependence of mass loss rates on 
stellar parameters.

The fundamental idea behind a so-called dust driven wind is that stellar photons, incident on dust grains, will lead to a transfer of momentum from 
the radiation to the atmospheric gas, which is dragged along by the dust grains. Models of this process have existed for quite some time and have
become increasingly more complex. Wood (1979) \nocite{Wood79} and Bowen (1988) made dynamic atmosphere models introducing a parameterised opacity 
to describe the effects of dust formation in the circumstellar envelope. Bowen \& Willson (1991) further elaborated on the implications of their models 
for the mass loss during AGB evolution, i.e., the development of a so-called "superwind" phase. Bl\"ocker (1995) then derived a modified version of 
Reimers' law based on the models of Bowen (1988) to prescribe mass loss in his calculations of stellar evolution on the AGB. This mass-loss 
prescription is still widely used, although it is entirely based on the early work by Bowen (1988). \nocite{Bowen88} 

Fleischer et al. (1992) \nocite{Fleischer92} presented self-consistent dynamic models of carbon rich AGB stars that include a time-dependent 
description of dust formation. Their work revealed phenomena like a discrete spatial structure of the circumstellar dust shell and multiperiodicity. 
A picture has emerged with dust-driven winds as a complex phenomenon, where the onset of mass loss requires particular circumstances. 
For about a decade, time-dependent dynamical models for stellar winds \cite[e.g.,][]{Hofner97, Winters00} were based on rather 
crude descriptions of radiative transfer and simplified micro-physics, which led to unrealistic atmospheric density-temperature 
structures.

H\"ofner et al. (2003) presented dynamic model atmospheres which couple time-dependent dynamics and frequency-dependent radiative transfer,  
including a non-grey description of the dust component. Comparing the observable properties calculated from these state-of-the-art dynamical models 
(spectra and their variations over time) with observations has shown drastic improvements compared to the synthetic spectra based on previous (grey) 
generations of models \cite[see, e.g.,][]{Gautschy04,Nowotny05a,Nowotny05b}. While further improvement may become possible in the future, a 
reasonable level of realism  has now been reached in the modelling to allow application to stellar evolution. The model used in the 
present study is identical to that used in Mattsson et al. (2007a) and contains only minor modifications of the H\"ofner et al. (2003) model. It is 
probably the most advanced theoretical tool available at present for studying how dust driven winds of carbon stars depend on stellar parameters.

Detailed numerical modelling provides valuable insights into the mass loss problem, since it can be used to constrain the actual physics involved in 
wind formation. In this paper, we explore the stellar parameter space using a numerical radiation hydrodynamic (RHD) model of carbon-star atmospheres,
including time-dependent dust formation and frequency-dependent radiative transfer, in order to determine how the mass-loss of carbon stars depends 
on stellar parameters. The results we give here reflect primarily the intense mass loss during the so-called thermal pulse phase, i.e., the very 
late stages of carbon-rich AGB evolution.

\section{Theory and methods}

\subsection{RHD atmosphere models}
  \label{dynamic}
  The model includes frequency-dependent radiative transfer and non-equilibrium dust formation, i.e., we solve the coupled 
  system of frequency-dependent radiation hydrodynamics (RHD) and time-dependent dust formation employing 
  an implicit numerical method and an adaptive grid \cite[cf.][]{Hofner03}.

  In our model, the stellar atmosphere and circumstellar envelope are described in terms of conservation 
  laws for the gas, the dust and the radiation field, expressed by the following set of coupled, nonlinear partial 
  differential equations (PDEs):

  \begin{itemize}
  \item The three equations describing conservation of mass, momentum and energy for the gas.
  \item The 0th and 1st moment equation of radiative transfer.
  \item Four moment equations of dust formation (cf. Section \ref{dustform}).
  \item The Poisson equation (self gravity).
  \end{itemize}

To our system of nonlinear partial differential equations describing the physics of a C-star atmosphere we add a so called "grid equation" which 
determines the locations of the grid points according to accuracy considerations (Dorfi \& Drury 1987) and an equation keeping track of the 
amount of condensible carbon, leaving us with a total of 12 partial differential equations (PDEs). This system of PDEs is then
solved implicitly using a Newton-Raphson scheme. All equations are discretised in a volume-integrated conservation form on 
a staggered mesh. The spatial discretisation of the advection term is a monotonic second-order advection scheme (van Leer 
1977). The same order of numerical precision is used for all PDEs. Details of the numerical method are discussed by Dorfi 
\& Feuchtinger (1995) and in several previous papers about dust-driven wind models (cf. H\"ofner et al. 1995 and references therein).

\subsection{General procedure}
The RHD computations are started from hydrostatic dust-free initial models providing the atmospheric structure at $t=0$.
When the dust equations are switched on, dust condensation starts and the resulting radiative acceleration creates an outward 
motion of the dust and the gas. In the first computational phase, the expansion of the atmospheric layers is followed by the
grid to about $20 - 30\,R_\star$ (usually around to $\sim 10^{15}$ cm). At this distance, the location of the outer boundary is
fixed, allowing outflow. The outflow model then evolves typically for more than 100 periods. To avoid a 
significant depletion of mass inside the computational domain, the model calculation is stopped after about  
$10^5$ time steps. This is important since we cannot (due to the computational method) allow for material to flow over the 
inner boundary. Due to this fact, a large enough mass loss will eventually lead to a depletion of mass in the considered 
part of the atmosphere and circumstellar envelope.

\subsection{Inner boundary: Stellar pulsations}
The dynamic atmosphere model does not include a physical model for the pulsation mechanism. The pulsation is modelled by a "piston boundary 
condition" \cite{Bowen88} located at $R_{\rm in} \approx 0.8 - 0.9 R_\star$ in order to be in the optically thick part of the atmosphere but outside 
the zone where the pulsations are excited. It should be noted that all results obtained in this paper are, to some extent, subject to this inner 
boundary condition, and as shown by Mattsson et al. (2008), the effect on the mass-loss rate is not negligible in the non-saturated
wind regime where the degree of dust condensation is affected by the input of kinetic energy by pulsations.

We use a harmonic piston boundary given by the expressions
\begin{equation}
R_{\rm in}(t) = R_{\rm in}(0) + {\Delta u_{\rm p}\over \omega}\sin\left(\omega t\right),
\end{equation}
and
\begin{equation}
u_{\rm in}(t) =  \Delta u_{\rm p}\cos\left(\omega t\right), \quad  \omega = {2\pi\over \mathcal{P}}, 
\end{equation}
where $\mathcal{P}$ is the period and $\Delta u_{\rm p}$ is the velocity amplitude of the pulsations.
To restrict the number of free parameters, we also employ an empirical period-luminosity relation \cite{Feast89} and 
keep the period $\mathcal{P}$ tied to the luminosity for all models. The piston amplitude $\Delta u_{\rm p}$, or the kinetic-energy
injection by pulsation (cf. the $q$-parameter in Mattsson et al. 2008)
\begin{equation}
\Delta \mathcal{E}_{\rm p} = {1\over 2}\rho_{\rm in}\Delta u_{\rm p}^2,  
\end{equation}
is essentially a free parameter in the model, but previous experience \cite[e.g.,][]{Hofner97, Mattsson07a} shows that only a limited range can be 
considered as reasonable. Nonetheless, we must consider the effects of picking different amplitudes, and we therefore consider the
cases $\Delta u_{\rm p}= 2.0, 4.0 \mbox{ and } 6.0$ km s$^{-1}$. The motion of the piston is accompanied by luminosity variations, 
since the radiative flux through the (moving) inner boundary is kept fixed, as described in previous papers, \cite[e.g.,][]{Hofner97}. In 
order to have luminosity variations that are consistent with observations, the velocity amplitudes given above turn out to be a
good choice. 

The inner boundary condition, i.e., the location of $R_{\rm in}$, was chosen so that all models have a similar kinetic-energy input due to pulsation 
for a given piston amplitude. This is critical in order to obtain a consistent picture of how the wind properties may vary with stellar parameters 
\cite{Mattsson08}.

\subsection{Outer boundary: Mass loss}
During the initial phase of the computation the expansion of the model is followed by the grid. When the outer-most point of the adaptive grid 
reaches some given distance from the inner boundary, the program automatically switches to a fixed outer boundary located at this point, allowing 
outflow over the outer boundary. There are no specific conditions posed on the location of the outer boundary, except that it must be at a large 
enough distance to ensure that the wind velocity at the outer boundary is close to the terminal velocity. At this point, the conditions
\begin{equation}
\left({\partial u\over \partial r}\right)_{\rm out} =0, \quad H_{\rm out}  = \bar{\mu} J_{\rm out},
\end{equation}
are applied, where $\bar{\mu}$ reflects the angular intensity distribution of the radiation field.

From the models described here we obtain the density and the wind velocity, both as functions of radius and time. The mass loss rate is then a 
function of time given by
\begin{equation}
\label{mlreq}
\dot{M}(t) = 4\pi R_{\rm out}^2\,\rho_{\rm out}(t)\,u_{\rm out}(t),
\end{equation}
where $\rho_{\rm out}$ and $u_{\rm out}$ are the density and wind velocity at $R_{\rm out}$, i.e. the radius where the outer boundary is 
fixed (usually around $10^{15}$ cm, or typically 20-30 stellar radii, see above). 
Since practically all momentum transfer from radiation to matter is due to the interaction with dust, we expect the 
wind velocity to be correlated with the degree of dust condensation. 

  \begin{figure}
  \resizebox{\hsize}{!}{
  \includegraphics{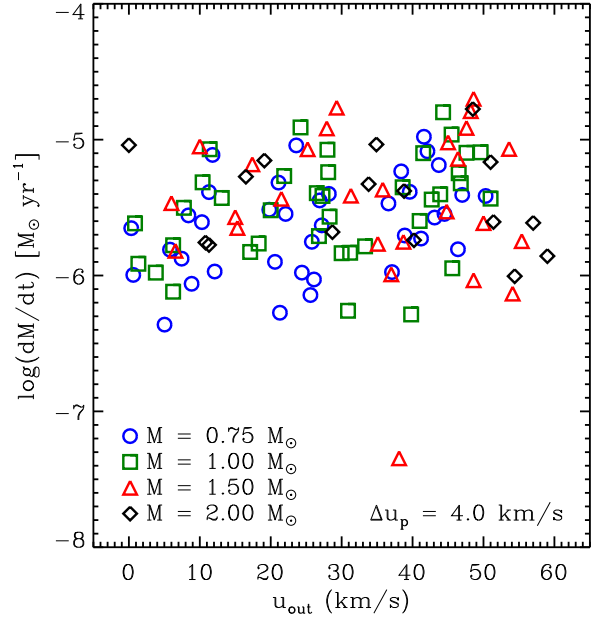}}
  \caption{  \label{mlr_v}
  Mass-loss rates as a function of wind speeds for models with $\Delta u_{\rm p} = 4$ km s$^{-1}$ and various values of $\log({\rm C-O})$.
  }
  \end{figure}
  
  \begin{figure}
  \resizebox{\hsize}{!}{
  \includegraphics{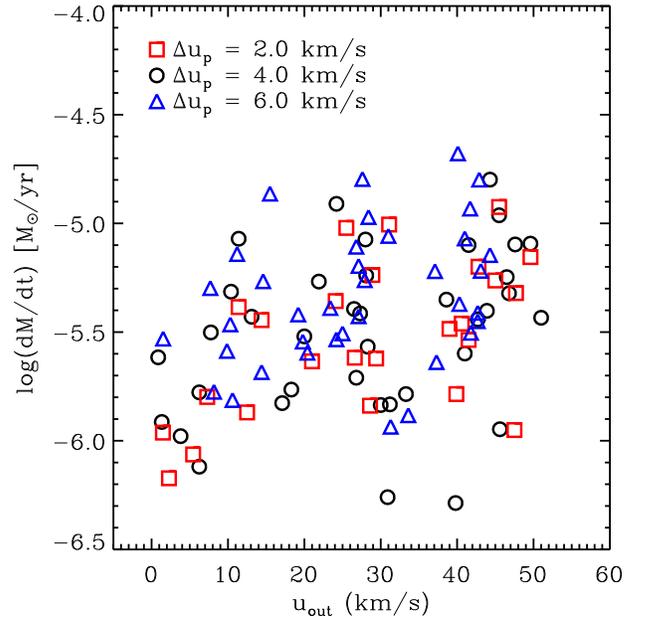}}
  \caption{  \label{piston}
  Mass-loss rate vs. wind speed for models with different $\Delta u_{\rm p}$, but the same carbon excess, $\log({\rm C-O})+12 = 8.80$.
  }
  \end{figure}

\subsection{Dust formation}
\label{dustform}
The models presented here include a time-dependent description of dust grain growth and evaporation using the moment method by \nocite{Gail88} Gail 
\& Sedlmayr (1988) and \nocite{Gauger90} Gauger et al. (1990). The dust component is described in terms of moments of the grain size distribution 
function weighted with a power $j$ of the grain radius. The zeroth moment is the total number density of grains (simply the integral of the size 
distribution function over all grain sizes), while the third moment is proportional to the average volume of the grains. The moment equations are 
solved simultaneously with the RHD equations.

In order to calculate how much momentum is transferred from photons to dust grains, we need to know the frequency-dependent opacity of these 
grains. This can be expressed in terms of the extinction efficiency, which is the ratio of the extinction cross section to the geometrical cross 
section of the grains. In the small-particle limit (which is used here), the dust grain opacity becomes a simple function of the grain radius. The 
dependence of the opacity on wavelength and grain size can thus be separated into two independent factors, which greatly simplifies the calculations. 
However, as we shall see, this approximation may need to be relaxed, since the carbon grains tend to grow quite rapidly. The models in this paper 
are calculated using the refractive index data of \nocite{Rouleau91} Rouleau \& Martin (1991) to obtain the dust extinction 
\cite[see][for further discussion]{Andersen03, Hofner03}. The intrinsic density of the grain material used in the model is set to 
$\rho_{\rm gr} = 1.85$ g cm$^{-3}$, 
which matches the material in Rouleau \& Martin (1991). \nocite{Rouleu91}

We assume that dust grains can be considered to be spherical, consisting of amorphous carbon only, since we are modelling carbon stars 
\cite[see][for details about these assumptions]{Andersen03}. The nucleation, growth and evaporation of grains is assumed to proceed by reactions 
involving ${\rm C}$, ${\rm C_2}$, ${\rm C_2H}$ and ${\rm C_2H_2}$. In this model of grain growth, a so called {\it sticking coefficient} (sometimes 
referred to as the reaction efficiency factor) is used that enters into the net growth rate of the dust grains. This parameter, $\alpha_{\rm S}$, 
is not definitely known unless we know the exact sequence of chemical reactions responsible for dust formation. However, Gail \& Sedlmayr (1988) 
argued that the sticking coefficient must be of the order of unity, mainly because it is expected that neutral radical reactions play a major role 
in the formation of carbon grains. We set $\alpha_{\rm S}=1$, but note that a set of test models with $\alpha_{\rm S}=0.5$ give similar results. 
%(see Fig. \ref{alpha_s}).

Dust grains in a stellar atmosphere influence both its energy and momentum balance. For simplicity we assume complete momentum and position coupling 
of gas and dust, i.e. the momentum gained by the dust from the radiation field is directly transferred to the gas and there is no drift between dust 
and gas. However, this strong coupling between the dust and the gas phase is not obvious. In a previous attempt to relax this phase coupling 
approximation, \nocite{Sandin03, Sandin04} Sandin \& H\"ofner (2003, 2004) found that the effects of decoupling the phases might be quite 
significant. The most striking feature is that the dust formation may increase significantly, but this does not necessarily increase the 
predicted mass loss rates for a given set of stellar parameters. A coupled solution of the detailed frequency-dependent equations of RHD, including 
dust formation and drift would increase the computing time for each model to an extent that would make it very laborious to compute a grid of the 
kind presented here. Thus, we chose to give frequency-dependent radiative transfer priority over relaxing the phase coupling approximation, even if 
this approximation cannot hold for very low gas densities, i.e. at very large distances from the star or for wind models with very low mass loss
rates. Furthermore, since transfer of internal energy between gas and dust is negligible compared to the interaction of each component with the 
radiative field \cite{Gauger90}, we assume radiative equilibrium for the dust. This allows us to estimate the grain temperatures from the radiation 
temperature, as we know the dust opacities.
  
  \begin{figure*}
  \resizebox{\hsize}{!}{
  \includegraphics{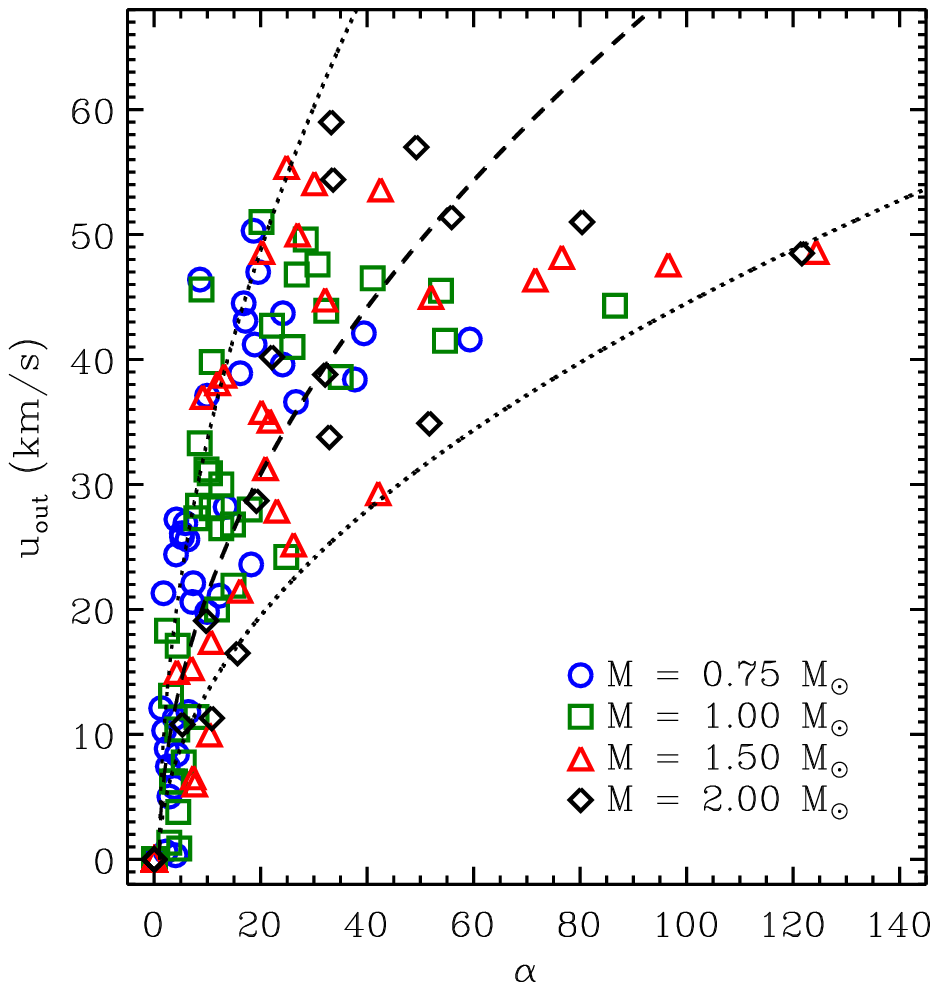}
  \includegraphics{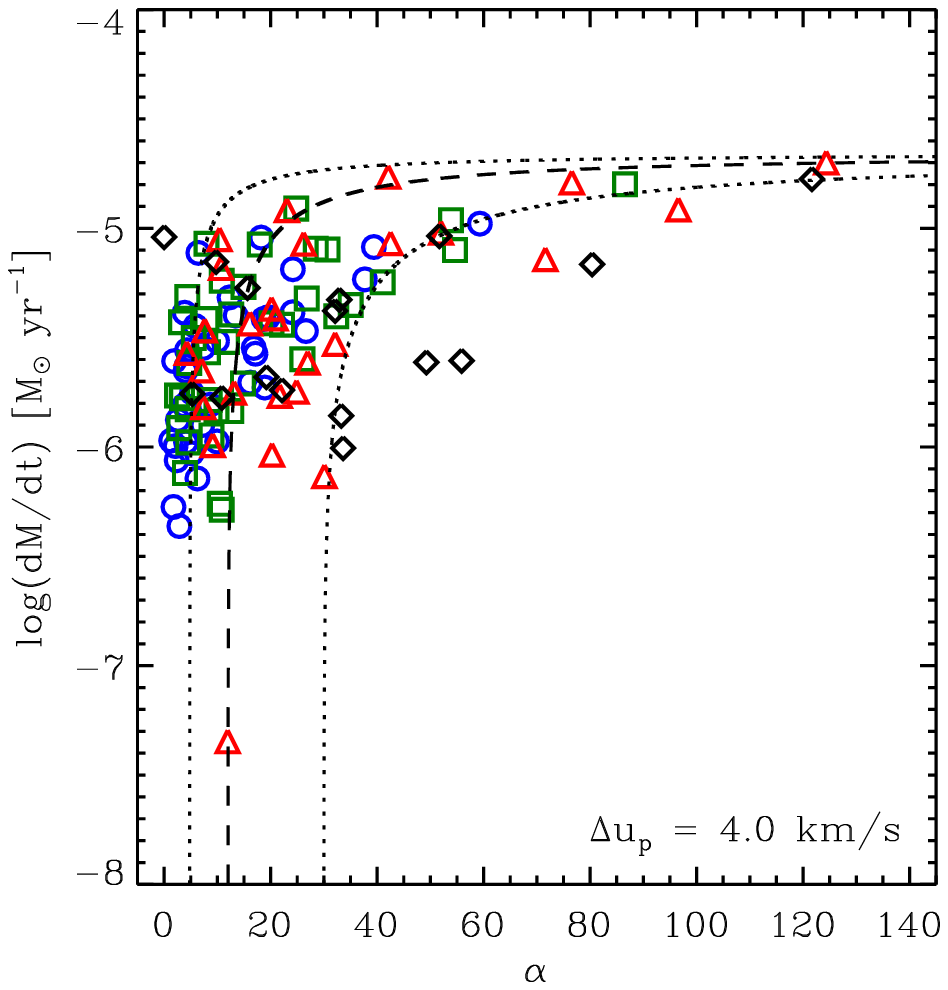}}
  \caption{  \label{AB}
  Wind speed (left) and mass-loss rate (right) as functions of the acceleration parameter $\alpha$ for models with 
  $\Delta u_{\rm p} = 4$ km s$^{-1}$. The dashed and dotted lines represent analytic models as described in Lamers \& Cassinelli (1999) and
  Groenewegen et al. (1998), assuming different $L_\star/M_\star$.}
  \end{figure*}
  
  \begin{table}
  \caption{Definition of the grid. $\Delta$ denotes the grid step. \label{griddef}  }
  \center
  \begin{tabular}{lllll}  
  \hline
  \hline
  $M_\star$  & $\log(L_\star)$ & $T_{\rm eff}$ & log(C-O)+12 & $\Delta u_{\rm p}$\\[1mm]
  [$M_\odot$] & [$L_\odot$] & [K] & & [km s$^{-1}$] \\[1mm]
  \hline
            &     $\Delta = 0.15$  & $\Delta = 200$ & $ \Delta =0.30$ & $\Delta = 2.0$\\[1mm]
  
  0.75 & 3.55 - 3.85 & 2400 - 3200 & 7.90 - 9.10 & 2.0 - 6.0 \\
  1.0  & 3.70 - 4.00 & 2400 - 3200 & 7.90 - 9.10 & 2.0 - 6.0 \\
  1.5  & 3.85 - 4.15 & 2400 - 3200 & 7.90 - 9.10 & 2.0 - 6.0 \\
  2.0  & 3.85 - 4.15 & 2400 - 3200 & 7.90 - 9.10 & 2.0 - 6.0 \\[1mm]
  \hline
  \end{tabular}
  \end{table}

\section{Definition of the grid}
Since one purpose of the grid is (among other things) to provide information about how the mass loss rate is affected by
changes in various stellar parameters as the star evolves on the AGB, it is wise to consider which values of the
fundamental parameters ($M,L_\star,T_{\rm eff}$) never appear in evolutionary tracks. Systematic studies by, e.g., Bl\"ocker
(1995) show for example that for carbon stars at solar metallicity we should not expect average luminosities above 
$2\cdot 10^4 L_\odot$ or masses above $2M_\odot$ on the AGB. The temperature of a carbon star lies approximately in the range $2000 - 3500$ K
according to both observations (e.g. Lambert et al 1986, Bergeat \& Chevallier 2005) and evolutionary models. Thus, the region in the 
$L_\star-T_{\rm eff}$ plane (HR-diagram) that we are interested in is quite small for a star of a given mass and initial 
metallicity. It is important, however, to stress that during its 
evolution an AGB star may temporarily have parameter configurations that do not 'agree' with typical observations. This is due to 
the fact that we are extremely unlikely to observe an AGB star as it undergoes, e.g., a thermal pulse, where the luminosity 
is very high and the temperature rather low (see Fig.~1 in Mattsson et al. 2007). Therefore, defining the grid according to parameter 
combinations found in evolutionary tracks will require a somewhat larger parameter space than observations may suggest is 
necessary (see Table \ref{griddef}).

As shown by Mattsson et al. (2008), the kinetic-energy injection by 
pulsation can affect the wind properties significantly. Placing the inner boundary at a fixed optical or geometric depth led 
to different energy injections due to different density structures, resulting from different chemical compositions. This fact
needs to be emphasised - theoretical work on AGB mass loss does not always include such "energy control" of the piston boundary,
which may affect the results. Here we have tried to keep the equilibrium density/pressure at the inner boundary roughly the
same for all models and used three different velocity amplitudes for the piston ($\Delta u_{\rm p} = 2.0, 4.0, 6.0 \mbox{ km s}^{-1}$).

Initially, parts of the grid were supposed to be computed using the solar composition by Grevesse \& Sauval (1998) \nocite{Grevesse98} 
as well as that of Asplund et al. (2005) \nocite{Asplund05}, which were first thought to give slightly different results. However, after some 
scrutiny, it became clear that the differences were due to a poorly constrained inner boundary condition \cite[see][]{Mattsson08}. The grid of
models presented here is computed using the Asplund et al. (2005) solar composition only.

\section{Results and discussion}

  \begin{table*}
  \caption{\label{parameters} Input parameters ($L_\star$, $T_{\rm eff}$, log(C-O), $P$) and the 
  resulting avergage mass loss rate, average wind speed and the mean degree of dust condensation at the outer boundary for a
  subset of models with $M_\star = 1 M_\odot$ and $\Delta u_{\rm p} = 4.0 \mbox{km s}^{-1}$. 
  The dust-to-gas mass ratio $\rho_{\rm d}/\rho_{\rm g}$ is calculated from $f_{\rm c}$ as described in H\"ofner \& Dorfi (1997) and
  the grain radius is defind as $a_{\rm gr}=r_0\,(K_1/K_0)$, where $K_0, K_1$ are the zeroth and first moment of the grain-size distribution, 
  respectively. Models with log(C-O)+12 = 7.90 are not included, since none of them produced any outflow.}
  \center
  \begin{tabular}{lcccccccccccc}  
  \hline
  \hline

  $\log(L_\star)$ & $T_\mathrm{eff}$  & log(C-O)+12 & $P$ & $\langle\dot{M}\rangle$ & $\langle u_{\rm out} \rangle$ & $\langle {\rm f_c} \rangle$ & 
  $\langle {\rho_{\rm d}/\rho_{\rm g}} \rangle$ & $\langle a_{\rm gr}\rangle$\\[1mm]
  [$L_{\odot}$] &  [$\mbox{K}$] & & [days] & [$M_\odot$ yr$^{-1}$] & [km  s$^{-1}$] & & & [cm]\\

  \hline
\\
    3.70   & 2400   &    8.20   &   295   &   -          &   -          &   -          &   -	      &   1.26E-04  \\
    3.70   & 2400   &    8.50   &   295   &   1.05E-06   &   3.80E+00   &   2.44E-01   &   6.61E-04   &   2.32E-05  \\
    3.70   & 2400   &    8.80   &   295   &   3.02E-06   &   2.00E+01   &   3.19E-01   &   1.73E-03   &   1.19E-05  \\
    3.70   & 2400   &    9.10   &   295   &   4.46E-06   &   3.86E+01   &   4.75E-01   &   5.13E-03   &   5.78E-06  \\
    3.85   & 2400   &    8.20   &   393   &   -          &   -          &   -          &   -	      &   1.08E-04  \\
    3.85   & 2400   &    8.50   &   393   &   3.15E-06   &   7.76E+00   &   2.13E-01   &   5.77E-04   &   2.32E-05  \\
    3.85   & 2400   &    8.80   &   393   &   5.40E-06   &   2.19E+01   &   2.85E-01   &   1.54E-03   &   1.15E-05  \\
    3.85   & 2400   &    9.10   &   393   &   7.95E-06   &   4.15E+01   &   5.24E-01   &   5.65E-03   &   6.19E-06  \\
    4.00   & 2400   &    8.20   &   524   &   2.42E-06   &   8.67E-01   &   2.54E-01   &   3.45E-04   &   8.30E-05  \\
    4.00   & 2400   &    8.50   &   524   &   8.50E-06   &   1.14E+01   &   2.15E-01   &   5.83E-04   &   2.48E-05  \\
    4.00   & 2400   &    8.80   &   524   &   1.23E-05   &   2.42E+01   &   3.36E-01   &   1.82E-03   &   1.26E-05  \\
    4.00   & 2400   &    9.10   &   524   &   1.59E-05   &   4.43E+01   &   5.88E-01   &   6.34E-03   &   6.70E-06  \\
    3.70   & 2600   &    8.20   &   295   &   -          &   -          &   -          &   -	      &   -         \\
    3.70   & 2600   &    8.50   &   295   &   7.60E-07   &   6.24E+00   &   2.11E-01   &   5.72E-04   &   -         \\
    3.70   & 2600   &    8.80   &   295   &   1.95E-06   &   2.68E+01   &   4.00E-01   &   2.16E-03   &   1.41E-05  \\
    3.70   & 2600   &    9.10   &   295   &   3.96E-06   &   4.39E+01   &   5.84E-01   &   6.30E-03   &   7.38E-06  \\
    3.85   & 2600   &    8.20   &   393   &   -          &   -          &   -          &   -	      &   1.05E-04  \\
    3.85   & 2600   &    8.50   &   393   &   1.67E-06   &   6.24E+00   &   1.71E-01   &   4.63E-04   &   1.87E-05  \\
    3.85   & 2600   &    8.80   &   393   &   4.04E-06   &   2.65E+01   &   3.22E-01   &   1.74E-03   &   1.24E-05  \\
    3.85   & 2600   &    9.10   &   393   &   5.66E-06   &   4.65E+01   &   5.25E-01   &   5.67E-03   &   6.20E-06  \\
    4.00   & 2600   &    8.20   &   524   &   1.22E-06   &   1.34E+00   &   2.06E-01   &   2.80E-04   &   5.52E-05  \\
    4.00   & 2600   &    8.50   &   524   &   4.85E-06   &   1.04E+01   &   1.57E-01   &   4.26E-04   &   2.10E-05  \\
    4.00   & 2600   &    8.80   &   524   &   8.43E-06   &   2.80E+01   &   3.25E-01   &   1.76E-03   &   1.18E-05  \\
    4.00   & 2600   &    9.10   &   524   &   1.09E-05   &   4.55E+01   &   4.88E-01   &   5.27E-03   &   6.26E-06  \\
    3.70   & 2800   &    8.20   &   295   &   -          &   -          &   -          &   -	      &   -         \\
    3.70   & 2800   &    8.50   &   295   &   -          &   -          &   -          &   -	      &   -         \\
    3.70   & 2800   &    8.80   &   295   &   1.46E-06   &   3.00E+01   &   4.55E-01   &   2.46E-03   &   1.66E-05  \\
    3.70   & 2800   &    9.10   &   295   &   2.52E-06   &   4.10E+01   &   4.69E-01   &   5.06E-03   &   6.94E-06  \\
    3.85   & 2800   &    8.20   &   393   &   -          &   -          &   -          &   -	      &   -         \\
    3.85   & 2800   &    8.50   &   393   &   1.49E-06   &   1.71E+01   &   2.25E-01   &   6.10E-04   &   -         \\
    3.85   & 2800   &    8.80   &   393   &   2.71E-06   &   2.83E+01   &   3.15E-01   &   1.70E-03   &   1.28E-05  \\
    3.85   & 2800   &    9.10   &   393   &   4.77E-06   &   4.68E+01   &   5.14E-01   &   5.55E-03   &   7.57E-06  \\
    4.00   & 2800   &    8.20   &   524   &   -          &   -          &   -          &   -	      &   5.71E-05  \\
    4.00   & 2800   &    8.50   &   524   &   3.72E-06   &   1.31E+01   &   1.72E-01   &   4.66E-04   &   2.15E-05  \\
    4.00   & 2800   &    8.80   &   524   &   5.76E-06   &   2.81E+01   &   2.94E-01   &   1.59E-03   &   1.33E-05  \\
    4.00   & 2800   &    9.10   &   524   &   8.07E-06   &   4.96E+01   &   3.87E-01   &   4.18E-03   &   6.12E-06  \\
    3.70   & 3000   &    8.20   &   295   &   -          &   -          &   -          &   -	      &   -         \\
    3.70   & 3000   &    8.50   &   295   &   -          &   -          &   -          &   -	      &   -         \\
    3.70   & 3000   &    8.80   &   295   &   -          &   -          &   -          &   -	      &   1.25E-05  \\
    3.70   & 3000   &    9.10   &   295   &   5.50E-07   &   3.09E+01   &   2.84E-01   &   3.06E-03   &   4.79E-06  \\
    3.85   & 3000   &    8.20   &   393   &   -          &   -          &   -          &   -	      &   -         \\
    3.85   & 3000   &    8.50   &   393   &   -          &   -          &   -          &   -	      &   -         \\
    3.85   & 3000   &    8.80   &   393   &   1.47E-06   &   3.12E+01   &   3.76E-01   &   2.03E-03   &   -         \\
    3.85   & 3000   &    9.10   &   393   &   3.62E-06   &   4.27E+01   &   4.25E-01   &   4.59E-03   &   6.55E-06  \\
    4.00   & 3000   &    8.20   &   524   &   -          &   -          &   -          &   -	      &   -         \\
    4.00   & 3000   &    8.50   &   524   &   1.72E-06   &   1.83E+01   &   1.75E-01   &   4.74E-04   &   2.66E-05  \\
    4.00   & 3000   &    8.80   &   524   &   3.85E-06   &   2.73E+01   &   2.86E-01   &   1.55E-03   &   1.21E-05  \\
    4.00   & 3000   &    9.10   &   524   &   8.00E-06   &   4.76E+01   &   5.56E-01   &   6.00E-03   &   7.43E-06  \\
    3.70   & 3200   &    8.20   &   295   &   -          &   -          &   -          &   -	      &   -         \\
    3.70   & 3200   &    8.50   &   295   &   -          &   -          &   -          &   -	      &   -         \\
    3.70   & 3200   &    8.80   &   295   &   -          &   -          &   -          &   -	      &   -         \\
    3.70   & 3200   &    9.10   &   295   &   1.13E-06   &   4.56E+01   &   3.22E-01   &   3.47E-03   &   5.64E-06  \\
    3.85   & 3200   &    8.20   &   393   &   -          &   -          &   -          &   -	      &   -         \\
    3.85   & 3200   &    8.50   &   393   &   -          &   -          &   -          &   -	      &   -         \\
    3.85   & 3200   &    8.80   &   393   &   -          &   -          &   -          &   -	      &   -         \\
    3.85   & 3200   &    9.10   &   393   &   5.17E-07   &   3.98E+01   &   2.77E-01   &   2.99E-03   &   5.03E-06  \\
    4.00   & 3200   &    8.20   &   524   &   -          &   -          &   -          &   -	      &   -         \\
    4.00   & 3200   &    8.50   &   524   &   -          &   -          &   -          &   -	      &   -         \\
    4.00   & 3200   &    8.80   &   524   &   1.64E-06   &   3.33E+01   &   3.09E-01   &   1.67E-03   &   1.47E-05  \\
    4.00   & 3200   &    9.10   &   524   &   3.68E-06   &   5.10E+01   &   3.65E-01   &   3.94E-03   &   5.75E-06  \\
\\
  \hline
\\
  \end{tabular}
  \label{models}
  \end{table*}

\subsection{General trends}
Wind velocities and mass loss rates are obviously correlated with some stellar parameters, such as luminosity, temperature and mass. 
The presence of trends with these parameters can be seen in observations, but in general the information is degenerate since the different
stellar parameters are very much coupled throughout the evolution of a star. This is why, for example, one finds a simple mass-loss-period law 
\citep[see, e.g.,][]{Vassiliadis93,Groenewegen93,Groenewegen95} which of course does not contain all the information needed for a proper treatment 
of mass loss in, e.g., stellar evolution modelling. Empirical results do not tell us directly how the mass loss rate is affected by changes in 
individual stellar parameters. 

The mass-loss rate and the wind speed is, from a theoretical point of view, often related to the ratio of radiative to gravitational acceleration,
\begin{equation}
\Gamma = {\kappa\,L_\star \over 4\pi c\,GM_\star}
\end{equation}
where $\kappa$ is the total flux-mean opacity and $c$ is the speed of light. Here we define a similar quantity,
\begin{equation}
\label{alpha}
\alpha \equiv {\rho_{\rm d} \over \rho_{\rm g}} {L_\star \over M_\star},
\end{equation}
where
\begin{equation}
{\rho_{\rm d} \over \rho_{\rm g}} = {m_{\rm C} \over 
m_{\rm H} + m_{\rm He} \varepsilon_{\rm He}}
\,\tilde{\varepsilon}_{\rm C}f_{\rm c} = 
{12\over 1.4}\left({\varepsilon_{\rm C}\over\varepsilon_{\rm O}}
-1\right)\,\varepsilon_{\rm O}f_{\rm c},
\end{equation}
in which $\tilde{\varepsilon}_{\rm C} = \varepsilon_{\rm C}-\varepsilon_{\rm O}$, i.e., the carbon excess (condensible carbon) and $f_{\rm c}$ is 
the degree of condensation. $\alpha$ is in principle proportional to $\Gamma$ and can be used in its place for most purposes. The reason we use 
$\alpha$ instead of $\Gamma$ is that $\alpha$ only contains the physical quantities studied in this paper. In Fig. \ref{AB} we show the wind speed 
and the mass-loss rate as functions of $\alpha$ for all models with $\Delta u_{\rm p} = 4.0\mbox{ km s}^{-1}$. The wind speed is relatively 
well-correlated with $\alpha$, which is expected from simple analytical arguments \cite[see][Chap. 4 \& 7]{Lamers99}. It is straight forward to 
show that, for a stationary\footnote{The time dependent case will, at least on average, resemble the stationary case in the sense that the wind 
speed is mainly due to the amount of dust opacity and how strong the gravitational field is. The main differences, both in snap-shots and average
structures, occur in the dust-free pulsating atmosphere, influencing the conditions in the dust-formation zone, therefore also $\Gamma$.} 
polytropic wind,
\begin{equation}
u_{\rm out}^2 \approx \left({2\over \gamma-1}\right)\,c_{\rm s}^2(R_{\rm c}) + (\Gamma-1)\,u_{\rm esc}^2(R_{\rm c})
\end{equation}
where $u_{\rm esc}$ is the escape speed, $c_{\rm s}$ is the sound speed, $R_{\rm c}$ is the condensation radius and $\gamma$ is the polytropic
index in a polytropic equation of state. For the mass-loss rate, the correlation with $\Gamma$ (or $\alpha$) is 
weaker, which is due to variations of the gas-density profiles of the wind regions between different models. However, the mass-loss rate
can (in the stationary case) be expressed analytically as (see Chap. 7 in Lamers \& Cassinelli 1999, and the Appendix in Groenewegen et al. 1998)
\begin{equation}
\dot{M} \propto \tau_{\rm w} {L_\star\over u_{\rm out} - u(R_{\rm c})} \left(1-{1\over \Gamma}\right),
\end{equation}
where $\tau_{\rm w}$ is the flux-mean optical depth far out in the wind, and all other symbols are as previously defined. By analysing the time 
development of several wind models, it becomes obvious that the density profiles are not directly a consequence of the momentum transfer, as the 
wind speed profiles are, but of a combination of momentum transfer and pulsation dynamics. Thus, one should not expect a strong correlation 
between mass-loss rate and $\alpha$ for the non-stationary case.

\begin{figure*}
  \resizebox{\hsize}{!}{
  \includegraphics{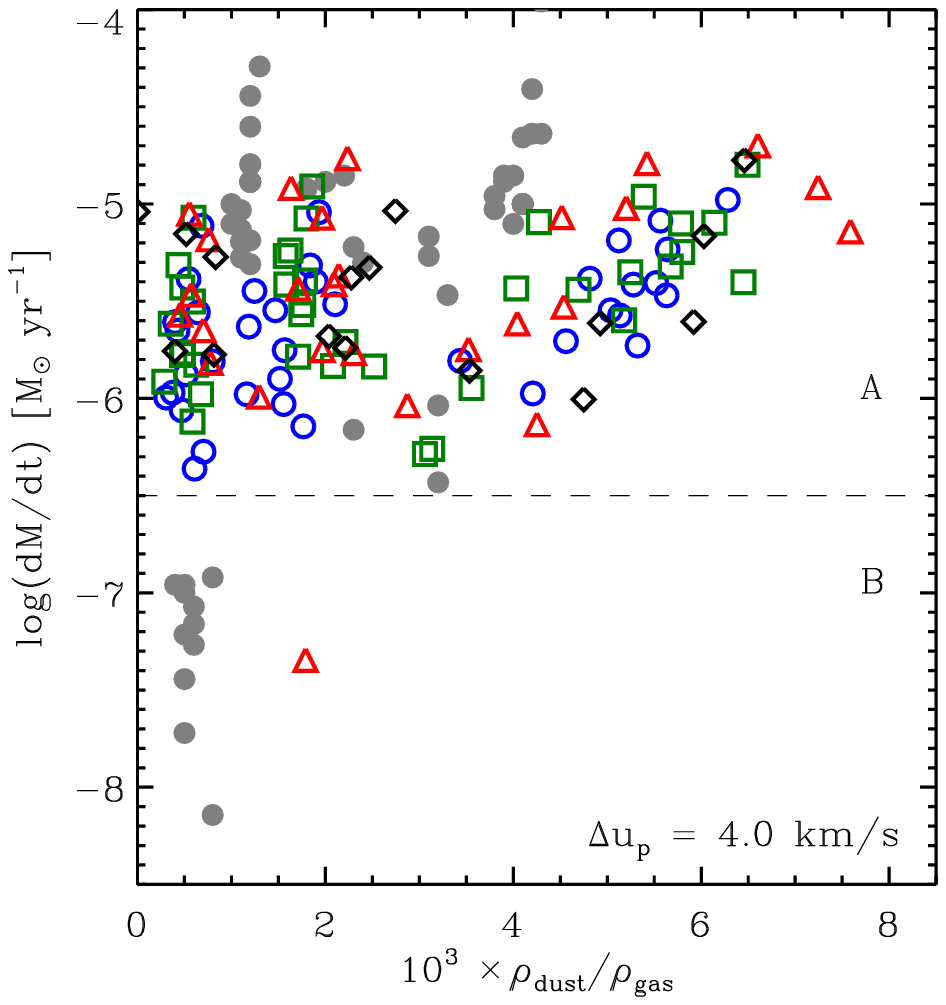}
  \includegraphics{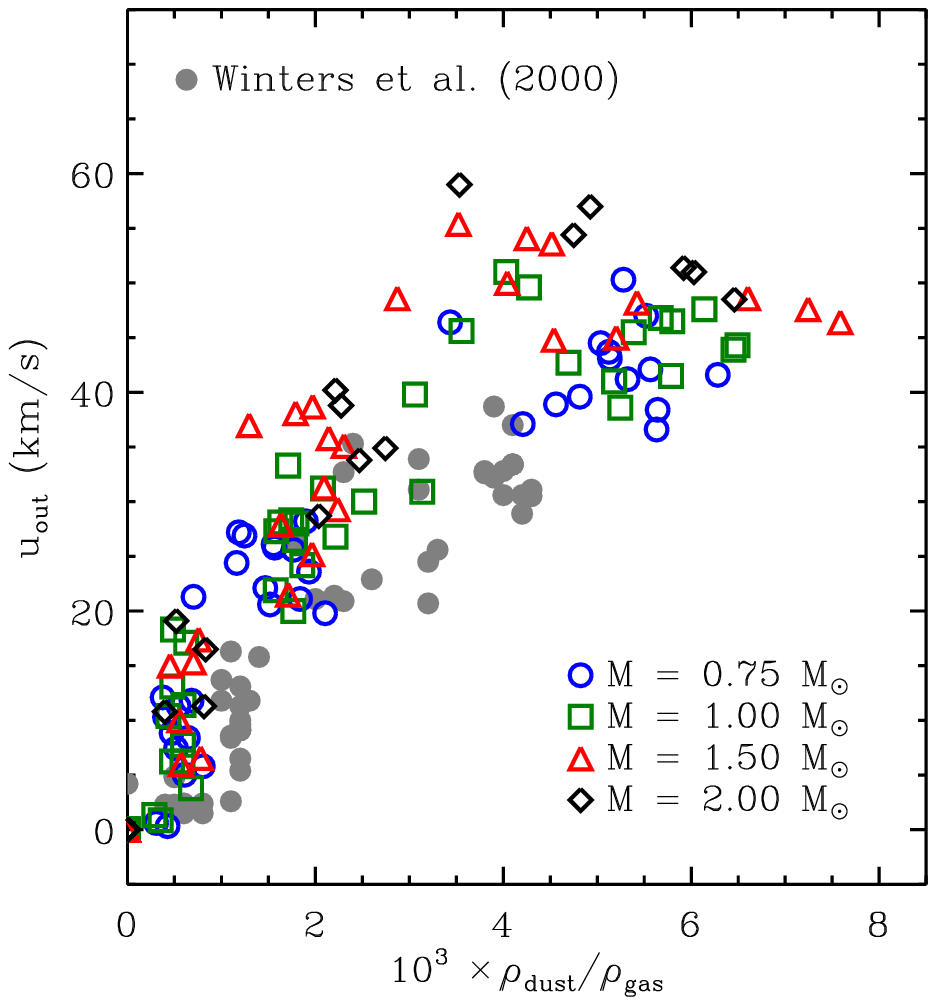}}
  \caption{  \label{winters}
  Wind speed (left) and mass-loss rate (right) as functions of the dust-to-gas ratio for models with 
  $\Delta u_{\rm p} = 4$ km s$^{-1}$. Grey dots shows modelling results by Winters et al. (2000) and the
  dashed line marks the division between A- and B-type models according to Winters et al. (2000).}
  \end{figure*}

Previous efforts to model the winds of carbon stars are in many cases not directly comparable to the present results. For instance, stationary
wind models with detailed dust formation should not be compared with DMAs, nor should models including drift but having grey radiative
transfer be compared with models not including drift but featuring frequency-dependent radiative transfer instead (as in the present work).
In comparison with existing theoretical work similar to the present \cite{Winters00,Wachter02}, our mass-loss rates show over-all trends that are
comparable, although individual combinations of stellar parameters in some cases lead to quite different results compared to their models.
More recent work by the Berlin group \cite{Wachter08} has been focusing on sub-solar metallicities and is not adequate for comparison.
The wind speeds that we obtain for the most carbon rich models are higher than the highest observed \cite[cf.][]{Schoier01}, but consistent with 
the results by Winters (2000) and Wachter et al. (2002) for similar dust-to-gas ratios (see Fig. \ref{winters}). 

\subsection{Wind velocity}
The wind velocity depends strongly on the carbon excess $\tilde{\varepsilon}_{\rm C}$, which appears as a steep slope in Fig.~\ref{models1}. 
This is expected, since the acceleration of the gas depends on the dust opacity, which in turn is proportional to the dust-to-gas ratio.
The strong correlation between dust abundance and wind velocity (see Fig. \ref{winters}) 
demonstrates how the wind speed is a direct consequence of momentum transfer from radiation to gas via dust grains. In addition, 
that the abrupt decline in the wind speeds at high temperatures (see Fig. \ref{models1}) is mostly due to inefficient dust formation. At 
high $T_{\rm eff}$ the condensation radius $R_{\rm c}$ is moved out to a distance too far away from the star for any significant amount 
of material to reach this radius. Consequently, almost no dust is formed and therefore no wind either.

\subsection{Mass-loss rate}
\label{massloss}
The classical Reimers (1975) \nocite{Reimers75} formula for the mass loss rate of red giants, 
\begin{equation}
\label{reimers}
\dot{M}_{\rm R} = 1.34\,10^{-5}\eta\, {L_\star^{3/2}\over M_\star T_{\rm eff}^2}, 
%4.0\,10^{-13}\eta\,{L_\star R_\star\over M_\star} = 
\end{equation}
where we have used $L_\star = 4\pi\sigma\,R_\star^2T_{\rm eff}^4$ to eliminate $R_\star$, can be obtained by relating the 
kinetic energy of the outflow to the radiative energy flux and the gravitational potential energy. The Reimers law
was derived strictly for red giants and cannot be applied to AGB stars without significant modifications \cite{Kudritzki78}. 
Fitting a simple function to a set of wind models reveals a stronger dependence on luminosity and effective temperature for carbon stars 
\cite[see, e.g.,][]{Arndt97, Wachter02}. In Fig. \ref{mlr_reimers} one may note a weak correlation between the mass-loss rates 
obtained from our detailed modelling and the Reimers formula, but since the Reimers formula (and commonly used modified 
versions of it, e.g., the formula suggested by Bl\"ocker 1995) \nocite{Blocker95} does not include any dependence on the free carbon abundance, it 
cannot easily be modified so that it would reproduce our results. Using a modified Reimers law to account for the mass-loss during the carbon star 
phase in stellar evolution models is therefore not to be recommended.

In general, the mass loss rate follows a slightly different pattern than the wind velocity. First of all there is an obvious anti-correlation 
between effective temperature $T_{\rm eff}$ and the mass loss rate (see Fig. \ref{models1}), which partly can be interpreted as an effect of the 
$T_{\rm eff}-\log(g)$ correlation. But this anti-correlation is not directly due to the effect of surface gravity, $\log(g)$, on the wind dynamics. 
With increasing $\log(g)$, the star becomes more compact and the outer atmospheric layers become thinner, which means that there will be less 
material available in the dust-formation zone around the condensation radius. With lower density where the dust forms, the mass loss rate may drop, 
since there will be less gas available in the wind forming region.

The parameter configurations we have used here give mass-loss rates mostly within the interval $10^{-7} - 10^{-5}M_\odot$ yr$^{-1}$, which is 
in agreement with mass loss rates derived from observations 
\cite[see, e.g.,][and Fig. \ref{mlr_v} in this paper]{Schoier01}, apart from the lack of models with mass loss rates below 
$10^{-7} M_\odot \mbox{yr}^{-1}$ and the correlation between mass-loss rate and wind speed obtained from observations. This correlation could be an 
effect of the coupling between stellar parameters as the stars evolve on the AGB. But a detailed 
discussion about how to compare the present model grid with observations so that biases from sparse grid sampling and observational selection 
effects are reduced to a minimum, is beyond the scope of this paper. However, we may note that stars will stay much longer in evolutionary 
states that correspond to low mass-loss rates than those that correspond the superwind phase (typically the tip of the AGB) during which the 
mass-loss rate is very high and the evolution is rapid. Hence, one is more likely to find a star with a low mass loss rate. The low mass-loss 
rates may correspond to the transition regions where the mass-loss thresholds appear, and these transitions from no (or very inefficient) mass 
loss to strong, dust driven winds are perhaps not adequately resolved, the way the grid is defined. We have explored some of these transition 
regions in more detail by taking smaller steps in effective temperature and luminosity, which showed that the wind appears to be "switched on" at 
more or less a specific parameter value (see Fig. \ref{threshold}). Whether this steep threshold is a real property of dust driven winds or a 
model artefact is not clear at present, although mass-loss thresholds should exist for physical reasons (see Sect. \ref{mlrthres}). 
It is possible that a smooth transition is more realistic and that the observed stars with low mass-loss rates are sitting in these 
transition regions, which may correspond to very restricted parameter intervals. We will return to this important issue in a forthcoming paper.

The energy injection by pulsations, or more precisely, the kinetic energy density at the inner boundary of the model atmosphere,
also plays an important r\^ole in the wind-formation process. A correlation between this quantity and the mass loss rate obtained for 
specific sets of stellar parameters was pointed out by Mattsson et al. (2008) \nocite{Mattsson08} and stems from the fact that the pulsations 
are needed to levitate the atmosphere such that dust formation can take place. The efficiency of dust formation is therefore more or less 
affected by the pulsation energy. Hence, for AGB stars, it is necessary to consider the kinetic energy from the pulsations as a part of 
the wind-driving mechanism, although it is the momentum transfer from radiation to dust that maintains the outflow. We have tried different 
piston-velocity amplitudes and, as expected, stronger pulsations seem to favour wind formation \cite[see][for further details]{Mattsson08}, 
although such effects of $\Delta u_{\rm p}$ are only important in the critical wind regime (near the mass-loss threshold). The most obvious 
effect of changing $\Delta u_{\rm p}$ is that reducing the strength of the pulsations means that a smaller region in stellar parameter space 
will correspond to stable outflows and vice versa, i.e., changing $\Delta u_{\rm p}$ can shift the locations of the mass-loss thresholds 
\cite[see Fig. 2 in][]{Mattsson07b}, but the typical mass-loss rate as well as the range of 
mass-loss rates in the wind-forming cases are not affected significantly (see Fig. \ref{piston}).

  \begin{figure}
  \resizebox{\hsize}{!}{
  \includegraphics{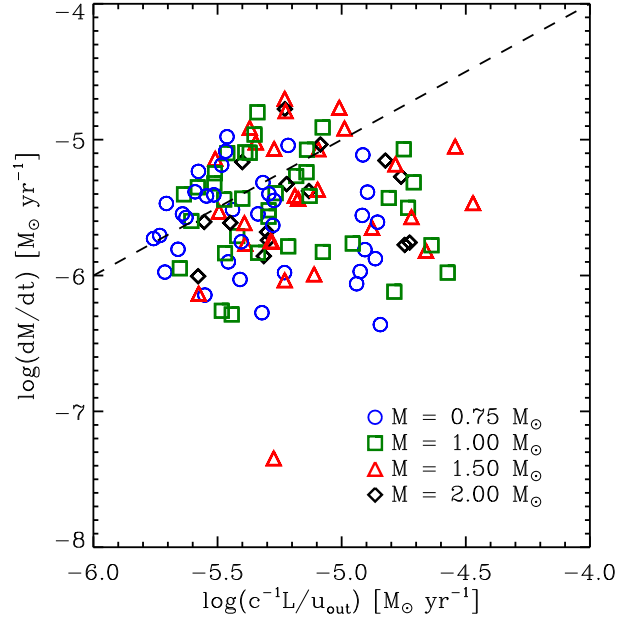}}
  \caption{  \label{sslplot}
  Modelled mass-loss rates vs. the "single-scattering limit" (see Eq. \ref{ssl}). The dashed line marks the case of a one-to-one correspondance. }
  \end{figure}

Comparing the mass-loss rates we obtain with the "single-scattering limit" mass loss, i.e.,
\begin{equation}
\label{ssl}
\dot{M}_0 \equiv {1\over c}{L_\star \over u_{\rm out}},
\end{equation}
where $c$ is the speed of light,
we find that a fair number of models where the mass-loss rate is higher than $\dot{M}_0$, which can be seen in Fig. \ref{sslplot}. The quantity 
$\dot{M}_0$ is, however, not a proper upper limit for "single scattering" due to the assumptions made in deriving it (see, e.g.,
Lamers \& Cassinelli 1999). The optical depth of the wind may very well be larger than unity around the flux peak in some 
cases, even if the total optical depth is not. Furthermore, it has been shown by Gail \& Sedlmayr (1986) that heavily obscured objects must have
mean optical depths that exceed unity, so the result in Fig. \ref{sslplot} is hardly surprising.
Furthermore, the time-dependent luminosity $L$ may not be represented properly by the luminosity $L_\star$ of the static start model used in Eq. 
(\ref{ssl}) and in Fig. \ref{sslplot}, since $L$ is phase dependent in the dynamical models and wind acceleration may occur within a limited
range of phases in a given model, thus changing the "effective luminosity.

\subsubsection{Dust-to-gas ratios}
\begin{figure*}
  \resizebox{\hsize}{!}{
  \includegraphics{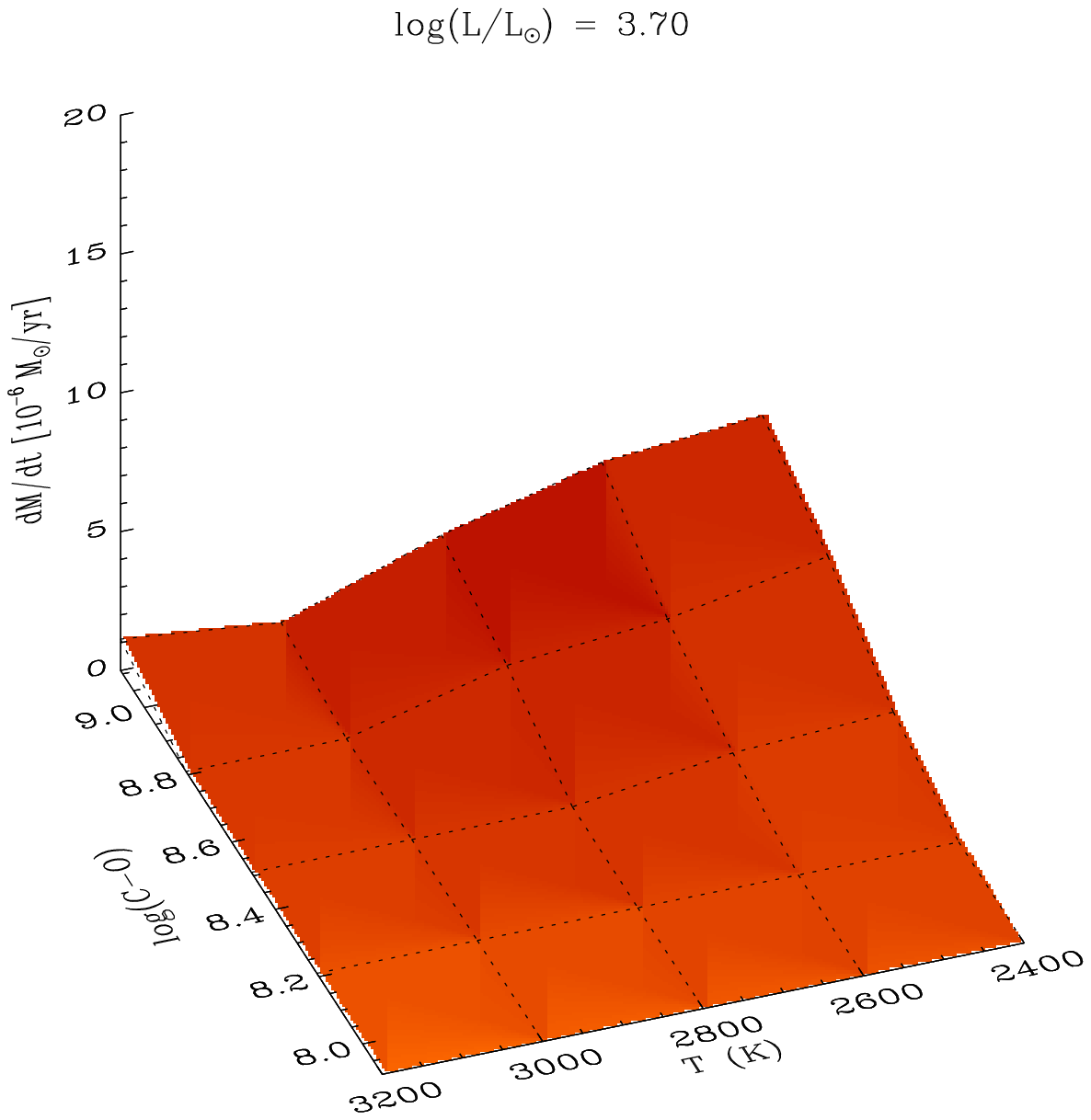}
  \includegraphics{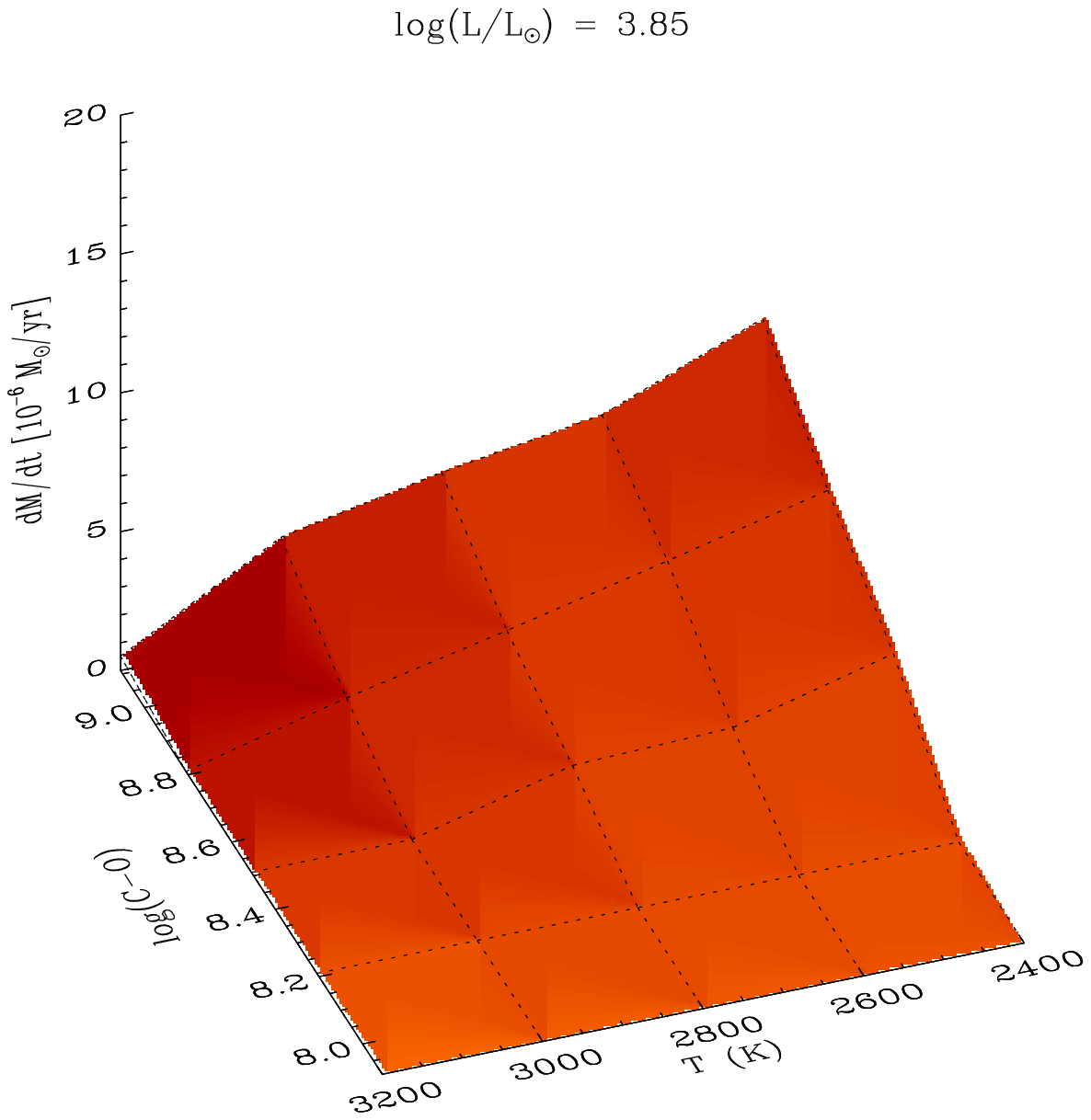}
  \includegraphics{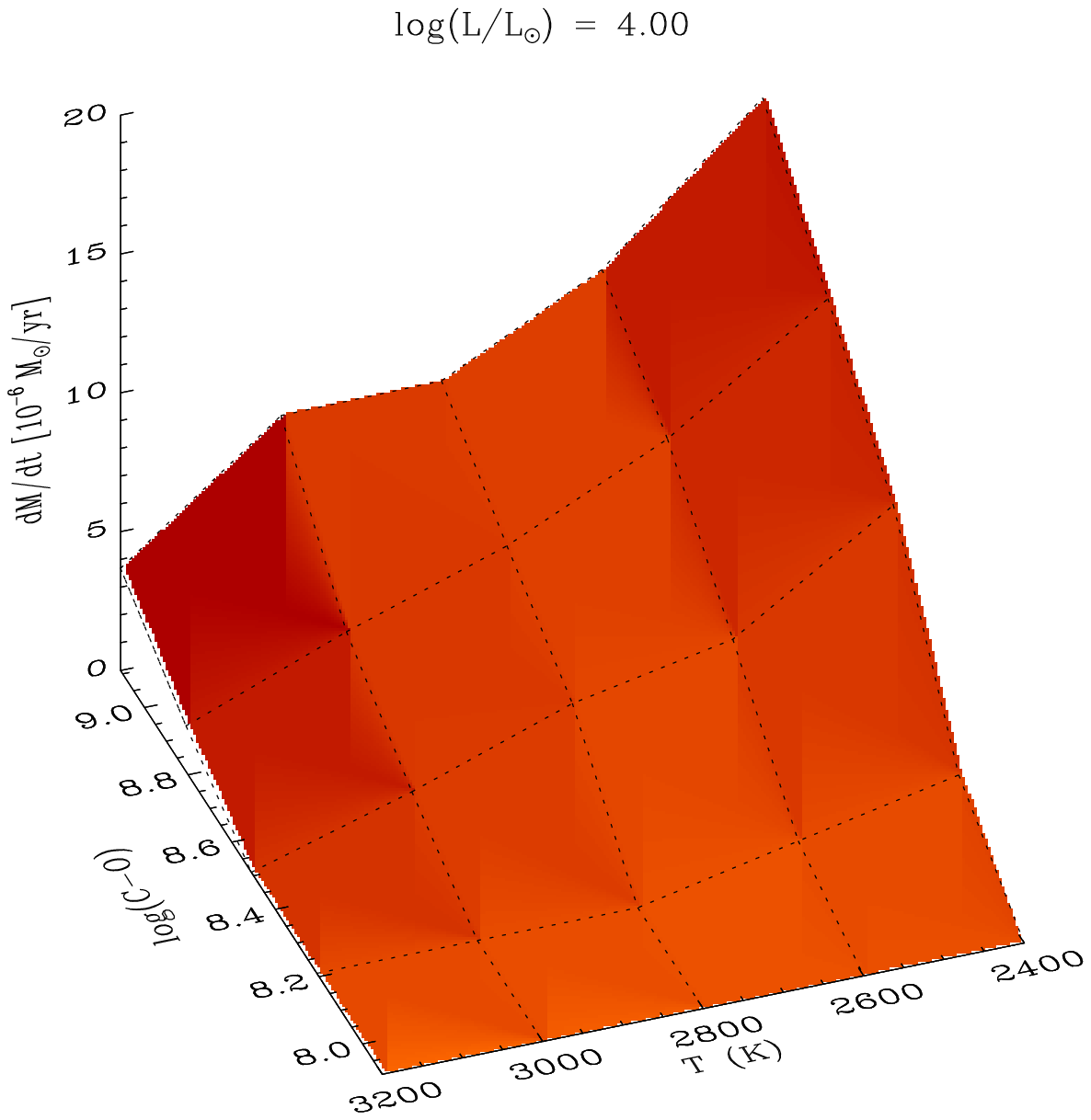}}
  \resizebox{\hsize}{!}{
  \includegraphics{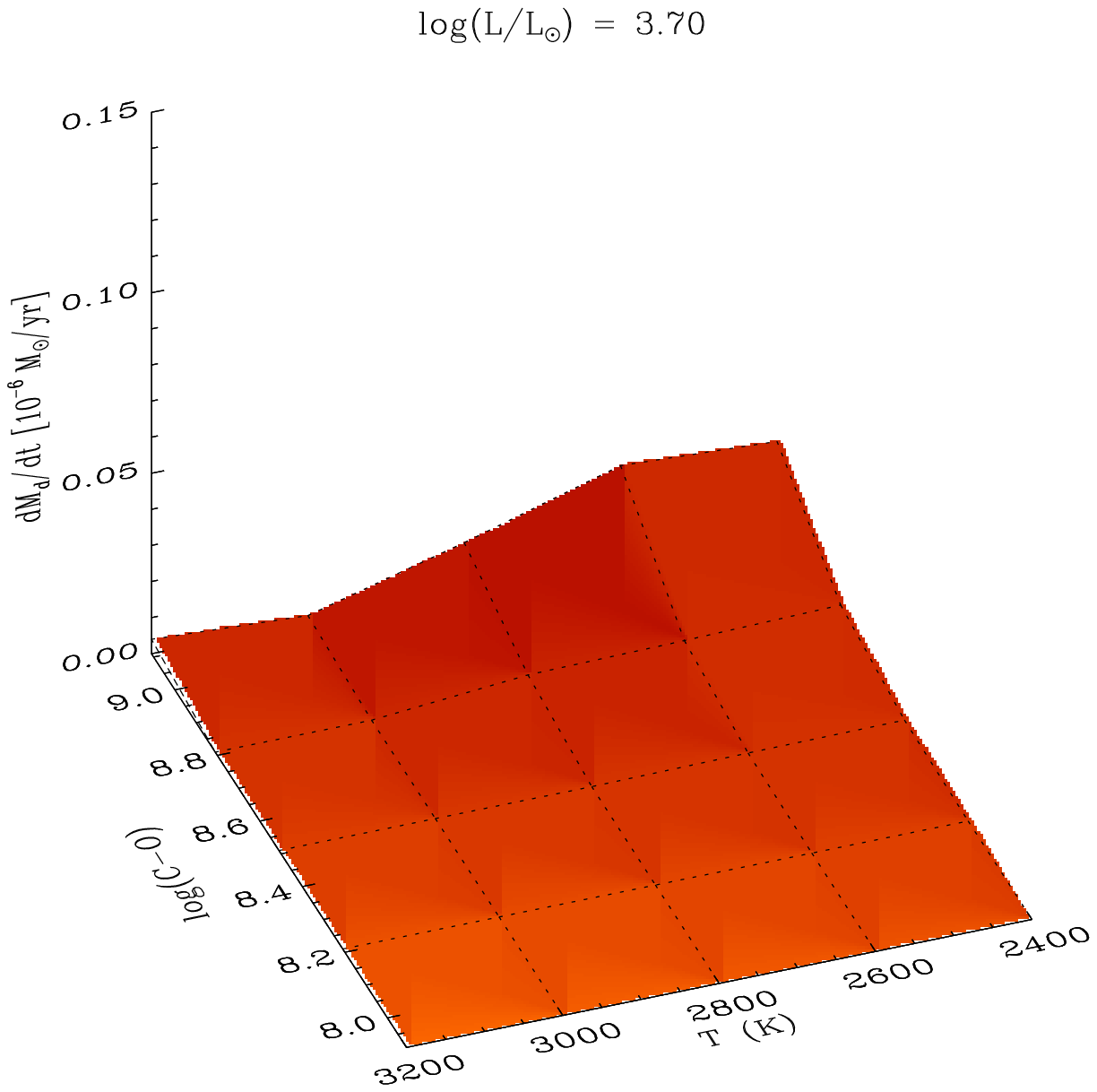}
  \includegraphics{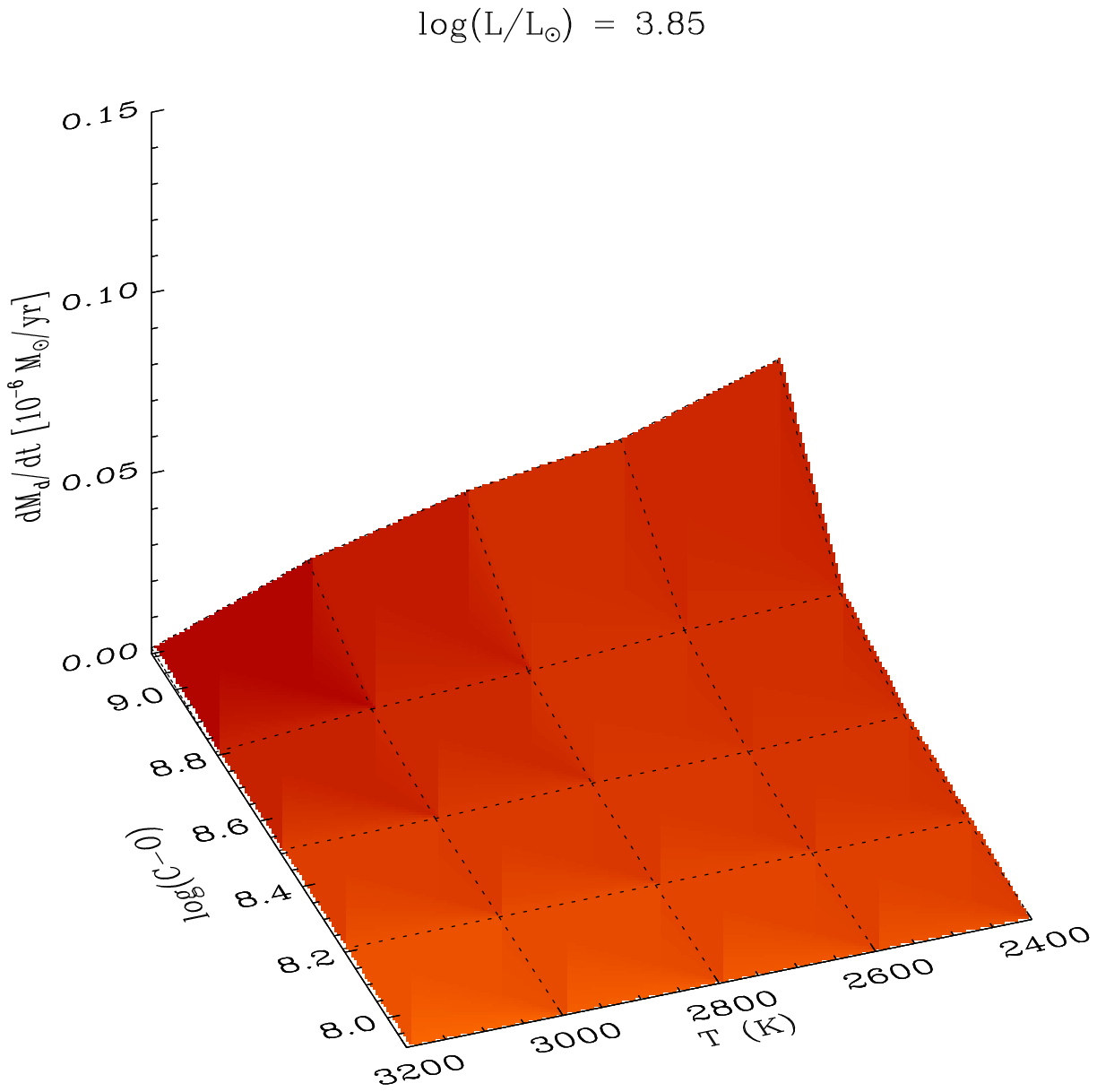}
  \includegraphics{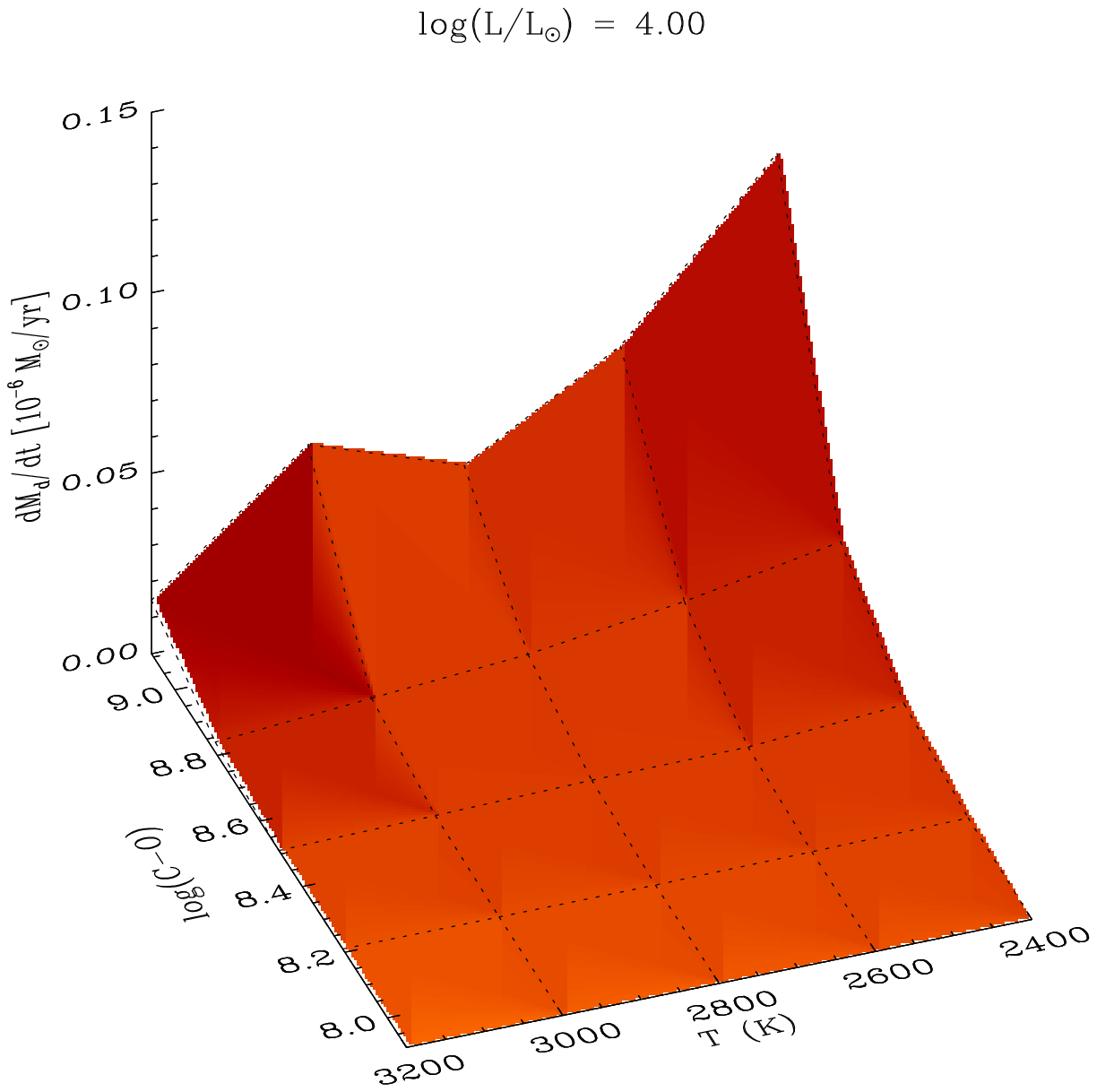}}
  \resizebox{\hsize}{!}{
  \includegraphics{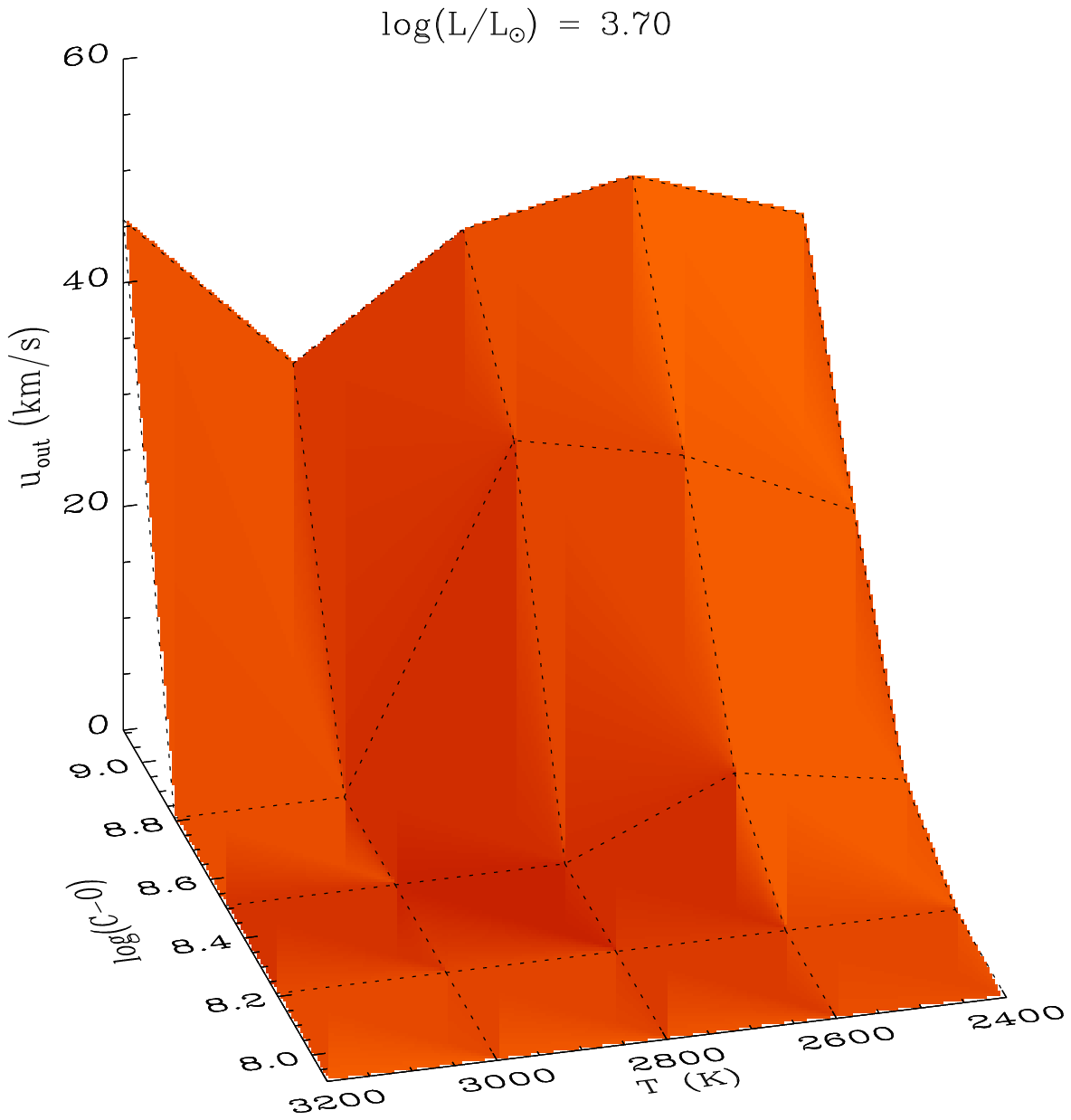}
  \includegraphics{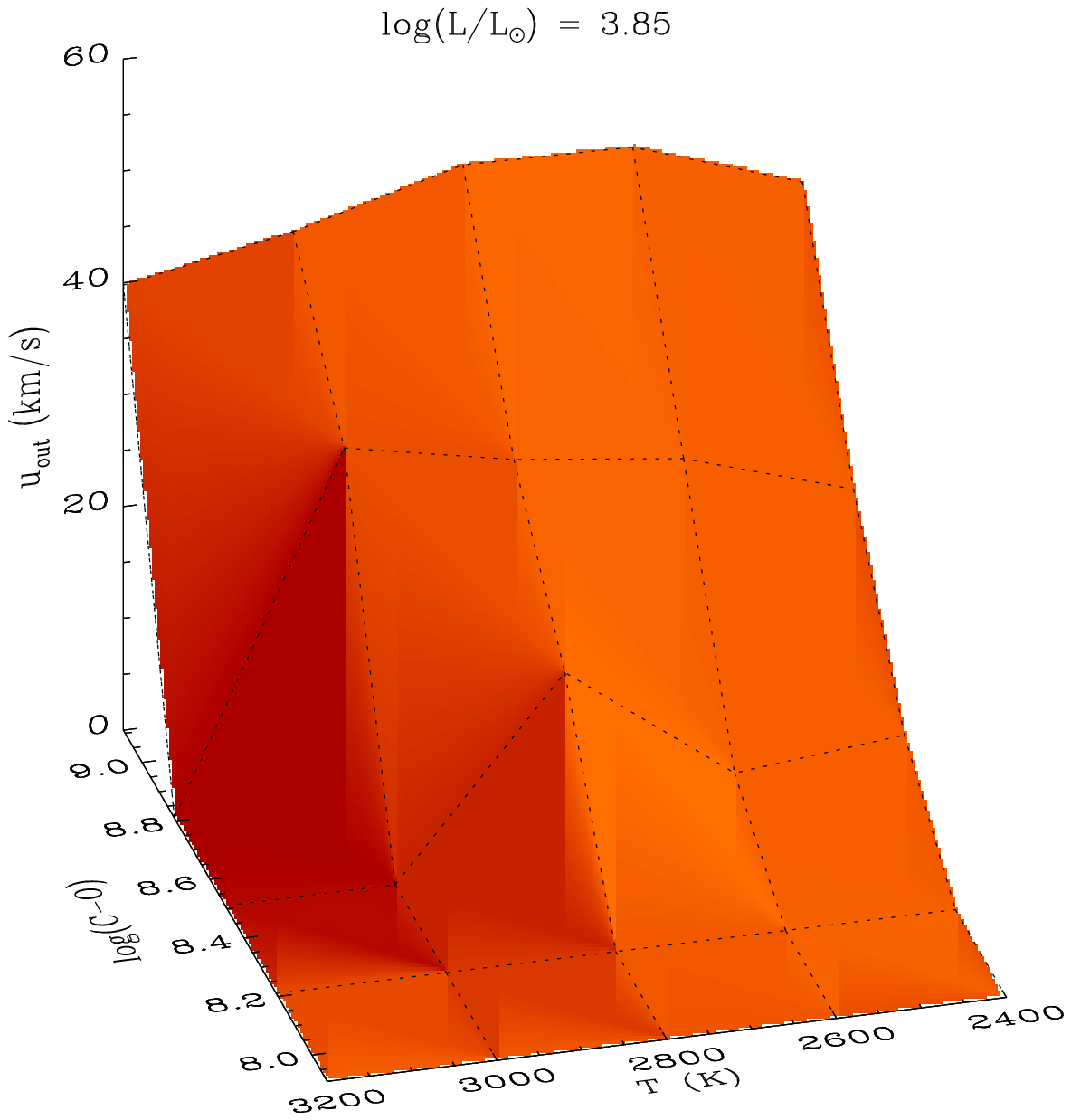}
  \includegraphics{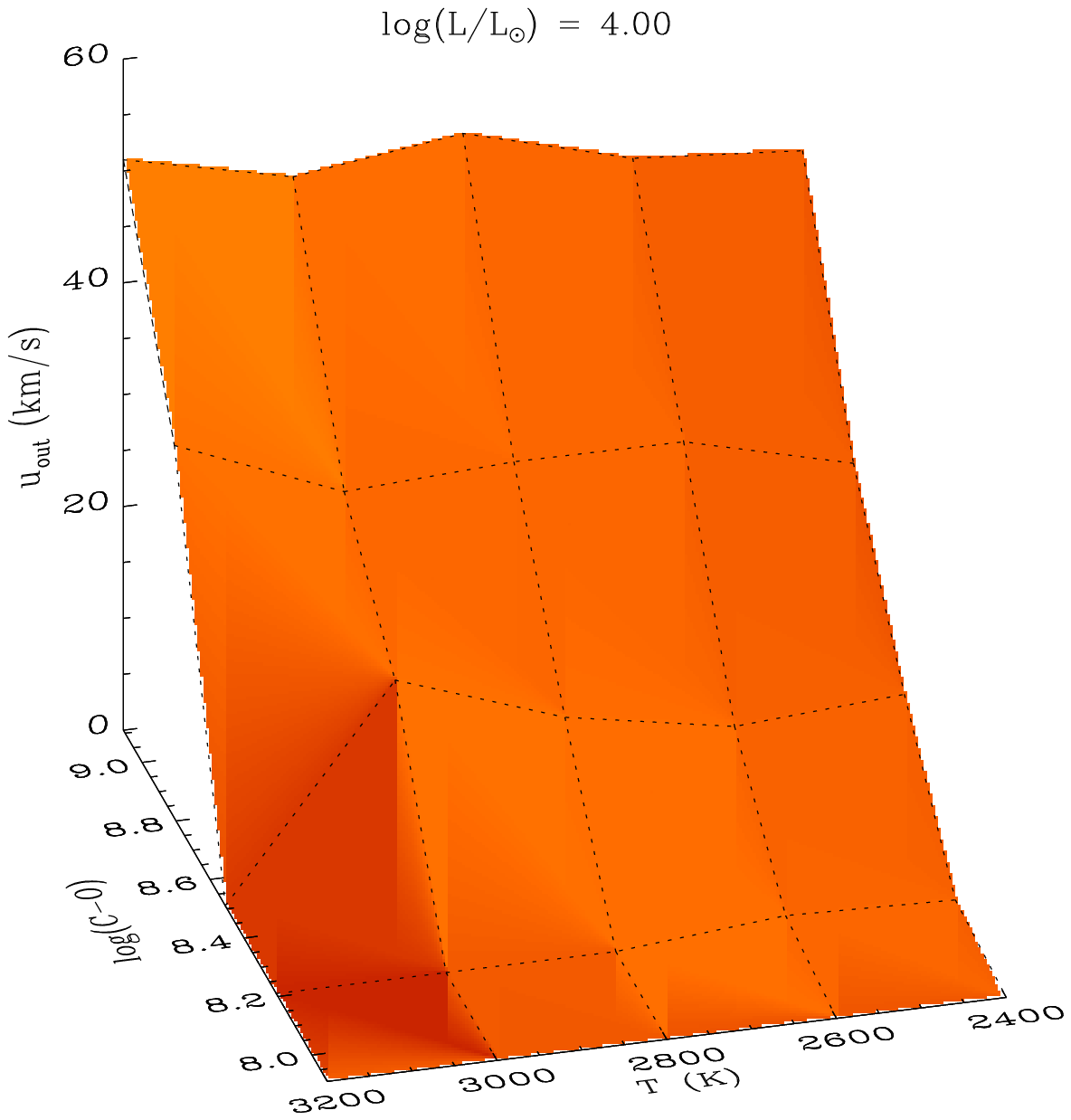}}
  \caption{  \label{models1}
  Mean mass-loss rate (upper panels), mean dust-loss rate (middle panels) and wind speed (lower panels) as functions of effective temperature and 
  the abundance of condensible carbon for models with $M_\star = 1.0 M_\odot$ and
  $\Delta u_{\rm p} = 4$ km s$^{-1}$.
  }
  \end{figure*}

\begin{figure*}
  \resizebox{\hsize}{!}{
  \includegraphics{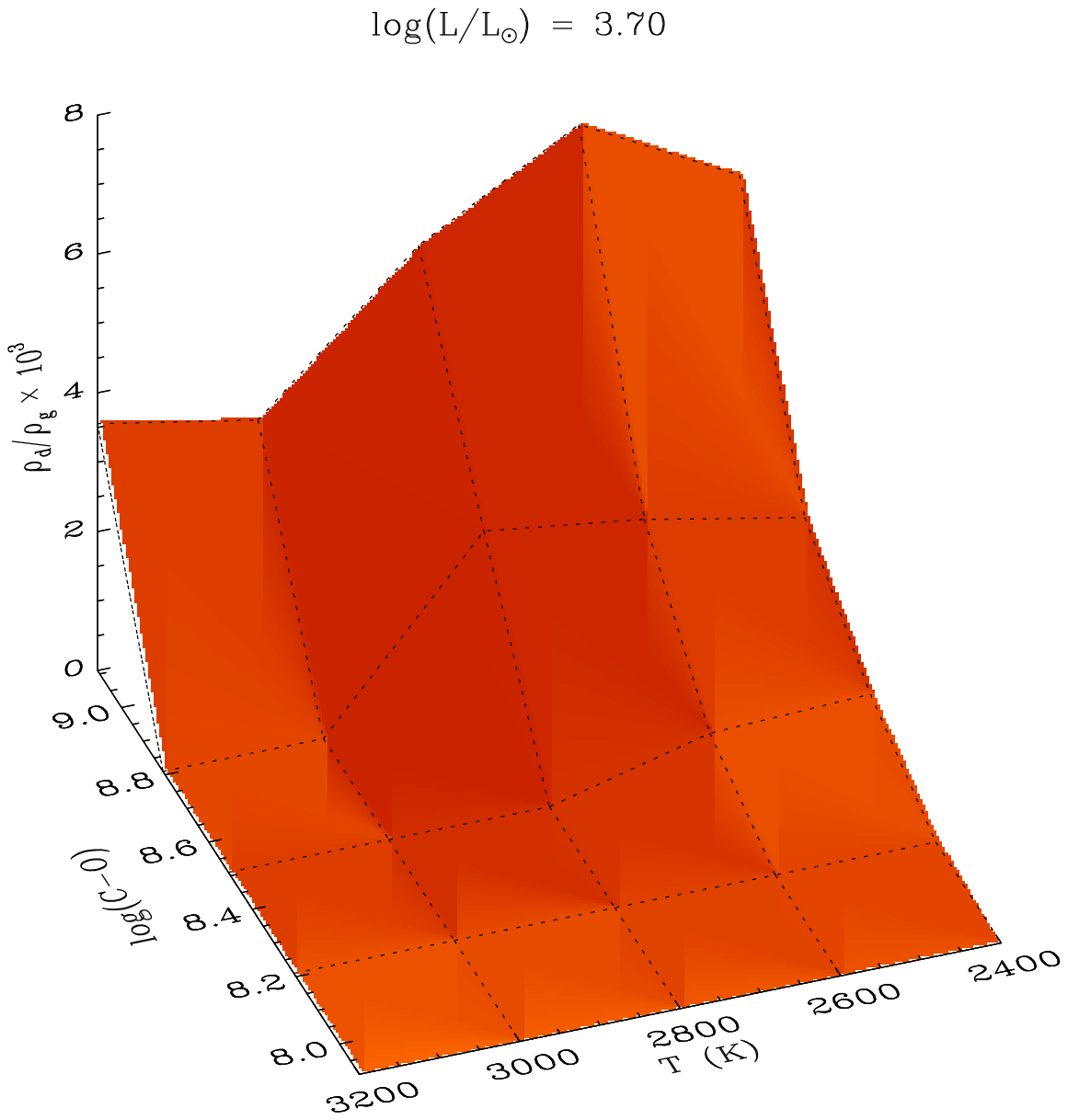}
  \includegraphics{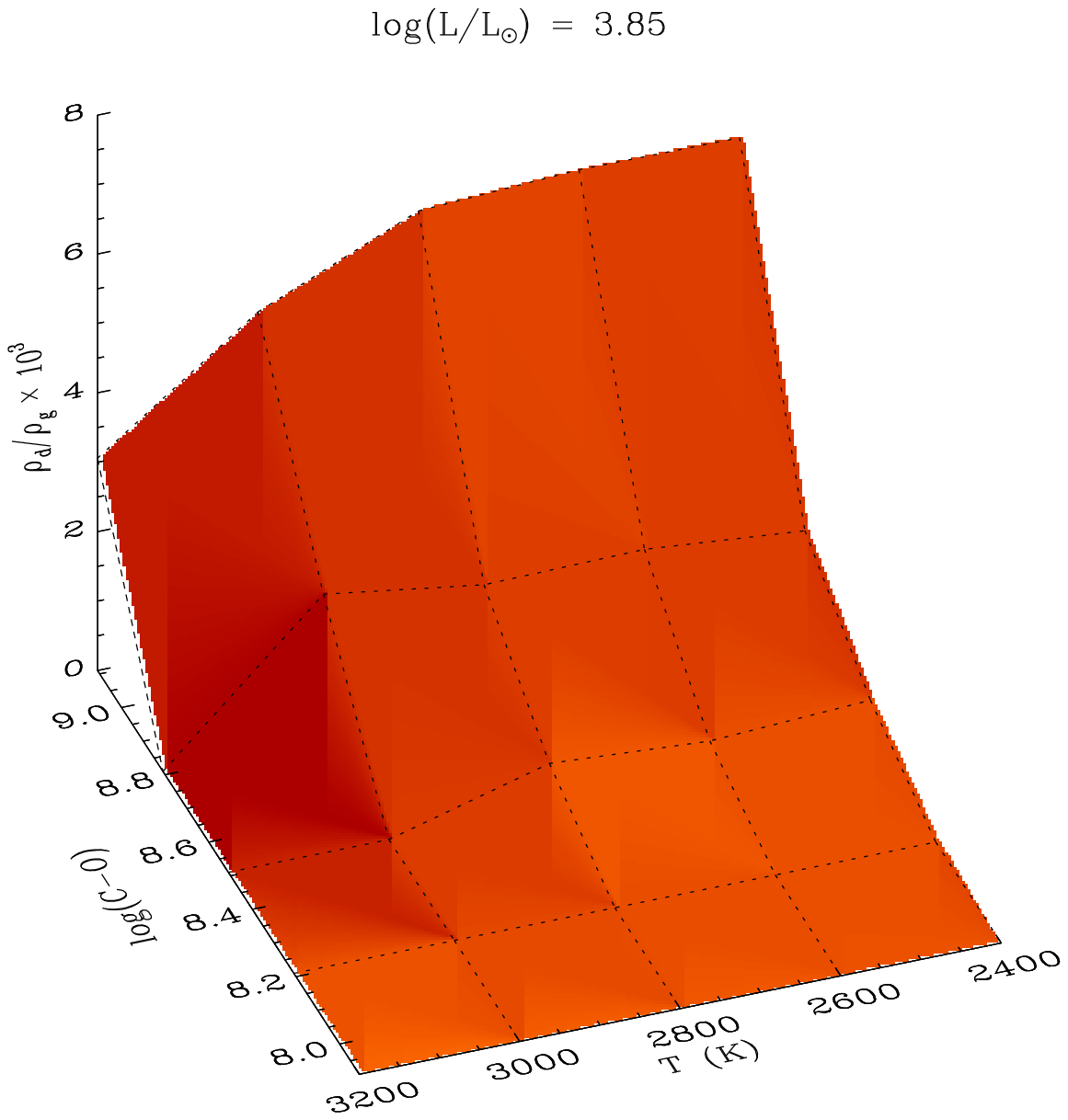}
  \includegraphics{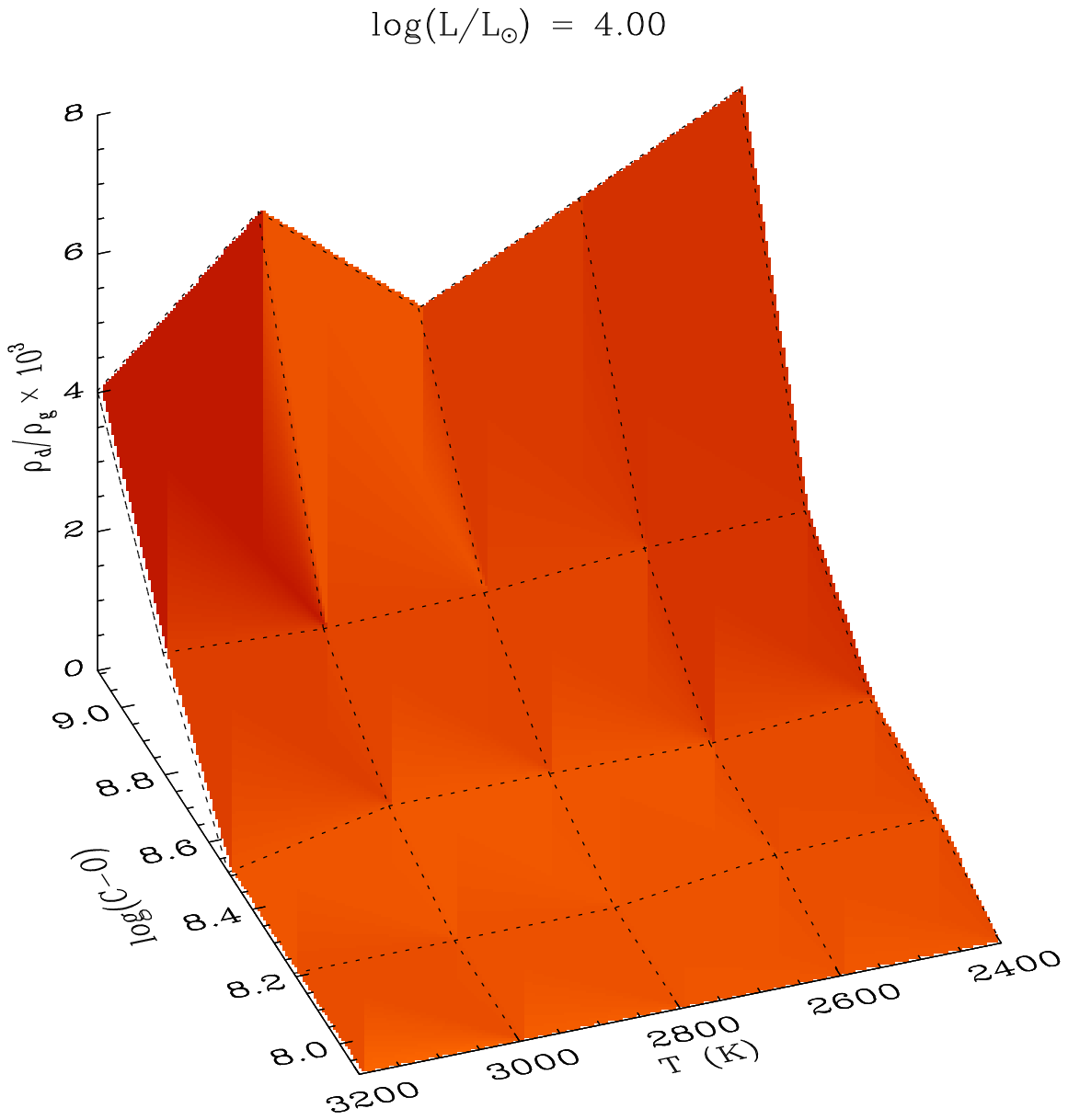}}
  \resizebox{\hsize}{!}{
  \includegraphics{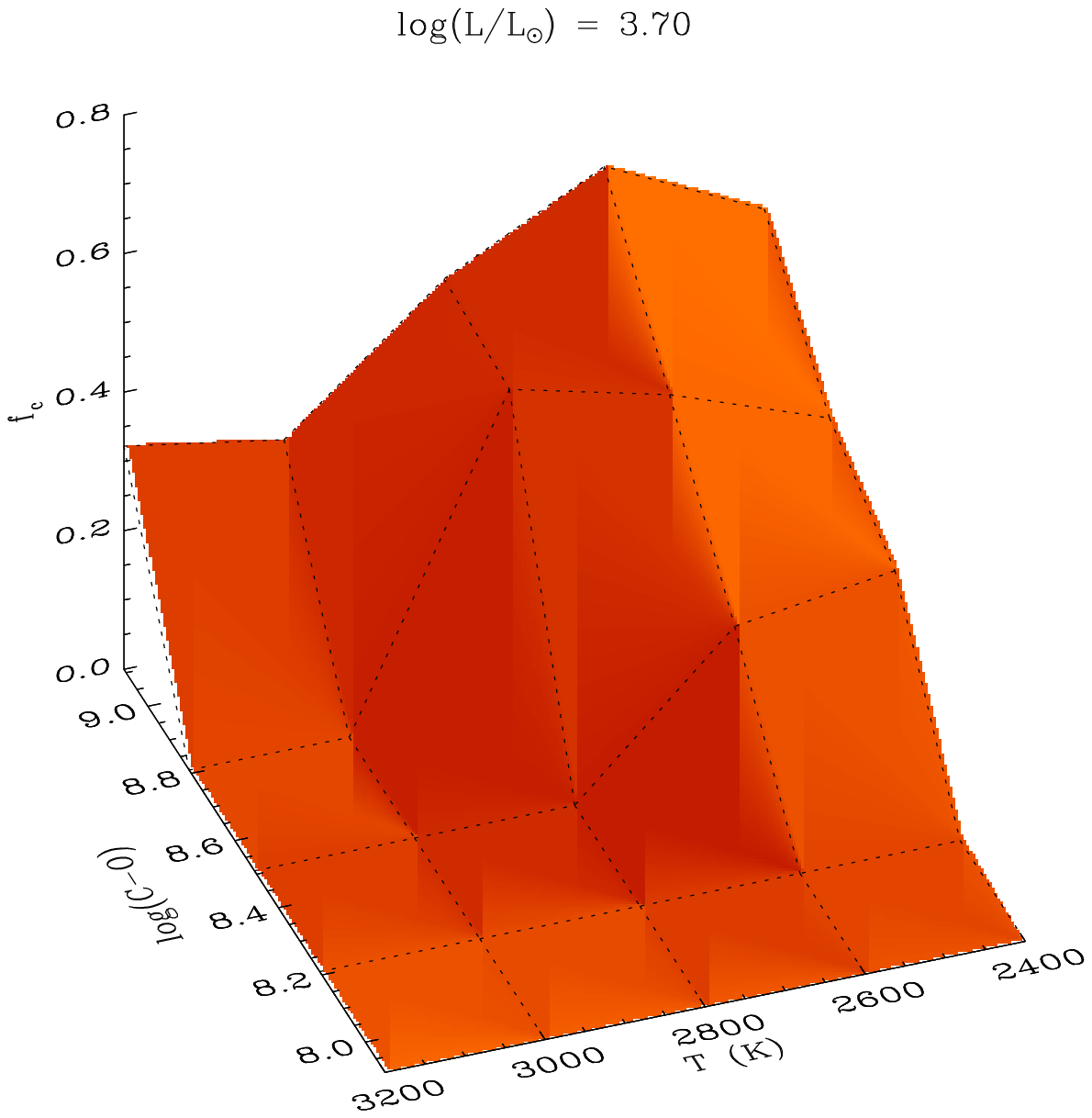}
  \includegraphics{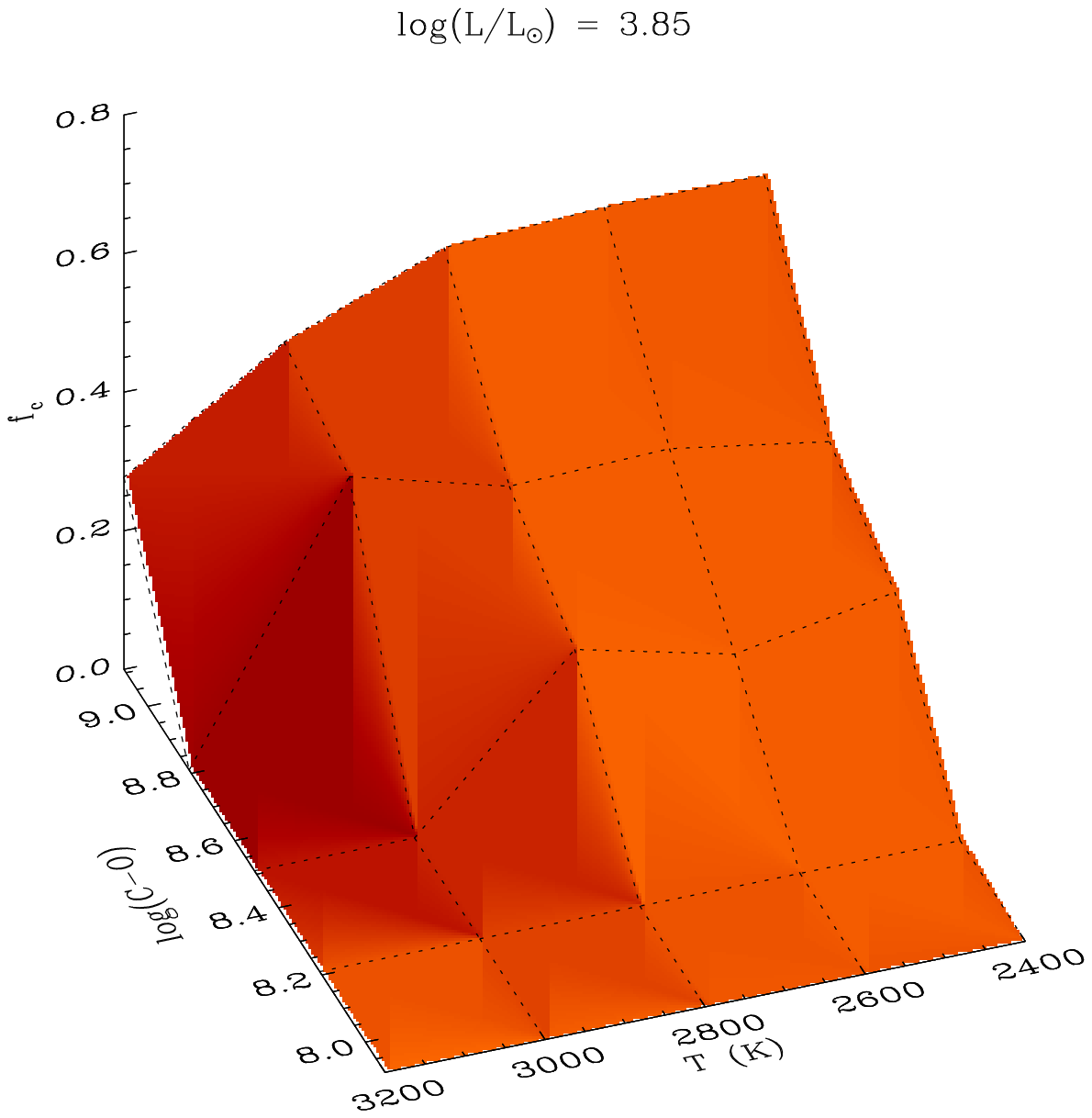}
  \includegraphics{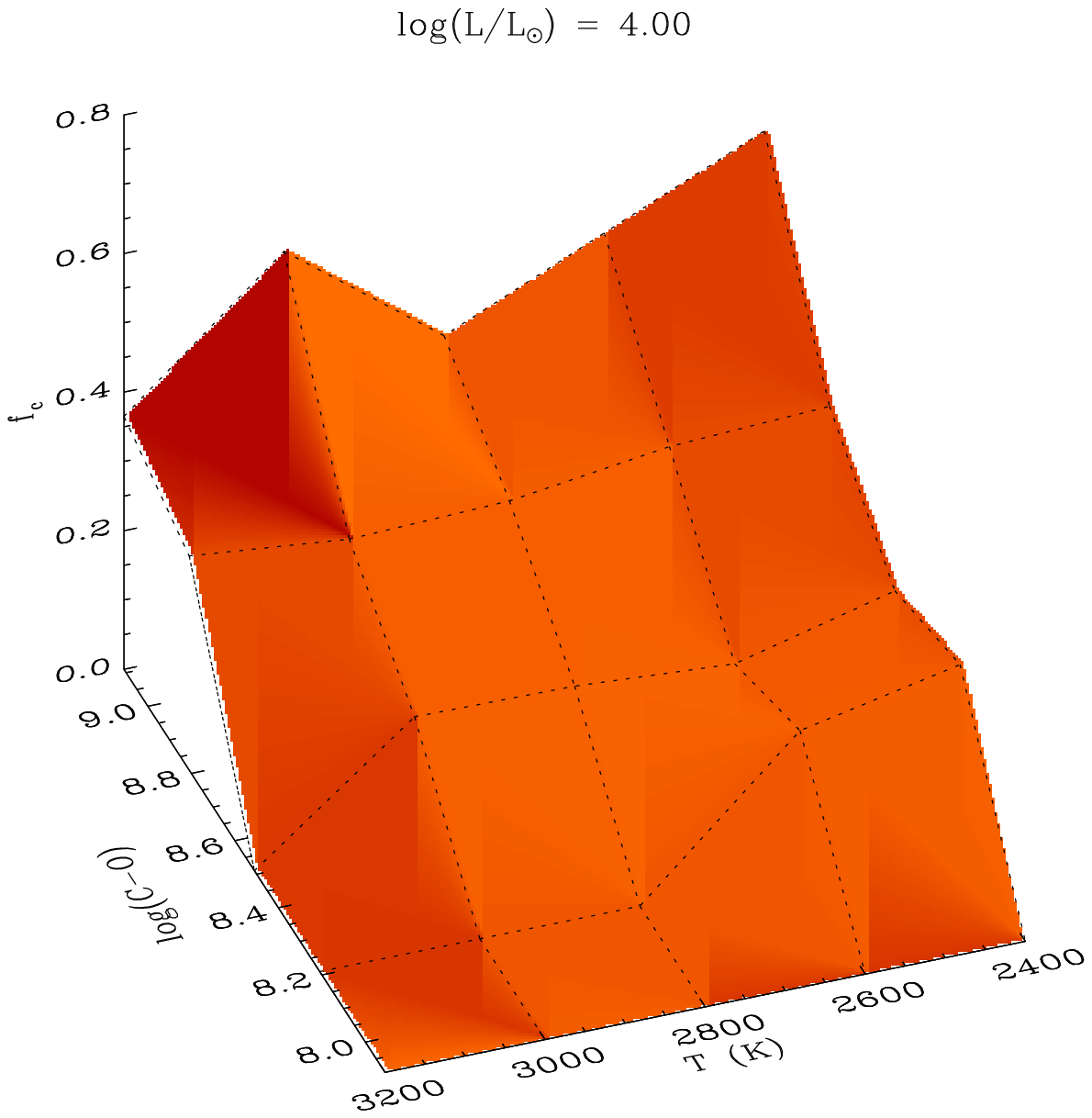}}
  \caption{  \label{models2}
  Mean dust-to-gas ratio (upper panels) and degree of dust condensation (lower panels) as functions of effective temperature and the 
  abundance of condensable carbon for models with $M_\star = 1.0 M_\odot$ and
  $\Delta u_{\rm p} = 4$ km s$^{-1}$.
  }
  \end{figure*}

The dust opacity scales with the relative dust abundance $\rho_{\rm d}/\rho_{\rm g}$. Thus, we should expect the wind 
velocity to behave in a way similar to $\rho_{\rm d}/\rho_{\rm g}$, which is actually the case if one compares Fig. \ref{models1} with 
Fig. \ref{models2}. According to observations \cite[e.g.,][]{Schoier01,Bergeat05}, the $\rho_{\rm d}/\rho_{\rm g}$-ratio is nearly 
independent of $T_{\rm eff}$. This fairly consistent with the weak $T_{\rm eff}$-dependence that we obtain. However, the clearly 
increasing trend with the abundance of condensible carbon $\tilde{\varepsilon}_{\rm C}$ is not seen in observations \cite[see, e.g.,][]{Bergeat05} 
although the existence of such a trend should be rather obvious since $\rho_{\rm d} \propto \tilde{\varepsilon}_{\rm C}$.

Because the stellar luminosity and mass is kept constant in Fig. \ref{models2}, the plotted dust-to-gas ratio will also tell us how the acceleration 
ratio $\alpha$ varies with $T_{\rm eff}$ and the abundance of free carbon, $\tilde{\varepsilon}_{\rm C}$. Below a certain $T_{\rm eff}$, the 
dust-to-gas ratio $\rho_{\rm d}/\rho_{\rm g}$ depends mainly on $\tilde{\varepsilon}_{\rm C}$, and we may thus conclude that $\alpha$ is to a 
large extent set by $\tilde{\varepsilon}_{\rm C}$ for given a mass and luminosity.  

\subsubsection{Average grain size}
We return now to the grain-size issue and whether the small-particle limit actually applies to the dusty atmospheres of carbon stars. The 
characteristic grain radius can be computed as
\begin{equation}
a_{\rm gr} = r_0 {K_1\over K_0},
\end{equation}
where $r_0$ is the monomer radius and $K_0$, $K_1$ are the zeroth and first moment of the grain size distribution function, respectively. Using this
definition of grain radius, we obtain a typical grain radius of $a_{\rm gr} \sim 10^{-5}$ cm in the set of models with $M_\star = 1M_\odot$, 
$\Delta u_{\rm p} = 4.0$ km s$^{-1}$, that produce winds. The small particle approximation is assumed to be applicable as long as the true radiation
pressure efficiency factor $Q_{\rm rp}$ does not deviate from that of the approximation by more than 10\% at $1\mu$m. Thus, the small-particle limit 
requires that $a_{\rm gr} < 10^{-5.4}$ cm, i.e, not larger than a few times $10^{-6}$ cm (see Fig. \ref{smallpart}), which indicates that the grains 
forming in our models are on average a little too large for $Q_{\rm rp}$ to be well-approximated by the small-particle limit. The wind speed is 
anti-correlated with dust-grain size (see Fig. \ref{agr}), which we interpret as a simple consequence of the fact that for a slow wind the grains 
stay longer in the dust-formation zone and thus they have time to grow bigger.

Since $Q_{\rm rp}$ (which describes the net effect of momentum transfer by absorption and scattering for dust) will level-out to a constant above 
a certain grain radius, i.e., $Q_{\rm rp}$ is only proportional to the grain size for small grains, our model (which is based on the small particle
approximation of $Q_{\rm rp}$, irrespective of grain size) will tend overestimate the dust opacity for really large particles. However, in the 
transition region, where $Q_{\rm rp}$ goes from being proportional to the grain size to being almost independent of the grain size, the 
small-particle approximation will instead {\it underestimate} the extinction by as much as a factor of five (again, see Fig. \ref{smallpart}). 
Since there is a competition between nucleation and grain growth, we expect this underestimation to be important only in critical cases with slow 
winds. The majority of the grid models, however, show grain sizes that correspond to the 
transition region where $Q_{\rm rp}$ is likely underestimated. Exactly how this may affect the mass-loss rates and other results in this paper 
is difficult to estimate and needs further investigation. 

  \begin{figure*}
  \resizebox{\hsize}{!}{
  \includegraphics{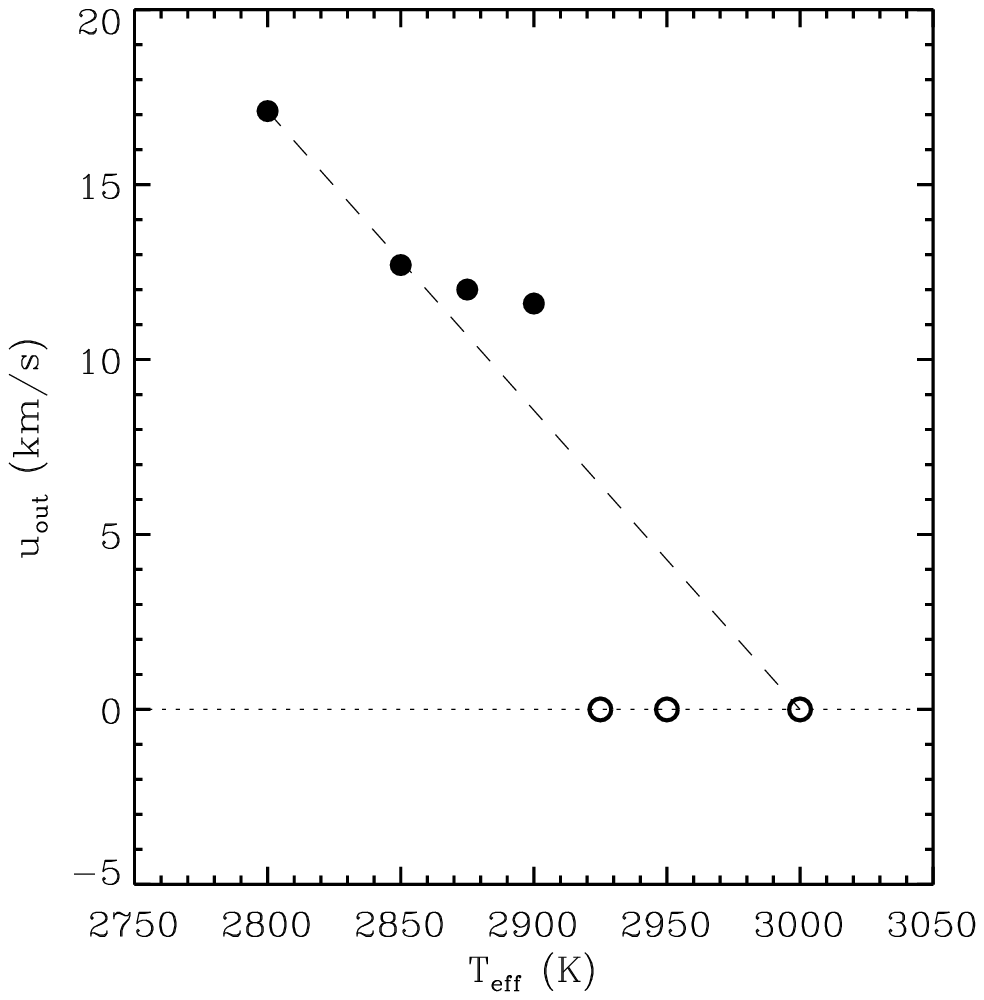}
  \includegraphics{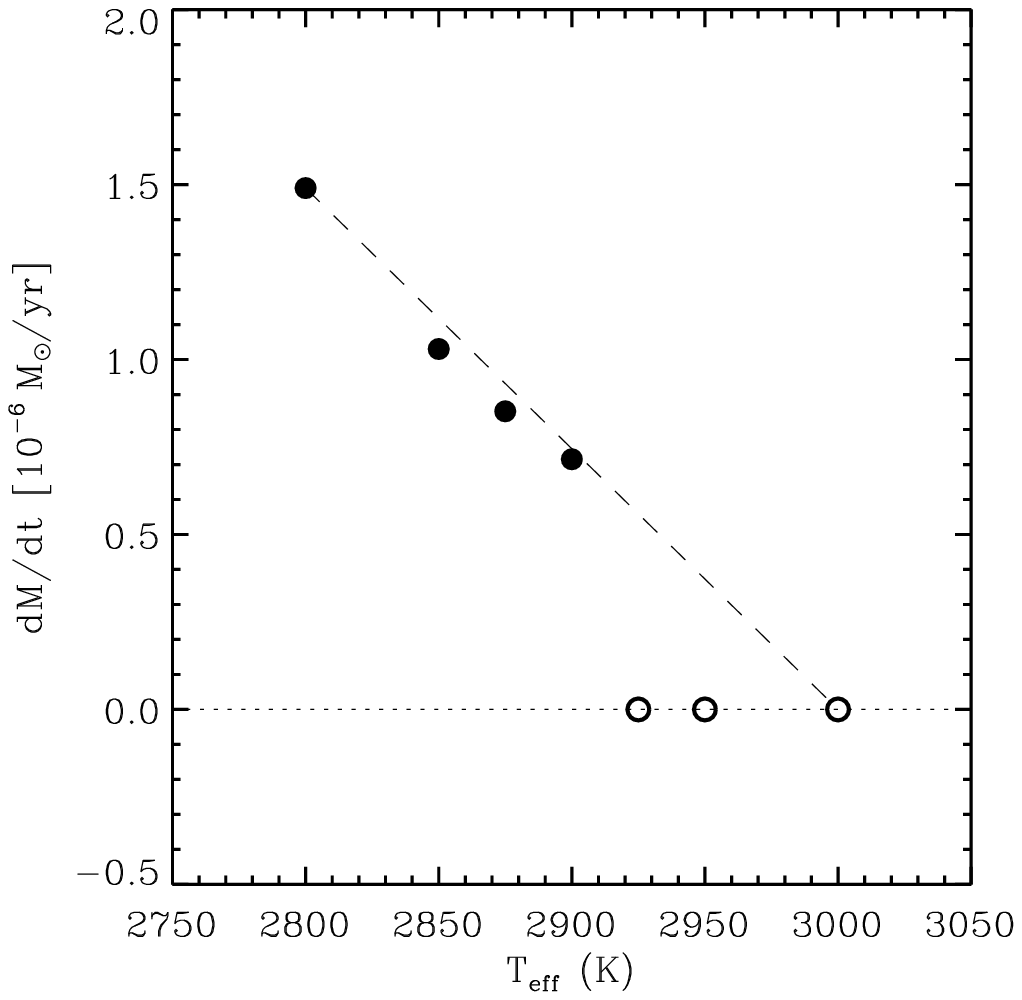}}
  \caption{  \label{threshold}
  Wind speed (left) and mass-loss rate (right) as functions of the effective temperature near a threshold point in the grid. Note that linear
  extrapolation of the mass-loss rate may be adequate, since the steep threshold could be an effect of the approximations and limitations 
  of the model.}
  \end{figure*}

  \begin{figure}
  \resizebox{\hsize}{!}{
  \includegraphics{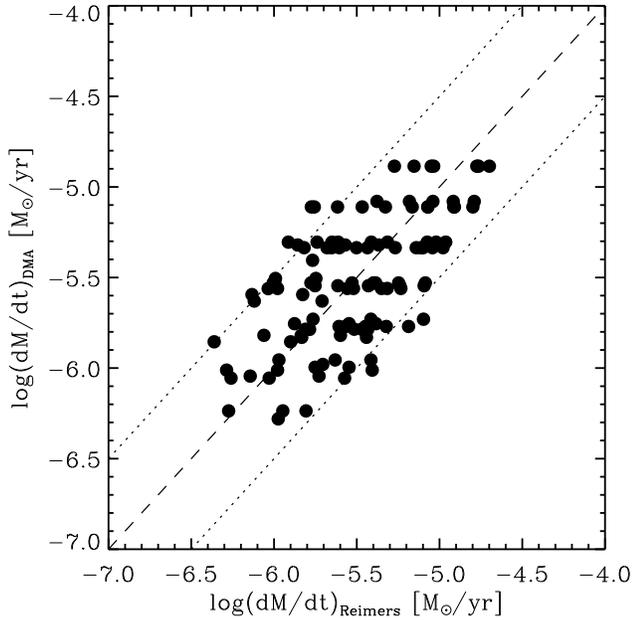}}
  \caption{  \label{mlr_reimers}
  Comparison between our mass-loss models (with $\Delta u_{\rm p} = 4.0$ km s$^{-1}$) and the Reimers (1975) formula with $\eta = 2.5$. The dashed 
  line marks a one-to-one relation and the dotted lines indicate a $\pm 0.5$ dex spread.}
  \end{figure}

\subsection{Mass-loss thresholds}
\label{mlrthres}
Mattsson et al. (2007b) \nocite{Mattsson07b} demonstrated the necessity of considering thresholds in mass-loss prescriptions for dust-driven winds.
As we mentioned above, a mass-loss threshold appears, as one would expect, and we find that 
for $\tilde{\varepsilon}_{\rm C} < 8.20$ and/or $T_{\rm eff} > 3200$ K a dust driven does not seem to form regardless of the combination of 
other stellar parameters (see Fig. \ref{models1} and Fig. \ref{models2}, as well as the additional online material). Since the surface gravity 
scales with $T_{\rm eff}$ and dust formation is a process highly sensitive to temperature, it is expected that the mass-loss rate is 
highly dependent on $T_{\rm eff}$ too. But, as shown above, there is a rather strong dependence on condensible carbon for both the wind velocity 
and the mass-loss rate, which is quite interesting in comparison with previous studies of this kind. Arndt et al. (1997) \nocite{Arndt97} as 
well as Wachter et al. (2002) \nocite{Wachter02} find a rather weak dependence on ${\rm C/O}$ (which is their choice of parameter). Their 
results stand in quite sharp contrast to the results presented here. However, our findings here are, qualitatively speaking, hardly a new 
discovery. H\"ofner and Dorfi (1997) \nocite{Hofner97} and Winters (2000) \nocite{Winters00} have already pointed out the C/O-dependence, 
especially in the critical wind regime, although this has not been widely recognised.

The mass-loss thresholds originate from the simple fact that there exists a critical acceleration ratio $\alpha_{\rm crit}$ and that there 
consequently also exist critical values for $\tilde{\varepsilon}_{\rm C}$, $T_{\rm eff}$ and $L_\star$ for which $\alpha \equiv \alpha_{\rm crit}$. 
The principles behind this may easily be understood from basic physics, but none of the existing parametric formulae for dust-driven mass loss 
(e.g. Arndt et al. 1997, Wachter et al. 2002) contain any thresholds. These transition regions are perhaps not adequately covered due to the
grid spacing we have chosen and the sudden onset of an outflow discussed earlier may partly be due to model assumptions. But, the existence of 
mass-loss thresholds as such is a fact, and should definitely be included in a mass-loss prescription for stellar evolution modelling, even 
if it is unlikely that one can find a sufficiently simple functional form that will cover all aspects of how mass loss depends on stellar 
parameters. 

The threshold due to the amount of carbon excess implies that Galactic carbon stars with low observed C/O ratios should typically show no or 
very little mass loss. Some support for this idea comes from the work by Lambert et al. (1986), where about 50\% of their sample of carbon stars
showed C/O~$<1.1$ (all stars in their sample have roughly solar oxygen abundances), and a fair number of those stars do not have any detected 
outflows. A puzzeling fact, however, is that many stars with C/O~$\sim 1$ do have rather strong winds \cite{Ramstedt06}, observations which cannot 
be properly explained at present.

\subsection{No simple formula!}
Consider two stars with the same mass, luminosity and effective temperature. Increasing the amount of 
condensible carbon $\tilde{\varepsilon}_{\rm C}$ by a factor of 5 may increase the mass loss rate by almost an order of magnitude 
(see Fig. \ref{models1}). This may not appear as a huge problem in a log-log-plot, as in Fig. \ref{mlr_reimers} where the grid models
are compared to a scaled Reimers (1975) law, even if $\tilde{\varepsilon}_{\rm C}$ is not a parameter in Reimers' formula. But note that there is 
a 0.5 dex scatter in Fig. \ref{mlr_reimers} (dotted lines) which actually makes a considerable difference for, e.g., stellar evolution modelling,
where $\tilde{\varepsilon}_{\rm C}$ changes during the carbon-star phase, causing a different mass-loss evolution and the occurrence of a superwind.
A direct fit to the whole set of data, including both $\tilde{\varepsilon}_{\rm C}$ and $\Delta u_{\rm p}$ as fitting parameters does not really
fix the problem, since statistically preferred correlations (formulae obtained through, e.g., least-square fitting or $\chi^2$-minimisation) only 
show a rather weak dependence on these two parameters \cite[see][]{Arndt97, Wachter02}. On the other hand, if one keeps all other parameters fixed 
and then varies $\tilde{\varepsilon}_{\rm C}$ or $\Delta u_{\rm p}$, the mass-loss rate is definitely strongly affected and it therefore seems 
dangerous to conclude that these two parameters can be omitted in the prescription as Arndt et al. (1997) and  Wachter et al. (2002) suggested.

For the reasons given above, {\it we do not recommend that formulae derived from the data given in this paper are used to replace the 
actual grid}. Instead, using the tables (on-line material) to create a look-up matrix for the mass loss\footnote{A FORTRAN routine for this purpose 
is available on-line at http://coolstars.astro.uu.se} (combined with multi-dimensional 
interpolation) in stellar evolution modelling will guarantee that detailed features (such as the increase of the mass loss rate as the amount of 
condensible carbon increases) will be properly included, while simplified formulae will not. As an example, Mattsson et al. (2007) showed that, 
as the amount of condensible carbon changes rapidly (relatively speaking, often more rapidly than luminosity, effective temperature etc.) during the 
TP-AGB phase, simple prescriptions of mass loss may therefore not be correct, since that parameter (condensible carbon) is usually not included. 
Note, however, that the mass-loss prescription we provide here best describes the superwind of carbon stars (at the tip of the AGB) and winds
corresponding to the early/mid AGB (where mass-loss rates are much lower) may not always be accurately covered (see Sect. \ref{mlrthres} and below).

  \begin{figure*}
  \resizebox{\hsize}{!}{
  \includegraphics{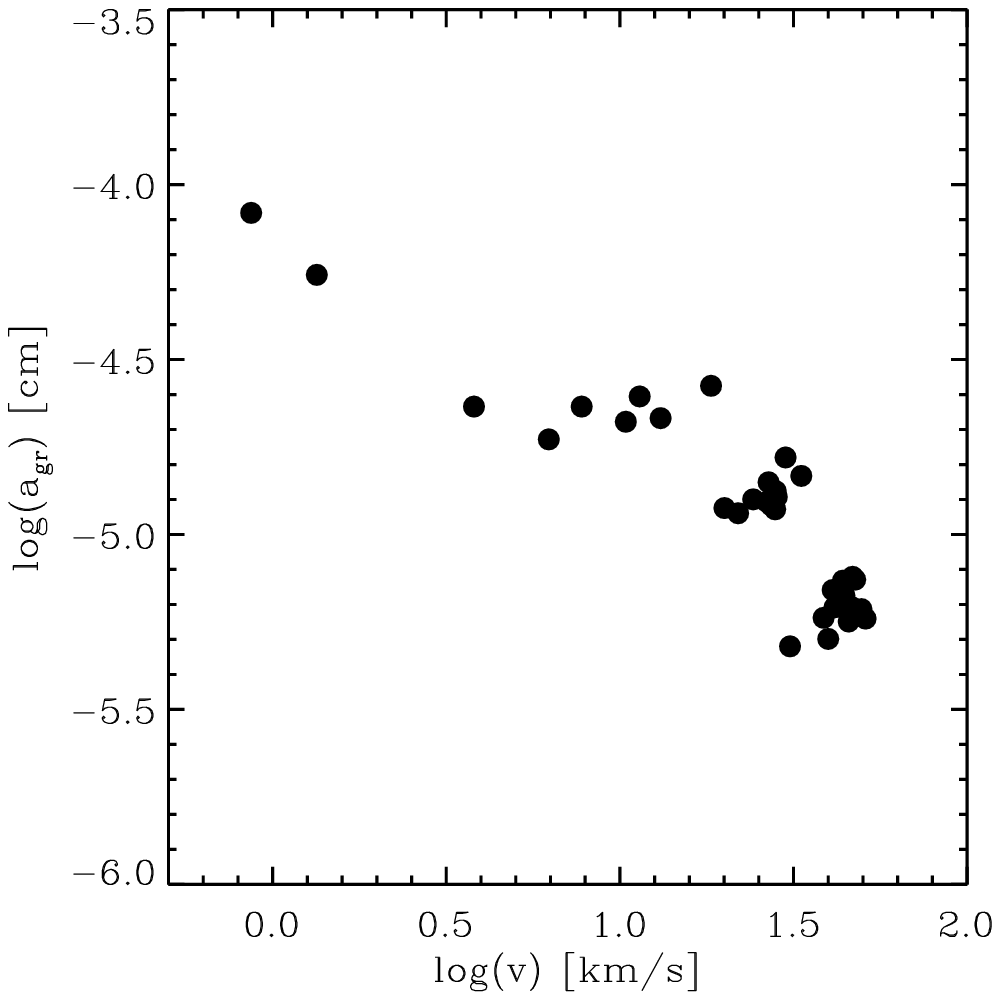}
  \includegraphics{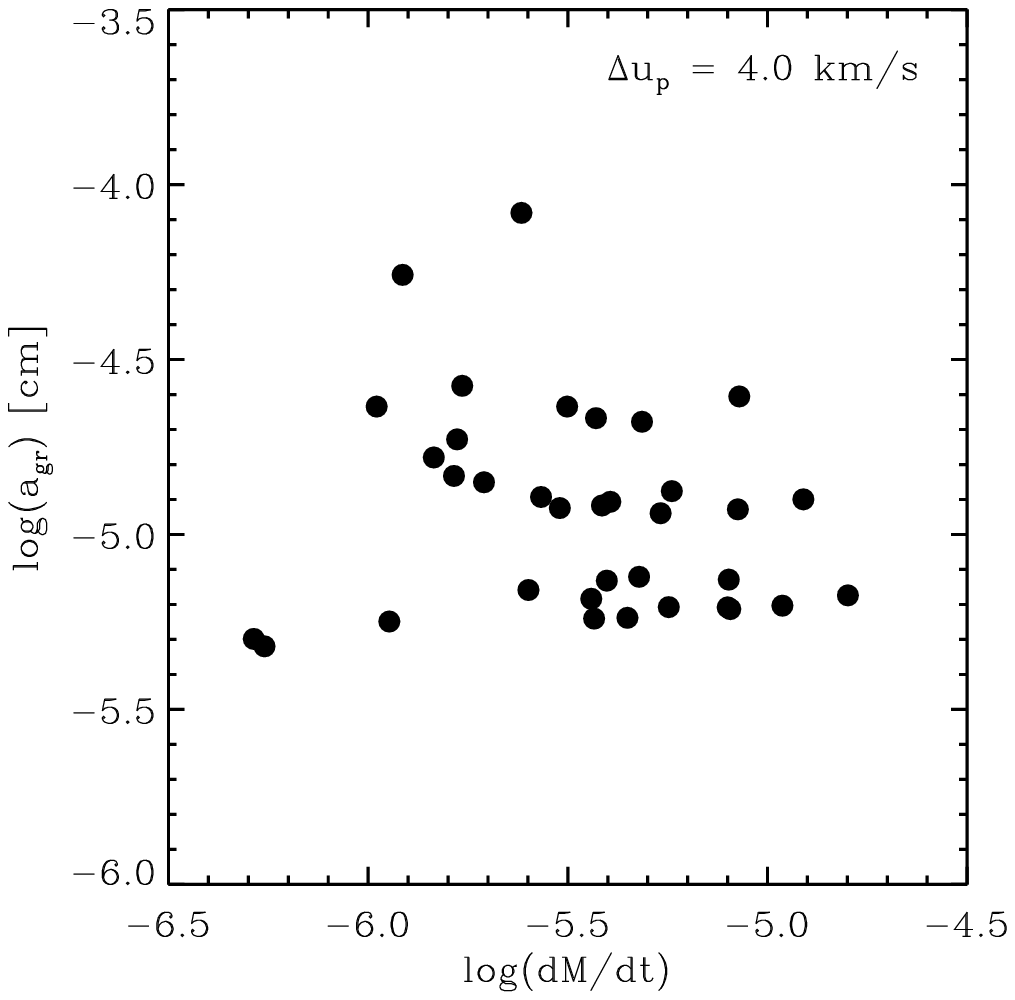}}
  \caption{  \label{agr}
  Wind speed (left) and mass-loss rate (right) versus the characteristic grain radii for models with 
  $\Delta u_{\rm p} = 4$ km s$^{-1}$ and $M_\star = 1 M_\odot$. Note the absence of any correlatiopn with mass-loss rate.}
  \end{figure*}

\subsection{Uncertainties}
Generally, numerical models are subject to various uncertainties due to simplifying assumptions, numerical limitations and the fact that some 
relevant aspects of the physical reality may not be included in the model. Several of the models presented here show intermittent, chaotic 
behaviour that may, or may not, reflect non-linearities in the physics of real dynamic stellar atmospheres. The existence of such non-linear features 
can affect the time averages of mass-loss rates, wind speeds and other wind properties that we have derived here. By necessity, the time series from 
which these averages are computed must be limited and the time series that we obtain for a specific set of stellar parameters over the 100-400 
pulsation periods the models cover may not always be representative. For example, the "deviating" mass-loss rate seen in Fig. 
\ref{models1} for $\log(L_\star/L_\odot) = 4.00$, $T = 3000$ K and $\log(\mbox{C-O})+12=9.10$ is probably an effect of dynamical non-linearities, 
which turn out to be favourable for dust formation. But it is 
hard to know for certain without a very long time series, and the models cannot be evolved for more than a few hundred periods, since it would lead 
to significant mass depletion in the modelled region. For the more "well-behaved" cases, the relative error in the time average due to time 
variations of the mass-loss rate, wind speed etc., is typically no bigger than 10\% (usually much less).

Another source of uncertainty (or bias) is our choices of stellar parameters, i.e., the definition of the grid. As previously discussed, the
threshold regions may not be adequately covered, which can affect the interpolations in the mass-loss prescription described above. As shown in
Fig. \ref{threshold}, linear interpolation is probably sufficiently accurate in most cases. But the very steep thresholds we find
when the grid spacing is decreased may not be reflecting the true nature of these critical regions. It cannot be excluded that one possible 
explanation for the steepness might be the resolution of the radiative transfer, which is relatively low (64 frequency points) in the present 
study. In critical cases, a higher resolution may slightly affect the momentum transfer efficiency (from radiation to dust). We have recently 
begun to study the effects of frequency resolution (i.e., computing models with high-resolution radiative transfer), but it is yet too early to 
draw general conclusions. It is also possible that the effects of dust grain sizes can affect the radiative transfer significantly in models near 
the thresholds. These effects are currently being investigated as well, and will be discussed in a forthcoming paper.

Finally, there are two, possibly important, ingredients that are not included the present DMAs: dynamical gas/dust decoupling (i.e., drift) and the
effects of the size distribution of dust grains on the momentum transfer efficiency. Grain size appears to be a critical issue for M-type stars 
\cite{Hofner08} and it may be of some importance also for the winds of carbon stars. It is known that drift does have an effect on the wind 
properties \cite[see, e.g.,][]{Sandin03,Sandin04}, but it has not been studied for the frequency-dependent case together with a detailed 
description of dust condensation in time-dependent wind models.

\section{Summary and conclusions}
The present grid of wind models has shown that mass-loss rates and wind speeds are clearly affected by the choice of stellar temperature, mass, luminosity 
and the abundance of available carbon. Furthermore, there are also inevitable mass-loss thresholds below which a dust-driven wind cannot form, in 
certain parts of the parameter space. Contrary to some previous studies, we find a strong dependence on the abundance of free carbon, which turns 
out to be a critical parameter. In a stellar evolution context, we expect that this dependence on the carbon excess will naturally lead to the 
development of a superwind after a few dredge-up events. Hence, the AGB would be terminated as soon as the atmosphere has become sufficiently
carbon rich, and the number of thermal pulses will be limited by a self-regulating mechanism: each thermal pulse increases the carbon excess, which 
in turn increases the mass loss. 

  \begin{figure*}
  \sidecaption 
  \includegraphics{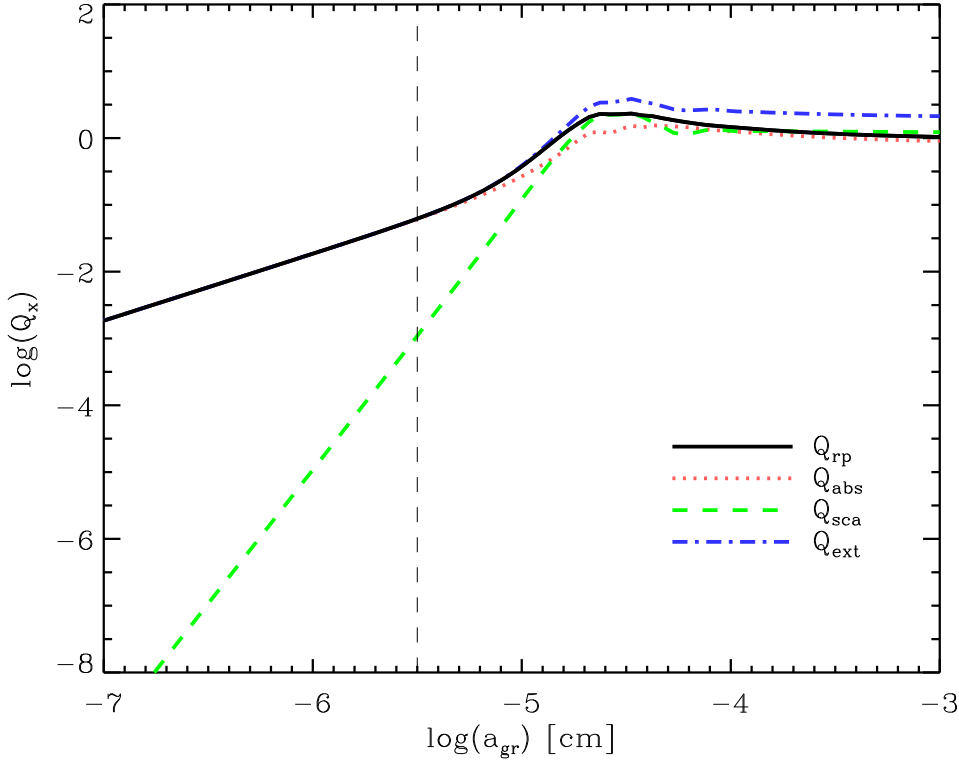}
  \caption{  \label{smallpart}
  The absorption, scattering and extinction efficiency, and the resultant efficiency factor for radiative pressure $Q_{\rm rp}$ as functions of 
  grain radius for amorphous carbon dust evaluated at $\lambda = 1\mu$m. The vertical, thin, dashed line shows where the small particle 
  approximation breaks down.
  }
  \end{figure*}

The set of models with $\Delta u_{\rm p} = 4.0$ km s$^{-1}$ show mass-loss rates that are typically a few times $10^{-6} M_\odot$ yr$^{-1}$ and
only one model has a really low mass-loss rate $\dot{M} < 10^{-7}$. 
With $\Delta u_{\rm p} = 2.0$ km s$^{-1}$ and $\Delta u_{\rm p} = 6.0$ km s$^{-1}$ the region in the stellar parameter space
where winds occur is somewhat smaller or larger, respectively. The typical mass-loss rates are otherwise similar to that of the 
$\Delta u_{\rm p} = 4.0$ km s$^{-1}$ subset. Models with a steady mass loss rate $\dot{M} < 10^{-7}$ appear to be difficult to produce regardless
of the combination of stellar parameters and  piston amplitude. The difficulty of producing really low mass-loss rates may represent a shortcoming 
of the present wind model.
It is yet unclear whether this is due to, e.g., the low-resolution radiative transfer of the models, or the phase coupling between gas and dust 
(i.e., absence of drift). A smaller number of test models with high resolution (440 frequency points instead of 64) has been computed,
but no dramatic differences were found. Further study of this matter is needed, however.

Winds of carbon stars are known to be relatively fast, compared to other long-period variables. The wind speeds that we obtain lie in the range 
$1-60 \mbox{km s}^{-1}$, where the fastest winds are found in the most carbon rich cases, i.e., carbon stars on the tip of the AGB. The lack of 
observed carbon stars with very high wind speeds ($u_{\rm out}> 40 \mbox{km s}^{-1}$) may be a selection effect, since the chance of finding a 
carbon star at the very final stage of its evolution must be quite small. The very low wind speeds seen in a few models 
($u_{\rm out} < 5 \mbox{km s}^{-1}$), but not in observed carbon stars, is an interesting prediction of the models. Such slow outflows may
exist, since a wind can, in theory, be maintained as long as the flow surpasses the escape speed at some point in the velocity profile.

The mean degree of dust condensation in the wind-producing models varies between $\langle f_c\rangle \approx 0.1$ and 
$\langle f_c\rangle \approx 0.7$, while the typical value is around 0.3. In the light of the recent models of M-type stars by H\"ofner (2008) 
\nocite{Hofner08}, we derived the characteristic grain sizes for a subset of models. The characteristic grain radius
in the the models that produce winds is $a_{\rm gr} \sim 10^{-5}$ cm, which is a somewhat too large for the small-particle limit to be a good 
approximation. The dust-grain size shows an anti-correlation with the wind speed, while the mass-loss rate seems totally uncorrelated. Large 
grains tend to form in slow winds, since they stay longer in the dust-formation zone and therefore have extra time to grow. 
However, the anti-correlation seen in Fig. \ref{agr} may not be entirely correct due to the fact that we compute the
dust extinction in the small-particle limit, which is not fully consistent with the grain sizes we obtain from the model of dust formation. 
The grain size is also anti-correlated with the abundance of condensible carbon, which is likely an effect of the competition between 
nucleation and grain growth.
 
Despite some shortcomings in reproducing low mass-loss rates, the present grid is still a significant step forward. This is the first systematic
study of mass loss and other wind properties as functions of stellar parameters that is based on a dynamic atmosphere model including both 
frequency-dependent radiative transfer and detailed, time-dependent dust formation, and is probably the most advanced theoretical constraint on the 
super-wind of carbon stars at present. Our results have shown that strong, dust driven winds cannot develop under arbitrary physical circumstances, 
i.e., in certain regions of (stellar) parameter space, carbon stars do not experience any significant dust-driven mass loss.

The thresholds that appear when, e.g., the effective temperature becomes too high or the abundance of condensible 
carbon too low, must be taken into account when constructing a mass-loss prescription to be used with other types of stellar models, such as 
stellar evolution and nucleosynthesis. Simple mass-loss formulae cannot reproduce all aspects of these thresholds and we have therefore not derived 
any such formula. Instead, we provide an easy-to-use FORTRAN-code that finds the corresponding mass-loss rate from interpolated data cubes. The 
code will soon be made available on-line.

\begin{acknowledgements} 
The authors wish to thank the referee, Martin Groenewegen, for his careful reading of the manuscript and constructive criticism that
helped to improve the clarity of this paper. 
B. Gustafsson, K. Eriksson and A. Wachter are thanked for their reading of the manuscript draft and for giving valuable suggestions.
This work was partly supported by the Swedish Research Council, {\em Vetenskapsr{\aa}det}.
\end{acknowledgements}

\bibliographystyle{aa}

\Online

  \begin{table*}
  \caption{\label{m075_u2} Input parameters ($L_\star$, $T_{\rm eff}$, C/O, $P$) and the 
  resulting mean mass loss rate, mean velocity at the outer boundary and mean degree of dust condensation at the outer boundary for a
  subset of models with $M_\star = 0.75 M_\odot$ and $\Delta u_{\rm p} = 2.0 \mbox{km s}^{-1}$. 
  The dust-to-gas mass ratio $\rho_{\rm dust}/\rho_{\rm gas}$ is calculated from $f_{\rm c}$ as described in H\"ofner \& Dorfi (1997).
  Where "***" is given instead of a number, no meaningful wind properties could be derived due to numerical problems.}
  \center
  \begin{tabular}{lcccccccccccc}  
  \hline
  \hline

  $\log(L_\star)$ & $T_\mathrm{eff}$  & log(C-O)+12 & $P$ & $\langle\dot{M}\rangle$ & $\langle u_{\rm out} \rangle$ & $\langle {\rm f_c} \rangle$ & 
  $\langle {\rho_{\rm d}/\rho_{\rm g}} \rangle$ \\[1mm]
  [$L_{\odot}$] &  [$\mbox{K}$] & & [days] & [$M_\odot$ yr$^{-1}$] & [km  s$^{-1}$] & & \\

  \hline
\\
 3.55   & 2400   & 8.20   &  221   &          -   &          -   &          -   &          -  \\
 3.55   & 2400   & 8.50   &  221   &   3.95E-07   &   4.46E+00   &   2.38E-01   &   6.45E-04  \\
 3.55   & 2400   & 8.80   &  221   &   1.62E-06   &   1.86E+01   &   2.89E-01   &   1.56E-03  \\
 3.55   & 2400   & 9.10   &  221   &   2.53E-06   &   3.70E+01   &   4.51E-01   &   4.87E-03  \\
 3.70   & 2400   & 8.20   &  295   &          -   &          -   &          -   &          -  \\
 3.70   & 2400   & 8.50   &  295   &   1.46E-06   &   9.01E+00   &   2.21E-01   &   5.99E-04  \\
 3.70   & 2400   & 8.80   &  295   &   3.62E-06   &   2.13E+01   &   3.11E-01   &   1.68E-03  \\
 3.70   & 2400   & 9.10   &  295   &   4.62E-06   &   3.93E+01   &   4.80E-01   &   5.18E-03  \\
 3.85   & 2400   & 8.20   &  393   &   1.06E-06   &   1.68E+00   &   2.28E-01   &   3.10E-04  \\
 3.85   & 2400   & 8.50   &  393   &   4.59E-06   &   1.29E+01   &   2.44E-01   &   6.61E-04  \\
 3.85   & 2400   & 8.80   &  393   &   6.81E-06   &   2.46E+01   &   3.37E-01   &   1.82E-03  \\
 3.85   & 2400   & 9.10   &  393   &   8.09E-06   &   4.31E+01   &   5.34E-01   &   5.76E-03  \\
 3.55   & 2600   & 8.20   &  221   &          -   &          -   &          -   &          -  \\
 3.55   & 2600   & 8.50   &  221   &   3.13E-07   &   6.79E+00   &   2.18E-01   &   5.91E-04  \\
 3.55   & 2600   & 8.80   &  221   &   9.67E-07   &   2.28E+01   &   2.96E-01   &   1.60E-03  \\
 3.55   & 2600   & 9.10   &  221   &   1.46E-06   &   4.10E+01   &   3.88E-01   &   4.19E-03  \\
 3.70   & 2600   & 8.20   &  295   &          -   &          -   &          -   &          -  \\
 3.70   & 2600   & 8.50   &  295   &   7.44E-07   &   6.86E+00   &   1.69E-01   &   4.58E-04  \\
 3.70   & 2600   & 8.80   &  295   &   2.34E-06   &   2.35E+01   &   2.70E-01   &   1.46E-03  \\
 3.70   & 2600   & 9.10   &  295   &   2.83E-06   &   4.24E+01   &   3.95E-01   &   4.26E-03  \\
 3.85   & 2600   & 8.20   &  393   &   4.18E-07   &   2.07E+00   &   2.01E-01   &   2.73E-04  \\
 3.85   & 2600   & 8.50   &  393   &   2.03E-06   &   1.01E+01   &   1.56E-01   &   4.23E-04  \\
 3.85   & 2600   & 8.80   &  393   &        ***   &        ***   &        ***   &        ***  \\
 3.85   & 2600   & 9.10   &  393   &   5.90E-06   &   4.31E+01   &   4.26E-01   &   4.60E-03  \\
 3.55   & 2800   & 8.20   &  221   &          -   &          -   &          -   &          -  \\
 3.55   & 2800   & 8.50   &  221   &          -   &          -   &          -   &          -  \\
 3.55   & 2800   & 8.80   &  221   &          -   &          -   &          -   &          -  \\
 3.55   & 2800   & 9.10   &  221   &   1.33E-06   &   3.80E+01   &   3.82E-01   &   4.12E-03  \\
 3.70   & 2800   & 8.20   &  295   &          -   &          -   &          -   &          -  \\
 3.70   & 2800   & 8.50   &  295   &          -   &          -   &          -   &          -  \\
 3.70   & 2800   & 8.80   &  295   &   8.62E-07   &   2.57E+01   &   2.24E-01   &   1.21E-03  \\
 3.70   & 2800   & 9.10   &  295   &   2.72E-06   &   4.33E+01   &   4.39E-01   &   4.74E-03  \\
 3.85   & 2800   & 8.20   &  393   &          -   &          -   &          -   &          -  \\
 3.85   & 2800   & 8.50   &  393   &   1.36E-06   &   1.13E+01   &   1.38E-01   &   3.74E-04  \\
 3.85   & 2800   & 8.80   &  393   &   2.59E-06   &   2.69E+01   &   2.48E-01   &   1.34E-03  \\
 3.85   & 2800   & 9.10   &  393   &   3.73E-06   &   4.70E+01   &   3.84E-01   &   4.14E-03  \\
 3.55   & 3000   & 8.20   &  221   &          -   &          -   &          -   &          -  \\
 3.55   & 3000   & 8.50   &  221   &          -   &          -   &          -   &          -  \\
 3.55   & 3000   & 8.80   &  221   &          -   &          -   &          -   &          -  \\
 3.55   & 3000   & 9.10   &  221   &          -   &          -   &          -   &          -  \\
 3.70   & 3000   & 8.20   &  295   &          -   &          -   &          -   &          -  \\
 3.70   & 3000   & 8.50   &  295   &          -   &          -   &          -   &          -  \\
 3.70   & 3000   & 8.80   &  295   &          -   &          -   &          -   &          -  \\
 3.70   & 3000   & 9.10   &  295   &   1.23E-06   &   3.80E+01   &   2.87E-01   &   3.10E-03  \\
 3.85   & 3000   & 8.20   &  393   &          -   &          -   &          -   &          -  \\
 3.85   & 3000   & 8.50   &  393   &          -   &          -   &          -   &          -  \\
 3.85   & 3000   & 8.80   &  393   &   1.14E-06   &   2.71E+01   &   2.21E-01   &   1.20E-03  \\
 3.85   & 3000   & 9.10   &  393   &   3.44E-06   &   4.59E+01   &   3.92E-01   &   4.23E-03  \\
 3.55   & 3200   & 8.20   &  221   &          -   &          -   &          -   &          -  \\
 3.55   & 3200   & 8.50   &  221   &          -   &          -   &          -   &          -  \\
 3.55   & 3200   & 8.80   &  221   &          -   &          -   &          -   &          -  \\
 3.55   & 3200   & 9.10   &  221   &          -   &          -   &          -   &          -  \\
 3.70   & 3200   & 8.20   &  295   &          -   &          -   &          -   &          -  \\
 3.70   & 3200   & 8.50   &  295   &          -   &          -   &          -   &          -  \\
 3.70   & 3200   & 8.80   &  295   &          -   &          -   &          -   &          -  \\
 3.70   & 3200   & 9.10   &  295   &   2.47E-08   &   1.26E+01   &   3.64E-01   &   3.93E-03  \\
 3.85   & 3200   & 8.20   &  393   &          -   &          -   &          -   &          -  \\
 3.85   & 3200   & 8.50   &  393   &          -   &          -   &          -   &          -  \\
 3.85   & 3200   & 8.80   &  393   &          -   &          -   &          -   &          -  \\
 3.85   & 3200   & 9.10   &  393   &          -   &          -   &          -   &          -  \\

\\
  \hline
\\
  \end{tabular}
  \label{models}
  \end{table*}

  \begin{table*}
  \caption{\label{m075_u4} Same as \ref{m075_u2}, but for $\Delta u_{\rm p} = 4.0 \mbox{km s}^{-1}$. 
  }
  \center
  \begin{tabular}{lcccccccccccc}  
  \hline
  \hline

  $\log(L_\star)$ & $T_\mathrm{eff}$  & log(C-O)+12 & $P$ & $\langle\dot{M}\rangle$ & $\langle u_{\rm out} \rangle$ & $\langle {\rm f_c} \rangle$ & 
  $\langle {\rho_{\rm d}/\rho_{\rm g}} \rangle$ \\[1mm]
  [$L_{\odot}$] &  [$\mbox{K}$] & & [days] & [$M_\odot$ yr$^{-1}$] & [km  s$^{-1}$] & & \\

  \hline
\\
 3.55   & 2400   & 8.20   &  221   &   -          &   -           &   -          &   -         \\
 3.55   & 2400   & 8.50   &  221   &   1.55E-06   &   5.80E+000   &   2.89E-01   &   7.83E-04  \\
 3.55   & 2400   & 8.80   &  221   &   3.05E-06   &   1.98E+001   &   3.80E-01   &   2.06E-03  \\
 3.55   & 2400   & 9.10   &  221   &   3.39E-06   &   3.66E+001   &   5.10E-01   &   5.50E-03  \\
 3.70   & 2400   & 8.20   &  295   &   -          &   -           &   -          &   -         \\
 3.70   & 2400   & 8.50   &  295   &   2.77E-06   &   8.40E+000   &   2.32E-01   &   6.29E-04  \\
 3.70   & 2400   & 8.80   &  295   &   4.83E-06   &   2.11E+001   &   3.32E-01   &   1.80E-03  \\
 3.70   & 2400   & 9.10   &  295   &   5.84E-06   &   3.84E+001   &   5.11E-01   &   5.51E-03  \\
 3.85   & 2400   & 8.20   &  393   &   2.23E-06   &   3.32E-01    &   3.07E-01   &   4.17E-04  \\
 3.85   & 2400   & 8.50   &  393   &   7.72E-06   &   1.18E+001   &   2.46E-01   &   6.67E-04  \\
 3.85   & 2400   & 8.80   &  393   &   9.07E-06   &   2.36E+001   &   3.49E-01   &   1.89E-03  \\
 3.85   & 2400   & 9.10   &  393   &   1.05E-05   &   4.16E+001   &   5.69E-01   &   6.14E-03  \\
 3.55   & 2600   & 8.20   &  221   &   -          &   -           &   -          &   -         \\
 3.55   & 2600   & 8.50   &  221   &   4.35E-07   &   5.03E+000   &   2.19E-01   &   5.94E-04  \\
 3.55   & 2600   & 8.80   &  221   &   1.26E-06   &   2.06E+001   &   2.74E-01   &   1.48E-03  \\
 3.55   & 2600   & 9.10   &  221   &   1.97E-06   &   3.89E+001   &   4.13E-01   &   4.46E-03  \\
 3.70   & 2600   & 8.20   &  295   &   -          &   -           &   -          &   -         \\
 3.70   & 2600   & 8.50   &  295   &   1.33E-06   &   7.44E+000   &   1.86E-01   &   5.04E-04  \\
 3.70   & 2600   & 8.80   &  295   &   2.84E-06   &   2.21E+001   &   2.65E-01   &   1.43E-03  \\
 3.70   & 2600   & 9.10   &  295   &   4.13E-06   &   3.96E+001   &   4.36E-01   &   4.70E-03  \\
 3.85   & 2600   & 8.20   &  393   &   1.01E-06   &   6.21E-01    &   2.19E-01   &   2.98E-04  \\
 3.85   & 2600   & 8.50   &  393   &   4.11E-06   &   1.13E+001   &   1.96E-01   &   5.31E-04  \\
 3.85   & 2600   & 8.80   &  393   &   3.99E-06   &   2.82E+001   &   3.43E-01   &   1.86E-03  \\
 3.85   & 2600   & 9.10   &  393   &   5.66E-06   &   4.65E+001   &   5.25E-01   &   5.67E-03  \\
 3.55   & 2800   & 8.20   &  221   &   -          &   -           &   -          &   -         \\
 3.55   & 2800   & 8.50   &  221   &   -          &   -           &   -          &   -         \\
 3.55   & 2800   & 8.80   &  221   &   7.18E-07   &   2.56E+001   &   3.19E-01   &   1.73E-03  \\
 3.55   & 2800   & 9.10   &  221   &   1.87E-06   &   4.12E+001   &   4.82E-01   &   5.20E-03  \\
 3.70   & 2800   & 8.20   &  295   &   -          &   -           &   -          &   -         \\
 3.70   & 2800   & 8.50   &  295   &   8.69E-07   &   8.85E+000   &   1.69E-01   &   4.58E-04  \\
 3.70   & 2800   & 8.80   &  295   &   1.77E-06   &   2.58E+001   &   2.83E-01   &   1.53E-03  \\
 3.70   & 2800   & 9.10   &  295   &   2.84E-06   &   4.45E+001   &   4.56E-01   &   4.92E-03  \\
 3.85   & 2800   & 8.20   &  393   &   -          &   -           &   -          &   -         \\
 3.85   & 2800   & 8.50   &  393   &   2.47E-06   &   1.03E+001   &   1.44E-01   &   3.90E-04  \\
 3.85   & 2800   & 8.80   &  393   &   3.57E-06   &   2.69E+001   &   2.25E-01   &   1.22E-03  \\
 3.85   & 2800   & 9.10   &  393   &   4.77E-06   &   4.68E+001   &   5.14E-01   &   5.55E-03  \\
 3.55   & 3000   & 8.20   &  221   &   -          &   -           &   -          &   -         \\
 3.55   & 3000   & 8.50   &  221   &   -          &   -           &   -          &   -         \\
 3.55   & 3000   & 8.80   &  221   &   -          &   -           &   -          &   -         \\
 3.55   & 3000   & 9.10   &  221   &   1.06E-06   &   3.71E+001   &   3.81E-01   &   4.11E-03  \\
 3.70   & 3000   & 8.20   &  295   &   -          &   -           &   -          &   -         \\
 3.70   & 3000   & 8.50   &  295   &   -          &   -           &   -          &   -         \\
 3.70   & 3000   & 8.80   &  295   &   9.35E-07   &   2.61E+001   &   2.81E-01   &   1.52E-03  \\
 3.70   & 3000   & 9.10   &  295   &   2.67E-06   &   4.31E+001   &   4.65E-01   &   5.02E-03  \\
 3.85   & 3000   & 8.20   &  393   &   -          &   -           &   -          &   -         \\
 3.85   & 3000   & 8.50   &  393   &   1.07E-06   &   1.21E+001   &   1.35E-01   &   3.66E-04  \\
 3.85   & 3000   & 8.80   &  393   &   2.34E-06   &   2.72E+001   &   2.14E-01   &   1.16E-03  \\
 3.85   & 3000   & 9.10   &  393   &   3.85E-06   &   5.03E+001   &   4.78E-01   &   5.16E-03  \\
 3.55   & 3200   & 8.20   &  221   &   -          &   -           &   -          &   -         \\
 3.55   & 3200   & 8.50   &  221   &   -          &   -           &   -          &   -         \\
 3.55   & 3200   & 8.80   &  221   &   -          &   -           &   -          &   -         \\
 3.55   & 3200   & 9.10   &  221   &   -          &   -           &   -          &   -         \\
 3.70   & 3200   & 8.20   &  295   &   -          &   -           &   -          &   -         \\
 3.70   & 3200   & 8.50   &  295   &   -          &   -           &   -          &   -         \\
 3.70   & 3200   & 8.80   &  295   &   5.32E-07   &   2.13E+001   &   1.27E-01   &   6.87E-04  \\
 3.70   & 3200   & 9.10   &  295   &   1.56E-06   &   4.64E+001   &   3.11E-01   &   3.36E-03  \\
 3.85   & 3200   & 8.20   &  393   &   -          &   -           &   -          &   -         \\
 3.85   & 3200   & 8.50   &  393   &   -          &   -           &   -          &   -         \\
 3.85   & 3200   & 8.80   &  393   &   1.05E-06   &   2.44E+001   &   2.10E-01   &   1.14E-03  \\
 3.85   & 3200   & 9.10   &  393   &   3.92E-06   &   4.70E+001   &   5.00E-01   &   5.40E-03  \\

\\
  \hline
\\
  \end{tabular}
  \label{models}
  \end{table*}

  \begin{table*}
  \caption{\label{m075_u6} Same as \ref{m075_u2}, but for $\Delta u_{\rm p} = 6.0 \mbox{km s}^{-1}$. 
  }
  \center
  \begin{tabular}{lcccccccccccc} 
  \hline 
  \hline

  $\log(L_\star)$ & $T_\mathrm{eff}$  & log(C-O)+12 & $P$ & $\langle\dot{M}\rangle$ & $\langle u_{\rm out} \rangle$ & $\langle {\rm f_c} \rangle$ & 
  $\langle {\rho_{\rm d}/\rho_{\rm g}} \rangle$ \\[1mm]
  [$L_{\odot}$] &  [$\mbox{K}$] & & [days] & [$M_\odot$ yr$^{-1}$] & [km  s$^{-1}$] & & \\

  \hline
\\
 3.55   & 2400   & 8.20   &  221   &   -          &   -           &   -          &   -         \\
 3.55   & 2400   & 8.50   &  221   &   1.35E-06   &   3.08E+000   &   2.58E-01   &   6.99E-04  \\
 3.55   & 2400   & 8.80   &  221   &   2.86E-06   &   1.71E+001   &   3.07E-01   &   1.66E-03  \\
 3.55   & 2400   & 9.10   &  221   &   4.11E-06   &   3.53E+001   &   5.27E-01   &   5.69E-03  \\
 3.70   & 2400   & 8.20   &  295   &   2.57E-06   &   9.87E-01    &   4.77E-01   &   6.48E-04  \\
 3.70   & 2400   & 8.50   &  295   &   3.99E-06   &   6.13E+000   &   2.22E-01   &   6.02E-04  \\
 3.70   & 2400   & 8.80   &  295   &   6.54E-06   &   2.06E+001   &   3.50E-01   &   1.89E-03  \\
 3.70   & 2400   & 9.10   &  295   &   6.99E-06   &   3.79E+001   &   5.50E-01   &   5.93E-03  \\
 3.85   & 2400   & 8.20   &  393   &   -          &   -           &   -          &   -         \\
 3.85   & 2400   & 8.50   &  393   &   1.05E-05   &   9.70E+000   &   2.11E-01   &   5.72E-04  \\
 3.85   & 2400   & 8.80   &  393   &   1.21E-05   &   2.65E+001   &   4.41E-01   &   2.39E-03  \\
 3.85   & 2400   & 9.10   &  393   &        ***   &        ***    &        ***   &        ***  \\
 3.55   & 2600   & 8.20   &  221   &   -          &   -           &   -          &   -         \\
 3.55   & 2600   & 8.50   &  221   &   7.04E-07   &   4.55E+000   &   2.27E-01   &   6.15E-04  \\
 3.55   & 2600   & 8.80   &  221   &   1.99E-06   &   1.95E+001   &   3.06E-01   &   1.65E-03  \\
 3.55   & 2600   & 9.10   &  221   &   2.96E-06   &   3.68E+001   &   4.69E-01   &   5.06E-03  \\
 3.70   & 2600   & 8.20   &  295   &   1.44E-06   &   1.38E+000   &   3.81E-01   &   5.18E-04  \\
 3.70   & 2600   & 8.50   &  295   &   2.32E-06   &   8.36E+000   &   2.04E-01   &   5.53E-04  \\
 3.70   & 2600   & 8.80   &  295   &   4.40E-06   &   2.20E+001   &   3.10E-01   &   1.68E-03  \\
 3.70   & 2600   & 9.10   &  295   &   5.68E-06   &   3.84E+001   &   5.05E-01   &   5.45E-03  \\
 3.85   & 2600   & 8.20   &  393   &   -          &   -           &   -          &   -         \\
 3.85   & 2600   & 8.50   &  393   &   6.00E-06   &   1.16E+001   &   2.14E-01   &   5.80E-04  \\
 3.85   & 2600   & 8.80   &  393   &        ***   &        ***    &        ***   &        ***  \\
 3.85   & 2600   & 9.10   &  393   &        ***   &        ***    &        ***   &        ***  \\
 3.55   & 2800   & 8.20   &  221   &   -          &   -           &   -          &   -         \\
 3.55   & 2800   & 8.50   &  221   &   4.11E-07   &   6.19E+000   &   2.08E-01   &   5.64E-04  \\
 3.55   & 2800   & 8.80   &  221   &   1.14E-06   &   2.27E+001   &   2.94E-01   &   1.59E-03  \\
 3.55   & 2800   & 9.10   &  221   &   1.94E-06   &   4.13E+001   &   4.93E-01   &   5.32E-03  \\
 3.70   & 2800   & 8.20   &  295   &   -          &   -           &   -          &   -         \\
 3.70   & 2800   & 8.50   &  295   &   1.91E-06   &   1.13E+001   &   2.21E-01   &   5.99E-04  \\
 3.70   & 2800   & 8.80   &  295   &   2.32E-06   &   2.57E+001   &   2.59E-01   &   1.40E-03  \\
 3.70   & 2800   & 9.10   &  295   &   3.78E-06   &   4.24E+001   &   4.99E-01   &   5.38E-03  \\
 3.85   & 2800   & 8.20   &  393   &   -          &   -           &   -          &   -         \\
 3.85   & 2800   & 8.50   &  393   &   3.79E-06   &   1.13E+001   &   1.67E-01   &   4.53E-04  \\
 3.85   & 2800   & 8.80   &  393   &   5.48E-06   &   2.85E+001   &   3.37E-01   &   1.82E-03  \\
 3.85   & 2800   & 9.10   &  393   &        ***   &        ***    &        ***   &        ***  \\
 3.55   & 3000   & 8.20   &  221   &   -          &   -           &   -          &   -         \\
 3.55   & 3000   & 8.50   &  221   &   4.48E-07   &   8.23E+000   &   2.10E-01   &   5.69E-04  \\
 3.55   & 3000   & 8.80   &  221   &   1.05E-06   &   2.49E+001   &   3.66E-01   &   1.98E-03  \\
 3.55   & 3000   & 9.10   &  221   &   2.10E-06   &   3.84E+001   &   5.10E-01   &   5.50E-03  \\
 3.70   & 3000   & 8.20   &  295   &   -          &   -           &   -          &   -         \\
 3.70   & 3000   & 8.50   &  295   &   4.19E-07   &   8.16E+000   &   1.55E-01   &   4.20E-04  \\
 3.70   & 3000   & 8.80   &  295   &   1.45E-06   &   2.43E+001   &   2.81E-01   &   1.52E-03  \\
 3.70   & 3000   & 9.10   &  295   &   2.46E-06   &   4.77E+001   &   5.05E-01   &   5.45E-03  \\
 3.85   & 3000   & 8.20   &  393   &   -          &   -           &   -          &   -         \\
 3.85   & 3000   & 8.50   &  393   &   1.80E-06   &   1.31E+001   &   1.56E-01   &   4.23E-04  \\
 3.85   & 3000   & 8.80   &  393   &   3.68E-06   &   3.04E+001   &   3.27E-01   &   1.77E-03  \\
 3.85   & 3000   & 9.10   &  393   &        ***   &        ***    &        ***   &        ***  \\
 3.55   & 3200   & 8.20   &  221   &   -          &   -           &   -          &   -         \\
 3.55   & 3200   & 8.50   &  221   &   -          &   -           &   -          &   -         \\
 3.55   & 3200   & 8.80   &  221   &   3.63E-07   &   2.77E+001   &   2.58E-01   &   1.40E-03  \\
 3.55   & 3200   & 9.10   &  221   &   1.03E-06   &   3.62E+001   &   3.98E-01   &   4.29E-03  \\
 3.70   & 3200   & 8.20   &  295   &   -          &   -           &   -          &   -         \\
 3.70   & 3200   & 8.50   &  295   &   9.51E-07   &   8.85E+000   &   1.48E-01   &   4.01E-04  \\
 3.70   & 3200   & 8.80   &  295   &   1.64E-06   &   2.56E+001   &   2.01E-01   &   1.09E-03  \\
 3.70   & 3200   & 9.10   &  295   &   3.11E-06   &   4.48E+001   &   3.79E-01   &   4.09E-03  \\
 3.85   & 3200   & 8.20   &  393   &   -          &   -           &   -          &   -         \\
 3.85   & 3200   & 8.50   &  393   &        ***   &        ***    &        ***   &        ***  \\
 3.85   & 3200   & 8.80   &  393   &        ***   &        ***    &        ***   &        ***  \\
 3.85   & 3200   & 9.10   &  393   &        ***   &        ***    &        ***   &        ***  \\

\\
  \hline
\\
  \end{tabular}
  \label{models}
  \end{table*}

  \begin{table*}
  \caption{\label{m1_u2} Input parameters ($L_\star$, $T_{\rm eff}$, C/O, $P$) and the 
  resulting mean mass loss rate, mean velocity at the outer boundary and mean degree of dust condensation at the outer boundary for a
  subset of models with $M_\star = 1.0 M_\odot$ and $\Delta u_{\rm p} = 2.0 \mbox{km s}^{-1}$. 
  The dust-to-gas mass ratio $\rho_{\rm dust}/\rho_{\rm gas}$ is calculated from $f_{\rm c}$ as described in H\"ofner \& Dorfi (1997).
  Where "***" is given instead of a number, no meaningful wind properties could be derived due to numerical problems.}
  \center
  \begin{tabular}{lcccccccccccc}  
  \hline
  \hline

  $\log(L_\star)$ & $T_\mathrm{eff}$  & log(C-O)+12 & $P$ & $\langle\dot{M}\rangle$ & $\langle u_{\rm out} \rangle$ & $\langle {\rm f_c} \rangle$ & 
  $\langle {\rho_{\rm d}/\rho_{\rm g}} \rangle$ \\[1mm]
  [$L_{\odot}$] &  [$\mbox{K}$] & & [days] & [$M_\odot$ yr$^{-1}$] & [km  s$^{-1}$] & & \\

  \hline
\\
 3.70   & 2400   & 8.20   &  295   &   -          &   -          &   -          &   -         \\
 3.70   & 2400   & 8.50   &  295   &   8.65E-07   &   5.44E+00   &   2.40E-01   &   6.51E-04  \\
 3.70   & 2400   & 8.80   &  295   &   2.32E-06   &   2.10E+01   &   2.91E-01   &   1.57E-03  \\
 3.70   & 2400   & 9.10   &  295   &   3.46E-06   &   4.06E+01   &   4.36E-01   &   4.70E-03  \\
 3.85   & 2400   & 8.20   &  393   &   -          &   -          &   -          &   -         \\
 3.85   & 2400   & 8.50   &  393   &   1.59E-06   &   7.28E+00   &   1.88E-01   &   5.10E-04  \\
 3.85   & 2400   & 8.80   &  393   &   4.39E-06   &   2.41E+01   &   2.95E-01   &   1.60E-03  \\
 3.85   & 2400   & 9.10   &  393   &   6.32E-06   &   4.28E+01   &   4.76E-01   &   5.14E-03  \\
 4.00   & 2400   & 8.20   &  524   &   1.09E-06   &   1.48E+00   &   2.26E-01   &   3.07E-04  \\
 4.00   & 2400   & 8.50   &  524   &   4.12E-06   &   1.14E+01   &   1.83E-01   &   4.96E-04  \\
 4.00   & 2400   & 8.80   &  524   &   9.53E-06   &   2.55E+01   &   3.08E-01   &   1.67E-03  \\
 4.00   & 2400   & 9.10   &  524   &   1.19E-05   &   4.55E+01   &   5.33E-01   &   5.75E-03  \\
 3.70   & 2600   & 8.20   &  295   &   -          &   -          &   -          &   -         \\
 3.70   & 2600   & 8.50   &  295   &   -          &   -          &   -          &   -         \\
 3.70   & 2600   & 8.80   &  295   &   1.45E-06   &   2.86E+01   &   3.59E-01   &   1.94E-03  \\
 3.70   & 2600   & 9.10   &  295   &   2.91E-06   &   4.15E+01   &   4.55E-01   &   4.91E-03  \\
 3.85   & 2600   & 8.20   &  393   &   -          &   -          &   -          &   -         \\
 3.85   & 2600   & 8.50   &  393   &   1.35E-06   &   1.25E+01   &   1.96E-01   &   5.31E-04  \\
 3.85   & 2600   & 8.80   &  393   &   2.41E-06   &   2.66E+01   &   2.69E-01   &   1.45E-03  \\
 3.85   & 2600   & 9.10   &  393   &   5.46E-06   &   4.50E+01   &   4.77E-01   &   5.15E-03  \\
 4.00   & 2600   & 8.20   &  524   &   6.72E-07   &   2.27E+00   &   1.99E-01   &   2.70E-04  \\
 4.00   & 2600   & 8.50   &  524   &   3.59E-06   &   1.44E+01   &   2.00E-01   &   5.42E-04  \\
 4.00   & 2600   & 8.80   &  524   &   5.79E-06   &   2.89E+01   &   2.81E-01   &   1.52E-03  \\
 4.00   & 2600   & 9.10   &  524   &   7.00E-06   &   4.96E+01   &   4.48E-01   &   4.83E-03  \\
 3.70   & 2800   & 8.20   &  295   &   -          &   -          &   -          &   -         \\
 3.70   & 2800   & 8.50   &  295   &   -          &   -          &   -          &   -         \\
 3.70   & 2800   & 8.80   &  295   &   -          &   -          &   -          &   -         \\
 3.70   & 2800   & 9.10   &  295   &   3.27E-06   &   3.90E+01   &   2.71E-01   &   2.92E-03  \\
 3.85   & 2800   & 8.20   &  393   &   -          &   -          &   -          &   -         \\
 3.85   & 2800   & 8.50   &  393   &   -          &   -          &   -          &   -         \\
 3.85   & 2800   & 8.80   &  393   &   9.88E-06   &   3.11E+01   &   2.64E-01   &   1.43E-03  \\
 3.85   & 2800   & 9.10   &  393   &   1.64E-06   &   3.99E+01   &   2.59E-01   &   2.79E-03  \\
 4.00   & 2800   & 8.20   &  524   &   -          &   -          &   -          &   -         \\
 4.00   & 2800   & 8.50   &  524   &   -          &   -          &   -          &   -         \\
 4.00   & 2800   & 8.80   &  524   &   2.39E-06   &   2.94E+01   &   2.38E-01   &   1.29E-03  \\
 4.00   & 2800   & 9.10   &  524   &   4.78E-06   &   4.77E+01   &   3.18E-01   &   3.43E-03  \\
 3.70   & 3000   & 8.20   &  295   &   -          &   -          &   -          &   -         \\
 3.70   & 3000   & 8.50   &  295   &   -          &   -          &   -          &   -         \\
 3.70   & 3000   & 8.80   &  295   &   -          &   -          &   -          &   -         \\
 3.70   & 3000   & 9.10   &  295   &   -          &   -          &   -          &   -         \\
 3.85   & 3000   & 8.20   &  393   &   -          &   -          &   -          &   -         \\
 3.85   & 3000   & 8.50   &  393   &   -          &   -          &   -          &   -         \\
 3.85   & 3000   & 8.80   &  393   &   -          &   -          &   -          &   -         \\
 3.85   & 3000   & 9.10   &  393   &   -          &   -          &   -          &   -         \\
 4.00   & 3000   & 8.20   &  524   &   -          &   -          &   -          &   -         \\
 4.00   & 3000   & 8.50   &  524   &   -          &   -          &   -          &   -         \\
 4.00   & 3000   & 8.80   &  524   &   -          &   -          &   -          &   -         \\
 4.00   & 3000   & 9.10   &  524   &   1.12E-06   &   4.75E+01   &   2.75E-01   &   2.97E-03  \\
 3.70   & 3200   & 8.20   &  295   &   -          &   -          &   -          &   -         \\
 3.70   & 3200   & 8.50   &  295   &   -          &   -          &   -          &   -         \\
 3.70   & 3200   & 8.80   &  295   &   -          &   -          &   -          &   -         \\
 3.70   & 3200   & 9.10   &  295   &   -          &   -          &   -          &   -         \\
 3.85   & 3200   & 8.20   &  393   &   -          &   -          &   -          &   -         \\
 3.85   & 3200   & 8.50   &  393   &   -          &   -          &   -          &   -         \\
 3.85   & 3200   & 8.80   &  393   &   -          &   -          &   -          &   -         \\
 3.85   & 3200   & 9.10   &  393   &   -          &   -          &   -          &             \\
 4.00   & 3200   & 8.20   &  524   &   -          &   -          &   -          &   -         \\
 4.00   & 3200   & 8.50   &  524   &   -          &   -          &   -          &   -         \\
 4.00   & 3200   & 8.80   &  524   &   -          &   -          &   -          &   -         \\
 4.00   & 3200   & 9.10   &  524   &   -          &   -          &   -          &   -         \\

\\
  \hline
\\
  \end{tabular}
  \label{models}
  \end{table*}

  \begin{table*}
  \caption{\label{m1_u4} Same as \ref{m1_u2}, but for $\Delta u_{\rm p} = 4.0 \mbox{km s}^{-1}$. 
  }
  \center
  \begin{tabular}{lcccccccccccc}  
  \hline
  \hline

  $\log(L_\star)$ & $T_\mathrm{eff}$  & log(C-O)+12 & $P$ & $\langle\dot{M}\rangle$ & $\langle u_{\rm out} \rangle$ & $\langle {\rm f_c} \rangle$ & 
  $\langle {\rho_{\rm d}/\rho_{\rm g}} \rangle$ \\[1mm]
  [$L_{\odot}$] &  [$\mbox{K}$] & & [days] & [$M_\odot$ yr$^{-1}$] & [km  s$^{-1}$] & & \\

  \hline
\\
 3.70   & 2400   & 8.20   &  295   &   -          &   -          &   -          &   -         \\
 3.70   & 2400   & 8.50   &  295   &   1.05E-06   &   3.80E+00   &   2.44E-01   &   6.61E-04  \\
 3.70   & 2400   & 8.80   &  295   &   3.02E-06   &   2.00E+01   &   3.19E-01   &   1.73E-03  \\
 3.70   & 2400   & 9.10   &  295   &   4.46E-06   &   3.86E+01   &   4.75E-01   &   5.13E-03  \\
 3.85   & 2400   & 8.20   &  393   &   -          &   -          &   -          &   -         \\
 3.85   & 2400   & 8.50   &  393   &   3.15E-06   &   7.76E+00   &   2.13E-01   &   5.77E-04  \\
 3.85   & 2400   & 8.80   &  393   &   5.40E-06   &   2.19E+01   &   2.85E-01   &   1.54E-03  \\
 3.85   & 2400   & 9.10   &  393   &   7.95E-06   &   4.15E+01   &   5.24E-01   &   5.65E-03  \\
 4.00   & 2400   & 8.20   &  524   &   2.42E-06   &   8.67E-01   &   2.54E-01   &   3.45E-04  \\
 4.00   & 2400   & 8.50   &  524   &   8.50E-06   &   1.14E+01   &   2.15E-01   &   5.83E-04  \\
 4.00   & 2400   & 8.80   &  524   &   1.23E-05   &   2.42E+01   &   3.36E-01   &   1.82E-03  \\
 4.00   & 2400   & 9.10   &  524   &   1.59E-05   &   4.43E+01   &   5.88E-01   &   6.34E-03  \\
 3.70   & 2600   & 8.20   &  295   &   -          &   -          &   -          &   -         \\
 3.70   & 2600   & 8.50   &  295   &   7.60E-07   &   6.24E+00   &   2.11E-01   &   5.72E-04  \\
 3.70   & 2600   & 8.80   &  295   &   1.95E-06   &   2.68E+01   &   4.00E-01   &   2.16E-03  \\
 3.70   & 2600   & 9.10   &  295   &   3.96E-06   &   4.39E+01   &   5.84E-01   &   6.30E-03  \\
 3.85   & 2600   & 8.20   &  393   &   -          &   -          &   -          &   -         \\
 3.85   & 2600   & 8.50   &  393   &   1.67E-06   &   6.24E+00   &   1.71E-01   &   4.63E-04  \\
 3.85   & 2600   & 8.80   &  393   &   4.04E-06   &   2.65E+01   &   3.22E-01   &   1.74E-03  \\
 3.85   & 2600   & 9.10   &  393   &   5.66E-06   &   4.65E+01   &   5.25E-01   &   5.67E-03  \\
 4.00   & 2600   & 8.20   &  524   &   1.22E-06   &   1.34E+00   &   2.06E-01   &   2.80E-04  \\
 4.00   & 2600   & 8.50   &  524   &   4.85E-06   &   1.04E+01   &   1.57E-01   &   4.26E-04  \\
 4.00   & 2600   & 8.80   &  524   &   8.43E-06   &   2.80E+01   &   3.25E-01   &   1.76E-03  \\
 4.00   & 2600   & 9.10   &  524   &   1.09E-05   &   4.55E+01   &   4.88E-01   &   5.27E-03  \\
 3.70   & 2800   & 8.20   &  295   &   -          &   -          &   -          &   -         \\
 3.70   & 2800   & 8.50   &  295   &   -          &   -          &   -          &   -         \\
 3.70   & 2800   & 8.80   &  295   &   1.46E-06   &   3.00E+01   &   4.55E-01   &   2.46E-03  \\
 3.70   & 2800   & 9.10   &  295   &   2.52E-06   &   4.10E+01   &   4.69E-01   &   5.06E-03  \\
 3.85   & 2800   & 8.20   &  393   &   -          &   -          &   -          &   -         \\
 3.85   & 2800   & 8.50   &  393   &   1.49E-06   &   1.71E+01   &   2.25E-01   &   6.10E-04  \\
 3.85   & 2800   & 8.80   &  393   &   2.71E-06   &   2.83E+01   &   3.15E-01   &   1.70E-03  \\
 3.85   & 2800   & 9.10   &  393   &   4.77E-06   &   4.68E+01   &   5.14E-01   &   5.55E-03  \\
 4.00   & 2800   & 8.20   &  524   &   -          &   -          &   -          &   -         \\
 4.00   & 2800   & 8.50   &  524   &   3.72E-06   &   1.31E+01   &   1.72E-01   &   4.66E-04  \\
 4.00   & 2800   & 8.80   &  524   &   5.76E-06   &   2.81E+01   &   2.94E-01   &   1.59E-03  \\
 4.00   & 2800   & 9.10   &  524   &   8.07E-06   &   4.96E+01   &   3.87E-01   &   4.18E-03  \\
 3.70   & 3000   & 8.20   &  295   &   -          &   -          &   -          &   -         \\
 3.70   & 3000   & 8.50   &  295   &   -          &   -          &   -          &   -         \\
 3.70   & 3000   & 8.80   &  295   &   -          &   -          &   -          &   -         \\
 3.70   & 3000   & 9.10   &  295   &   5.50E-07   &   3.09E+01   &   2.84E-01   &   3.06E-03  \\
 3.85   & 3000   & 8.20   &  393   &   -          &   -          &   -          &   -         \\
 3.85   & 3000   & 8.50   &  393   &   -          &   -          &   -          &   -         \\
 3.85   & 3000   & 8.80   &  393   &   1.47E-06   &   3.12E+01   &   3.76E-01   &   2.03E-03  \\
 3.85   & 3000   & 9.10   &  393   &   3.62E-06   &   4.27E+01   &   4.25E-01   &   4.59E-03  \\
 4.00   & 3000   & 8.20   &  524   &   -          &   -          &   -          &   -         \\
 4.00   & 3000   & 8.50   &  524   &   1.72E-06   &   1.83E+01   &   1.75E-01   &   4.74E-04  \\
 4.00   & 3000   & 8.80   &  524   &   3.85E-06   &   2.73E+01   &   2.86E-01   &   1.55E-03  \\
 4.00   & 3000   & 9.10   &  524   &   8.00E-06   &   4.76E+01   &   5.56E-01   &   6.00E-03  \\
 3.70   & 3200   & 8.20   &  295   &   -          &   -          &   -          &   -         \\
 3.70   & 3200   & 8.50   &  295   &   -          &   -          &   -          &   -         \\
 3.70   & 3200   & 8.80   &  295   &   -          &   -          &   -          &   -         \\
 3.70   & 3200   & 9.10   &  295   &   1.13E-06   &   4.56E+01   &   3.22E-01   &   3.47E-03  \\
 3.85   & 3200   & 8.20   &  393   &   -          &   -          &   -          &   -         \\
 3.85   & 3200   & 8.50   &  393   &   -          &   -          &   -          &   -         \\
 3.85   & 3200   & 8.80   &  393   &   -          &   -          &   -          &   -         \\
 3.85   & 3200   & 9.10   &  393   &   5.17E-07   &   3.98E+01   &   2.77E-01   &   2.99E-03  \\
 4.00   & 3200   & 8.20   &  524   &   -          &   -          &   -          &   -         \\
 4.00   & 3200   & 8.50   &  524   &   -          &   -          &   -          &   -         \\
 4.00   & 3200   & 8.80   &  524   &   1.64E-06   &   3.33E+01   &   3.09E-01   &   1.67E-03  \\
 4.00   & 3200   & 9.10   &  524   &   3.68E-06   &   5.10E+01   &   3.65E-01   &   3.94E-03  \\

\\
  \hline
\\
  \end{tabular}
  \label{models}
  \end{table*}

  \begin{table*}
  \caption{\label{m1_u6} Same as \ref{m1_u2}, but for $\Delta u_{\rm p} = 6.0 \mbox{km s}^{-1}$. 
  }
  \center
  \begin{tabular}{lcccccccccccc}  
  \hline
  \hline

  $\log(L_\star)$ & $T_\mathrm{eff}$  & log(C-O)+12 & $P$ & $\langle\dot{M}\rangle$ & $\langle u_{\rm out} \rangle$ & $\langle {\rm f_c} \rangle$ & 
  $\langle {\rho_{\rm d}/\rho_{\rm g}} \rangle$ \\[1mm]
  [$L_{\odot}$] &  [$\mbox{K}$] & & [days] & [$M_\odot$ yr$^{-1}$] & [km  s$^{-1}$] & & \\

  \hline
\\
 3.70   & 2400   & 8.20   &  295   &   -          &   -          &   -          &   -         \\
 3.70   & 2400   & 8.50   &  295   &   1.63E-06   &   1.45E+04   &   2.55E-01   &   6.91E-04  \\
 3.70   & 2400   & 8.80   &  295   &   3.81E-06   &   1.92E+01   &   3.22E-01   &   1.74E-03  \\
 3.70   & 2400   & 9.10   &  295   &   6.03E-06   &   3.71E+01   &   5.25E-01   &   5.67E-03  \\
 3.85   & 2400   & 8.20   &  393   &   -          &   -          &   -          &   -         \\
 3.85   & 2400   & 8.50   &  393   &   5.04E-06   &   7.69E+00   &   2.08E-01   &   5.64E-04  \\
 3.85   & 2400   & 8.80   &  393   &   7.81E-06   &   2.68E+01   &   4.47E-01   &   2.42E-03  \\
 3.85   & 2400   & 9.10   &  393   &   1.17E-05   &   4.17E+01   &   5.45E-01   &   5.88E-03  \\
 4.00   & 2400   & 8.20   &  524   &   -          &   -          &   -          &   -         \\
 4.00   & 2400   & 8.50   &  524   &   1.37E-05   &   1.55E+01   &   2.66E-01   &   7.21E-04  \\
 4.00   & 2400   & 8.80   &  524   &   1.60E-05   &   2.76E+01   &   4.94E-01   &   2.67E-03  \\
 4.00   & 2400   & 9.10   &  524   &   2.10E-05   &   4.01E+01   &   7.19E-01   &   7.76E-03  \\
 3.70   & 2600   & 8.20   &  295   &   -          &   -          &   -          &   -         \\
 3.70   & 2600   & 8.50   &  295   &   1.68E-06   &   8.16E+00   &   2.46E-01   &   6.67E-04  \\
 3.70   & 2600   & 8.80   &  295   &   2.93E-06   &   2.42E+01   &   4.03E-01   &   2.18E-03  \\
 3.70   & 2600   & 9.10   &  295   &   3.88E-06   &   4.27E+01   &   5.54E-01   &   5.98E-03  \\
 3.85   & 2600   & 8.20   &  393   &   -          &   -          &   -          &   -         \\
 3.85   & 2600   & 8.50   &  393   &   3.43E-06   &   1.03E+01   &   2.20E-01   &   5.96E-04  \\
 3.85   & 2600   & 8.80   &  393   &   5.48E-06   &   2.79E+01   &   4.16E-01   &   2.25E-03  \\
 3.85   & 2600   & 9.10   &  393   &   8.52E-06   &   4.10E+01   &   5.37E-01   &   5.79E-03  \\
 4.00   & 2600   & 8.20   &  524   &   2.95E-06   &   1.51E+00   &   2.26E-01   &   3.07E-04  \\
 4.00   & 2600   & 8.50   &  524   &   7.23E-06   &   1.12E+01   &   1.77E-01   &   4.80E-04  \\
 4.00   & 2600   & 8.80   &  524   &   1.07E-05   &   2.84E+01   &   3.67E-01   &   1.98E-03  \\
 4.00   & 2600   & 9.10   &  524   &   1.59E-05   &   4.29E+01   &   5.65E-01   &   6.10E-03  \\
 3.70   & 2800   & 8.20   &  295   &   -          &   -          &   -          &   -         \\
 3.70   & 2800   & 8.50   &  295   &   1.54E-06   &   1.06E+01   &   2.83E-01   &   7.67E-04  \\
 3.70   & 2800   & 8.80   &  295   &   3.12E-06   &   2.50E+01   &   4.33E-01   &   2.34E-03  \\
 3.70   & 2800   & 9.10   &  295   &   4.26E-06   &   4.03E+01   &   5.60E-01   &   6.04E-03  \\
 3.85   & 2800   & 8.20   &  393   &   -          &   -          &   -          &   -         \\
 3.85   & 2800   & 8.50   &  393   &   2.07E-06   &   1.44E+01   &   2.37E-01   &   6.42E-04  \\
 3.85   & 2800   & 8.80   &  393   &   3.74E-06   &   2.71E+01   &   3.15E-01   &   1.70E-03  \\
 3.85   & 2800   & 9.10   &  393   &   7.15E-06   &   4.43E+01   &   5.82E-01   &   6.28E-03  \\
 4.00   & 2800   & 8.20   &  524   &   -          &   -          &   -          &   -         \\
 4.00   & 2800   & 8.50   &  524   &   5.42E-06   &   1.46E+01   &   2.14E-01   &   5.80E-04  \\
 4.00   & 2800   & 8.80   &  524   &   8.76E-06   &   3.10E+01   &   4.11E-01   &   2.22E-03  \\
 4.00   & 2800   & 9.10   &  524   &        ***   &        ***   &        ***   &        ***  \\
 3.70   & 3000   & 8.20   &  295   &   -          &   -          &   -          &   -         \\
 3.70   & 3000   & 8.50   &  295   &   -          &   -          &   -          &   -         \\
 3.70   & 3000   & 8.80   &  295   &   1.16E-06   &   3.13E+01   &   4.49E-01   &   2.43E-03  \\
 3.70   & 3000   & 9.10   &  295   &   2.30E-06   &   3.73E+01   &   4.36E-01   &   4.70E-03  \\
 3.85   & 3000   & 8.20   &  393   &   -          &   -          &   -          &   -         \\
 3.85   & 3000   & 8.50   &  393   &   2.54E-06   &   2.04E+01   &   3.09E-01   &   8.38E-04  \\
 3.85   & 3000   & 8.80   &  393   &        ***   &        ***   &        ***   &        ***  \\
 3.85   & 3000   & 9.10   &  393   &   6.05E-06   &   4.31E+01   &   5.76E-01   &   6.22E-03  \\
 4.00   & 3000   & 8.20   &  524   &   -          &   -          &   -          &   -         \\
 4.00   & 3000   & 8.50   &  524   &   2.86E-06   &   1.98E+01   &   2.48E-01   &   6.72E-04  \\
 4.00   & 3000   & 8.80   &  524   &   6.37E-06   &   2.71E+01   &   3.21E-01   &   1.74E-03  \\
 4.00   & 3000   & 9.10   &  524   &        ***   &        ***   &        ***   &        ***  \\
 3.70   & 3200   & 8.20   &  295   &   -          &   -          &   -          &   -         \\
 3.70   & 3200   & 8.50   &  295   &   -          &   -          &   -          &   -         \\
 3.70   & 3200   & 8.80   &  295   &   -          &   -          &   -          &   -         \\
 3.70   & 3200   & 9.10   &  295   &   3.14E-06   &   4.18E+01   &   4.00E-01   &   4.32E-03  \\
 3.85   & 3200   & 8.20   &  393   &   -          &   -          &   -          &   -         \\
 3.85   & 3200   & 8.50   &  393   &   -          &   -          &   -          &   -         \\
 3.85   & 3200   & 8.80   &  393   &   1.31E-06   &   3.36E+01   &   3.81E-01   &   2.06E-03  \\
 3.85   & 3200   & 9.10   &  393   &   3.55E-06   &   4.27E+01   &   4.80E-01   &   5.18E-03  \\
 4.00   & 3200   & 8.20   &  524   &   -          &   -          &   -          &   -         \\
 4.00   & 3200   & 8.50   &  524   &   2.59E-06   &   9.87E+00   &   3.97E-01   &   1.08E-03  \\
 4.00   & 3200   & 8.80   &  524   &   4.08E-06   &   2.34E+01   &   3.39E-01   &   1.83E-03  \\
 4.00   & 3200   & 9.10   &  524   &        ***   &        ***   &        ***   &        ***  \\

\\
  \hline
\\
  \end{tabular}
  \label{models}
  \end{table*}

  \begin{table*}
  \caption{\label{m15_u2} Input parameters ($L_\star$, $T_{\rm eff}$, C/O, $P$) and the 
  resulting mean mass loss rate, mean velocity at the outer boundary and mean degree of dust condensation at the outer boundary for a
  subset of models with $M_\star = 1.5 M_\odot$ and $\Delta u_{\rm p} = 2.0 \mbox{km s}^{-1}$. 
  The dust-to-gas mass ratio $\rho_{\rm dust}/\rho_{\rm gas}$ is calculated from $f_{\rm c}$ as described in H\"ofner \& Dorfi (1997).
  Where "***" is given instead of a number, no meaningful wind properties could be derived due to numerical problems.}
  \center
  \begin{tabular}{lcccccccccccc}  
  \hline
  \hline

  $\log(L_\star)$ & $T_\mathrm{eff}$  & log(C-O)+12 & $P$ & $\langle\dot{M}\rangle$ & $\langle u_{\rm out} \rangle$ & $\langle {\rm f_c} \rangle$ & 
  $\langle {\rho_{\rm d}/\rho_{\rm g}} \rangle$ \\[1mm]
  [$L_{\odot}$] &  [$\mbox{K}$] & & [days] & [$M_\odot$ yr$^{-1}$] & [km  s$^{-1}$] & & \\

  \hline
\\
 3.85   & 2400   & 8.20   &  393   &   -          &   -          &   -          &   -         \\
 3.85   & 2400   & 8.50   &  393   &   5.40E-07   &   7.95E+00   &   2.53E-01   &   6.86E-04  \\
 3.85   & 2400   & 8.80   &  393   &   2.06E-06   &   2.60E+01   &   2.96E-01   &   1.60E-03  \\
 3.85   & 2400   & 9.10   &  393   &   4.07E-06   &   4.64E+01   &   4.92E-01   &   5.31E-03  \\
 4.00   & 2400   & 8.20   &  524   &   -          &   -          &   -          &   -         \\
 4.00   & 2400   & 8.50   &  524   &   2.23E-06   &   1.17E+01   &   2.30E-01   &   6.23E-04  \\
 4.00   & 2400   & 8.80   &  524   &   4.95E-06   &   2.96E+01   &   3.59E-01   &   1.94E-03  \\
 4.00   & 2400   & 9.10   &  524   &   8.75E-06   &   4.73E+01   &   5.18E-01   &   5.59E-03  \\
 4.15   & 2400   & 8.20   &  699   &   -          &   -          &   -          &   -         \\
 4.15   & 2400   & 8.50   &  699   &   5.26E-06   &   1.83E+01   &   2.68E-01   &   7.26E-04  \\
 4.15   & 2400   & 8.80   &  699   &   9.90E-06   &   3.43E+01   &   4.04E-01   &   2.18E-03  \\
 4.15   & 2400   & 9.10   &  699   &   1.50E-05   &   5.11E+01   &   5.55E-01   &   5.99E-03  \\
 3.85   & 2600   & 8.20   &  393   &   -          &   -          &   -          &   -         \\
 3.85   & 2600   & 8.50   &  393   &   -          &   -          &   -          &   -         \\
 3.85   & 2600   & 8.80   &  393   &   -          &   -          &   -          &   -         \\
 3.85   & 2600   & 9.10   &  393   &        ***   &        ***   &        ***   &        ***  \\
 4.00   & 2600   & 8.20   &  524   &   -          &   -          &   -          &   -         \\
 4.00   & 2600   & 8.50   &  524   &   -          &   -          &   -          &   -         \\
 4.00   & 2600   & 8.80   &  524   &   1.86E-06   &   3.51E+01   &   2.96E-01   &   1.60E-03  \\
 4.00   & 2600   & 9.10   &  524   &   2.47E-06   &   5.48E+01   &   3.77E-01   &   4.07E-03  \\
 4.15   & 2600   & 8.20   &  699   &   -          &   -          &   -          &   -         \\
 4.15   & 2600   & 8.50   &  699   &   1.50E-06   &   1.63E+01   &   1.89E-01   &   5.12E-04  \\
 4.15   & 2600   & 8.80   &  699   &   4.70E-06   &   3.29E+01   &   2.64E-01   &   1.43E-03  \\
 4.15   & 2600   & 9.10   &  699   &   5.04E-06   &   5.26E+01   &   3.39E-01   &   3.66E-03  \\
 3.85   & 2800   & 8.20   &  393   &   -          &   -          &   -          &   -         \\
 3.85   & 2800   & 8.50   &  393   &   -          &   -          &   -          &   -         \\
 3.85   & 2800   & 8.80   &  393   &   -          &   -          &   -          &   -         \\
 3.85   & 2800   & 9.10   &  393   &   -          &   -          &   -          &   -         \\
 4.00   & 2800   & 8.20   &  524   &   -          &   -          &   -          &   -         \\
 4.00   & 2800   & 8.50   &  524   &   -          &   -          &   -          &   -         \\
 4.00   & 2800   & 8.80   &  524   &   -          &   -          &   -          &   -         \\
 4.00   & 2800   & 9.10   &  524   &   9.63E-07   &   4.75E+01   &   2.64E-01   &   2.85E-03  \\
 4.15   & 2800   & 8.20   &  699   &   -          &   -          &   -          &   -         \\
 4.15   & 2800   & 8.50   &  699   &   -          &   -          &   -          &   -         \\
 4.15   & 2800   & 8.80   &  699   &   1.64E-06   &   3.77E+01   &   2.70E-01   &   1.46E-03  \\
 4.15   & 2800   & 9.10   &  699   &   2.24E-06   &   5.47E+01   &   2.86E-01   &   3.09E-03  \\
 3.85   & 3000   & 8.20   &  393   &   -          &   -          &   -          &   -         \\
 3.85   & 3000   & 8.50   &  393   &   -          &   -          &   -          &   -         \\
 3.85   & 3000   & 8.80   &  393   &   -          &   -          &   -          &   -         \\
 3.85   & 3000   & 9.10   &  393   &   -          &   -          &   -          &   -         \\
 4.00   & 3000   & 8.20   &  524   &   -          &   -          &   -          &   -         \\
 4.00   & 3000   & 8.50   &  524   &   -          &   -          &   -          &   -         \\
 4.00   & 3000   & 8.80   &  524   &   -          &   -          &   -          &   -         \\
 4.00   & 3000   & 9.10   &  524   &   -          &   -          &   -          &   -         \\
 4.15   & 3000   & 8.20   &  699   &   -          &   -          &   -          &   -         \\
 4.15   & 3000   & 8.50   &  699   &   -          &   -          &   -          &   -         \\
 4.15   & 3000   & 8.80   &  699   &   -          &   -          &   -          &   -         \\
 4.15   & 3000   & 9.10   &  699   &   7.98E-07   &   4.62E+01   &   1.98E-01   &   2.14E-03  \\
 3.85   & 3200   & 8.20   &  393   &   -          &   -          &   -          &   -         \\
 3.85   & 3200   & 8.50   &  393   &   -          &   -          &   -          &   -         \\
 3.85   & 3200   & 8.80   &  393   &   -          &   -          &   -          &   -         \\
 3.85   & 3200   & 9.10   &  393   &   -          &   -          &   -          &   -         \\
 4.00   & 3200   & 8.20   &  524   &   -          &   -          &   -          &   -         \\
 4.00   & 3200   & 8.50   &  524   &   -          &   -          &   -          &   -         \\
 4.00   & 3200   & 8.80   &  524   &   -          &   -          &   -          &   -         \\
 4.00   & 3200   & 9.10   &  524   &   -          &   -          &   -          &   -         \\
 4.15   & 3200   & 8.20   &  699   &   -          &   -          &   -          &   -         \\
 4.15   & 3200   & 8.50   &  699   &   -          &   -          &   -          &   -         \\
 4.15   & 3200   & 8.80   &  699   &   -          &   -          &   -          &   -         \\
 4.15   & 3200   & 9.10   &  699   &   -          &   -          &   -          &   -         \\

\\
  \hline
\\
  \end{tabular}
  \label{models}
  \end{table*}

  \begin{table*}
  \caption{\label{m15_u4} Same as \ref{m15_u2}, but for $\Delta u_{\rm p} = 4.0 \mbox{km s}^{-1}$. 
  }
  \center
  \begin{tabular}{lcccccccccccc}  
  \hline
  \hline

  $\log(L_\star)$ & $T_\mathrm{eff}$  & log(C-O)+12 & $P$ & $\langle\dot{M}\rangle$ & $\langle u_{\rm out} \rangle$ & $\langle {\rm f_c} \rangle$ & 
  $\langle {\rho_{\rm d}/\rho_{\rm g}} \rangle$ \\[1mm]
  [$L_{\odot}$] &  [$\mbox{K}$] & & [days] & [$M_\odot$ yr$^{-1}$] & [km  s$^{-1}$] & & \\

  \hline
\\
 3.85   & 2400   & 8.20   &  393   &   -          &   -          &   -          &   -         \\
 3.85   & 2400   & 8.50   &  393   &   1.52E-06   &   6.57E+00   &   2.81E-01   &   7.62E-04  \\
 3.85   & 2400   & 8.80   &  393   &   3.67E-06   &   2.15E+01   &   3.09E-01   &   1.67E-03  \\
 3.85   & 2400   & 9.10   &  393   &   7.23E-06   &   4.64E+01   &   6.87E-01   &   7.41E-03  \\
 4.00   & 2400   & 8.20   &  524   &   -          &   -          &   -          &   -         \\
 4.00   & 2400   & 8.50   &  524   &   3.41E-06   &   6.00E+00   &   2.06E-01   &   5.58E-04  \\
 4.00   & 2400   & 8.80   &  524   &   8.50E-06   &   2.52E+01   &   3.55E-01   &   1.92E-03  \\
 4.00   & 2400   & 9.10   &  524   &   1.22E-05   &   4.76E+01   &   6.56E-01   &   7.08E-03  \\
 4.15   & 2400   & 8.20   &  699   &   -          &   -          &   -          &   -         \\
 4.15   & 2400   & 8.50   &  699   &   8.91E-06   &   1.00E+01   &   1.99E-01   &   5.39E-04  \\
 4.15   & 2400   & 8.80   &  699   &   1.72E-05   &   2.93E+01   &   4.04E-01   &   2.18E-03  \\
 4.15   & 2400   & 9.10   &  699   &   2.00E-05   &   4.86E+01   &   5.98E-01   &   6.45E-03  \\
 3.85   & 2600   & 8.20   &  393   &   -          &   -          &   -          &   -         \\
 3.85   & 2600   & 8.50   &  393   &   -          &   -          &   -          &   -         \\
 3.85   & 2600   & 8.80   &  393   &   1.71E-06   &   3.51E+01   &   4.16E-01   &   2.25E-03  \\
 3.85   & 2600   & 9.10   &  393   &   2.96E-06   &   4.48E+01   &   4.11E-01   &   4.44E-03  \\
 4.00   & 2600   & 8.20   &  524   &   -          &   -          &   -          &   -         \\
 4.00   & 2600   & 8.50   &  524   &   2.24E-06   &   1.53E+01   &   2.53E-01   &   6.86E-04  \\
 4.00   & 2600   & 8.80   &  524   &   3.87E-06   &   3.13E+01   &   3.78E-01   &   2.04E-03  \\
 4.00   & 2600   & 9.10   &  524   &   9.54E-06   &   4.50E+01   &   4.71E-01   &   5.08E-03  \\
 4.15   & 2600   & 8.20   &  699   &   -          &   -          &   -          &   -         \\
 4.15   & 2600   & 8.50   &  699   &   6.58E-06   &   1.74E+01   &   2.73E-01   &   7.40E-04  \\
 4.15   & 2600   & 8.80   &  699   &   1.21E-05   &   2.79E+01   &   2.95E-01   &   1.60E-03  \\
 4.15   & 2600   & 9.10   &  699   &   1.62E-05   &   4.82E+01   &   4.91E-01   &   5.30E-03  \\
 3.85   & 2800   & 8.20   &  393   &   -          &   -          &   -          &   -         \\
 3.85   & 2800   & 8.50   &  393   &   -          &   -          &   -          &   -         \\
 3.85   & 2800   & 8.80   &  393   &   -          &   -          &   -          &   -         \\
 3.85   & 2800   & 9.10   &  393   &   7.36E-07   &   5.41E+01   &   3.85E-01   &   4.15E-03  \\
 4.00   & 2800   & 8.20   &  524   &   -          &   -          &   -          &   -         \\
 4.00   & 2800   & 8.50   &  524   &   -          &   -          &   -          &   -         \\
 4.00   & 2800   & 8.80   &  524   &   1.77E-06   &   3.87E+01   &   3.56E-01   &   1.93E-03  \\
 4.00   & 2800   & 9.10   &  524   &   2.43E-06   &   5.00E+01   &   3.66E-01   &   3.95E-03  \\
 4.15   & 2800   & 8.20   &  699   &   -          &   -          &   -          &   -         \\
 4.15   & 2800   & 8.50   &  699   &   2.69E-06   &   1.50E+01   &   1.63E-01   &   4.42E-04  \\
 4.15   & 2800   & 8.80   &  699   &   4.29E-06   &   3.58E+01   &   3.87E-01   &   2.09E-03  \\
 4.15   & 2800   & 9.10   &  699   &   8.54E-06   &   5.36E+01   &   4.09E-01   &   4.41E-03  \\
 3.85   & 3000   & 8.20   &  393   &   -          &   -          &   -          &   -         \\
 3.85   & 3000   & 8.50   &  393   &   -          &   -          &   -          &   -         \\
 3.85   & 3000   & 8.80   &  393   &   -          &   -          &   -          &   -         \\
 3.85   & 3000   & 9.10   &  393   &   -          &   -          &   -          &   -         \\
 4.00   & 3000   & 8.20   &  524   &   -          &   -          &   -          &   -         \\
 4.00   & 3000   & 8.50   &  524   &   -          &   -          &   -          &   -         \\
 4.00   & 3000   & 8.80   &  524   &   -          &   -          &   -          &   -         \\
 4.00   & 3000   & 9.10   &  524   &   4.51E-08   &   3.81E+01   &   1.62E-01   &   1.75E-03  \\
 4.15   & 3000   & 8.20   &  699   &   -          &   -          &   -          &   -         \\
 4.15   & 3000   & 8.50   &  699   &   -          &   -          &   -          &   -         \\
 4.15   & 3000   & 8.80   &  699   &   1.02E-06   &   3.70E+01   &   2.34E-01   &   1.27E-03  \\
 4.15   & 3000   & 9.10   &  699   &   1.79E-06   &   5.54E+01   &   3.19E-01   &   3.44E-03  \\
 3.85   & 3200   & 8.20   &  393   &   -          &   -          &   -          &   -         \\
 3.85   & 3200   & 8.50   &  393   &   -          &   -          &   -          &   -         \\
 3.85   & 3200   & 8.80   &  393   &   -          &   -          &   -          &   -         \\
 3.85   & 3200   & 9.10   &  393   &   -          &   -          &   -          &   -         \\
 4.00   & 3200   & 8.20   &  524   &   -          &   -          &   -          &   -         \\
 4.00   & 3200   & 8.50   &  524   &   -          &   -          &   -          &   -         \\
 4.00   & 3200   & 8.80   &  524   &   -          &   -          &   -          &   -         \\
 4.00   & 3200   & 9.10   &  524   &   -          &   -          &   -          &   -         \\
 4.15   & 3200   & 8.20   &  699   &   -          &   -          &   -          &   -         \\
 4.15   & 3200   & 8.50   &  699   &   -          &   -          &   -          &   -         \\
 4.15   & 3200   & 8.80   &  699   &   -          &   -          &   -          &   -         \\
 4.15   & 3200   & 9.10   &  699   &   9.22E-07   &   4.86E+01   &   2.60E-01   &   2.81E-03  \\

\\
  \hline
\\
  \end{tabular}
  \label{models}
  \end{table*}

  \begin{table*}
  \caption{\label{m15_u6} Same as \ref{m15_u2}, but for $\Delta u_{\rm p} = 6.0 \mbox{km s}^{-1}$. 
  }
  \center
  \begin{tabular}{lcccccccccccc}  
  \hline
  \hline

  $\log(L_\star)$ & $T_\mathrm{eff}$  & log(C-O)+12 & $P$ & $\langle\dot{M}\rangle$ & $\langle u_{\rm out} \rangle$ & $\langle {\rm f_c} \rangle$ & 
  $\langle {\rho_{\rm d}/\rho_{\rm g}} \rangle$ \\[1mm]
  [$L_{\odot}$] &  [$\mbox{K}$] & & [days] & [$M_\odot$ yr$^{-1}$] & [km  s$^{-1}$] & & \\

  \hline
\\
 3.85   & 2400   & 8.20   &  393   &   1.25E-06   &   1.23E-01   &   7.61E-01   &   1.03E-03  \\
 3.85   & 2400   & 8.50   &  393   &   4.52E-06   &   1.51E+01   &   3.93E-01   &   1.07E-03  \\
 3.85   & 2400   & 8.80   &  393   &   5.01E-06   &   2.76E+01   &   4.65E-01   &   2.51E-03  \\
 3.85   & 2400   & 9.10   &  393   &   9.82E-06   &   3.92E+01   &   7.73E-01   &   8.34E-03  \\
 4.00   & 2400   & 8.20   &  524   &   -          &   -          &   -          &   -         \\
 4.00   & 2400   & 8.50   &  524   &   7.15E-06   &   1.64E+01   &   3.68E-01   &   9.97E-04  \\
 4.00   & 2400   & 8.80   &  524   &   1.11E-05   &   3.05E+01   &   5.17E-01   &   2.80E-03  \\
 4.00   & 2400   & 9.10   &  524   &   1.47E-05   &   4.55E+01   &   6.10E-01   &   6.58E-03  \\
 4.15   & 2400   & 8.20   &  699   &   2.54E-05   &   8.25E+00   &   3.03E-01   &   4.12E-04  \\
 4.15   & 2400   & 8.50   &  699   &   2.29E-05   &   1.73E+01   &   4.16E-01   &   1.13E-03  \\
 4.15   & 2400   & 8.80   &  699   &   2.81E-05   &   3.18E+01   &   6.57E-01   &   3.55E-03  \\
 4.15   & 2400   & 9.10   &  699   &   3.34E-05   &   4.53E+01   &   6.37E-01   &   6.87E-03  \\
 3.85   & 2600   & 8.20   &  393   &   -          &   -          &   -          &   -         \\
 3.85   & 2600   & 8.50   &  393   &   2.66E-06   &   1.11E+01   &   3.06E-01   &   8.29E-04  \\
 3.85   & 2600   & 8.80   &  393   &   4.33E-06   &   2.74E+01   &   5.40E-01   &   2.92E-03  \\
 3.85   & 2600   & 9.10   &  393   &   6.22E-06   &   4.52E+01   &   6.70E-01   &   7.23E-03  \\
 4.00   & 2600   & 8.20   &  524   &   -          &   -          &   -          &   -         \\
 4.00   & 2600   & 8.50   &  524   &   6.83E-06   &   2.10E+01   &   4.71E-01   &   1.28E-03  \\
 4.00   & 2600   & 8.80   &  524   &   1.13E-05   &   2.82E+01   &   4.69E-01   &   2.54E-03  \\
 4.00   & 2600   & 9.10   &  524   &   1.36E-05   &   4.44E+00   &   6.30E-01   &   6.80E-03  \\
 4.15   & 2600   & 8.20   &  699   &   -          &   -          &   -          &   -         \\
 4.15   & 2600   & 8.50   &  699   &   1.42E-05   &   2.04E+01   &   3.77E-01   &   1.02E-03  \\
 4.15   & 2600   & 8.80   &  699   &        ***   &        ***   &        ***   &        ***  \\
 4.15   & 2600   & 9.10   &  699   &   2.03E-05   &   4.66E+01   &   5.50E-01   &   5.93E-03  \\
 3.85   & 2800   & 8.20   &  393   &   -          &   -          &   -          &   -         \\
 3.85   & 2800   & 8.50   &  393   &   -          &   -          &   -          &   -         \\
 3.85   & 2800   & 8.80   &  393   &   1.23E-06   &   3.80E+01   &   4.04E-01   &   2.18E-03  \\
 3.85   & 2800   & 9.10   &  393   &   2.07E-06   &   4.47E+01   &   3.84E-01   &   4.14E-03  \\
 4.00   & 2800   & 8.20   &  524   &   -          &   -          &   -          &   -         \\
 4.00   & 2800   & 8.50   &  524   &   2.20E-06   &   1.37E+01   &   2.16E-01   &   5.85E-04  \\
 4.00   & 2800   & 8.80   &  524   &   4.04E-06   &   3.05E+01   &   4.88E-01   &   2.64E-03  \\
 4.00   & 2800   & 9.10   &  524   &   8.55E-06   &   4.24E+01   &   4.18E-01   &   4.51E-03  \\
 4.15   & 2800   & 8.20   &  699   &        ***   &        ***   &        ***   &        ***  \\
 4.15   & 2800   & 8.50   &  699   &        ***   &        ***   &        ***   &        ***  \\
 4.15   & 2800   & 8.80   &  699   &   8.17E-06   &   3.43E+01   &   5.01E-01   &   2.71E-03  \\
 4.15   & 2800   & 9.10   &  699   &        ***   &        ***   &        ***   &        ***  \\
 3.85   & 3000   & 8.20   &  393   &   -          &   -          &   -          &   -         \\
 3.85   & 3000   & 8.50   &  393   &   -          &   -          &   -          &   -         \\
 3.85   & 3000   & 8.80   &  393   &   -          &   -          &   -          &   -         \\
 3.85   & 3000   & 9.10   &  393   &   2.04E-07   &   4.54E+01   &   3.30E-01   &   3.56E-03  \\
 4.00   & 3000   & 8.20   &  524   &   -          &   -          &   -          &   -         \\
 4.00   & 3000   & 8.50   &  524   &   -          &   -          &   -          &   -         \\
 4.00   & 3000   & 8.80   &  524   &   1.25E-06   &   3.99E+01   &   3.55E-01   &   1.92E-03  \\
 4.00   & 3000   & 9.10   &  524   &        ***   &        ***   &        ***   &        ***  \\
 4.15   & 3000   & 8.20   &  699   &        ***   &        ***   &        ***   &        ***  \\
 4.15   & 3000   & 8.50   &  699   &        ***   &        ***   &        ***   &        ***  \\
 4.15   & 3000   & 8.80   &  699   &        ***   &        ***   &        ***   &        ***  \\
 4.15   & 3000   & 9.10   &  699   &        ***   &        ***   &        ***   &        ***  \\
 3.85   & 3200   & 8.20   &  393   &   -          &   -          &   -          &   -         \\
 3.85   & 3200   & 8.50   &  393   &   -          &   -          &   -          &   -         \\
 3.85   & 3200   & 8.80   &  393   &   -          &   -          &   -          &   -         \\
 3.85   & 3200   & 9.10   &  393   &        ***   &        ***   &        ***   &        ***  \\
 4.00   & 3200   & 8.20   &  524   &   -          &   -          &   -          &   -         \\
 4.00   & 3200   & 8.50   &  524   &   -          &   -          &   -          &   -         \\
 4.00   & 3200   & 8.80   &  524   &        ***   &        ***   &        ***   &        ***  \\
 4.00   & 3200   & 9.10   &  524   &        ***   &        ***   &        ***   &        ***  \\
 4.15   & 3200   & 8.20   &  699   &        ***   &        ***   &        ***   &        ***  \\
 4.15   & 3200   & 8.50   &  699   &        ***   &        ***   &        ***   &        ***  \\
 4.15   & 3200   & 8.80   &  699   &        ***   &        ***   &        ***   &        ***  \\
 4.15   & 3200   & 9.10   &  699   &        ***   &        ***   &        ***   &        ***  \\

\\
  \hline
\\
  \end{tabular}
  \label{models}
  \end{table*}

  \begin{table*}
  \caption{\label{m2_u2} Input parameters ($L_\star$, $T_{\rm eff}$, C/O, $P$) and the 
  resulting mean mass loss rate, mean velocity at the outer boundary and mean degree of dust condensation at the outer boundary for a
  subset of models with $M_\star = 2.0 M_\odot$ and $\Delta u_{\rm p} = 2.0 \mbox{km s}^{-1}$. 
  The dust-to-gas mass ratio $\rho_{\rm dust}/\rho_{\rm gas}$ is calculated from $f_{\rm c}$ as described in H\"ofner \& Dorfi (1997).}
  \center
  \begin{tabular}{lcccccccccccc}  
  \hline
  \hline

  $\log(L_\star)$ & $T_\mathrm{eff}$  & log(C-O)+12 & $P$ & $\langle\dot{M}\rangle$ & $\langle u_{\rm out} \rangle$ & $\langle {\rm f_c} \rangle$ & 
  $\langle {\rho_{\rm d}/\rho_{\rm g}} \rangle$ \\[1mm]
  [$L_{\odot}$] &  [$\mbox{K}$] & & [days] & [$M_\odot$ yr$^{-1}$] & [km  s$^{-1}$] & & \\

  \hline
\\
 3.85   & 2400   & 8.20   &  393   &   -          &   -          &   -          &   -         \\
 3.85   & 2400   & 8.50   &  393   &   -          &   -          &   -          &   -         \\
 3.85   & 2400   & 8.80   &  393   &   -          &   -          &   -          &   -         \\
 3.85   & 2400   & 9.10   &  393   &   1.57E-06   &   5.38E+01   &   4.34E-01   &   4.68E-03  \\
 4.00   & 2400   & 8.20   &  524   &   -          &   -          &   -          &   -         \\
 4.00   & 2400   & 8.50   &  524   &   -          &   -          &   -          &   -         \\
 4.00   & 2400   & 8.80   &  524   &   2.05E-06   &   3.29E+01   &   3.23E-01   &   1.75E-03  \\
 4.00   & 2400   & 9.10   &  524   &        ***   &        ***   &        ***   &        ***  \\
 4.15   & 2400   & 8.20   &  699   &   -          &   -          &   -          &   -         \\
 4.15   & 2400   & 8.50   &  699   &        ***   &        ***   &        ***   &        ***  \\
 4.15   & 2400   & 8.80   &  699   &   4.13E-06   &   3.68E+01   &   3.52E-01   &   1.90E-03  \\
 4.15   & 2400   & 9.10   &  699   &   6.13E-06   &   5.59E+01   &   4.28E-01   &   4.62E-03  \\
 3.85   & 2600   & 8.20   &  393   &   -          &   -          &   -          &   -         \\
 3.85   & 2600   & 8.50   &  393   &   -          &   -          &   -          &   -         \\
 3.85   & 2600   & 8.80   &  393   &   -          &   -          &   -          &   -         \\
 3.85   & 2600   & 9.10   &  393   &   -          &   -          &   -          &   -         \\
 4.00   & 2600   & 8.20   &  524   &   -          &   -          &   -          &   -         \\
 4.00   & 2600   & 8.50   &  524   &   -          &   -          &   -          &   -         \\
 4.00   & 2600   & 8.80   &  524   &   -          &   -          &   -          &   -         \\
 4.00   & 2600   & 9.10   &  524   &   1.27E-06   &   5.09E+01   &   3.08E-01   &   3.32E-03  \\
 4.15   & 2600   & 8.20   &  699   &   -          &   -          &   -          &   -         \\
 4.15   & 2600   & 8.50   &  699   &   -          &   -          &   -          &   -         \\
 4.15   & 2600   & 8.80   &  699   &   -          &   -          &   -          &   -         \\
 4.15   & 2600   & 9.10   &  699   &   2.33E-06   &   5.29E+01   &   2.83E-01   &   3.05E-03  \\
 3.85   & 2800   & 8.20   &  393   &   -          &   -          &   -          &   -         \\
 3.85   & 2800   & 8.50   &  393   &   -          &   -          &   -          &   -         \\
 3.85   & 2800   & 8.80   &  393   &   -          &   -          &   -          &   -         \\
 3.85   & 2800   & 9.10   &  393   &   -          &   -          &   -          &   -         \\
 4.00   & 2800   & 8.20   &  524   &   -          &   -          &   -          &   -         \\
 4.00   & 2800   & 8.50   &  524   &   -          &   -          &   -          &   -         \\
 4.00   & 2800   & 8.80   &  524   &   -          &   -          &   -          &   -         \\
 4.00   & 2800   & 9.10   &  524   &   -          &   -          &   -          &   -         \\
 4.15   & 2800   & 8.20   &  699   &   -          &   -          &   -          &   -         \\
 4.15   & 2800   & 8.50   &  699   &   -          &   -          &   -          &   -         \\
 4.15   & 2800   & 8.80   &  699   &   -          &   -          &   -          &   -         \\
 4.15   & 2800   & 9.10   &  699   &   -          &   -          &   -          &   -         \\
 3.85   & 3000   & 8.20   &  393   &   -          &   -          &   -          &   -         \\
 3.85   & 3000   & 8.50   &  393   &   -          &   -          &   -          &   -         \\
 3.85   & 3000   & 8.80   &  393   &   -          &   -          &   -          &   -         \\
 3.85   & 3000   & 9.10   &  393   &   -          &   -          &   -          &   -         \\
 4.00   & 3000   & 8.20   &  524   &   -          &   -          &   -          &   -         \\
 4.00   & 3000   & 8.50   &  524   &   -          &   -          &   -          &   -         \\
 4.00   & 3000   & 8.80   &  524   &   -          &   -          &   -          &   -         \\
 4.00   & 3000   & 9.10   &  524   &   -          &   -          &   -          &   -         \\
 4.15   & 3000   & 8.20   &  699   &   -          &   -          &   -          &   -         \\
 4.15   & 3000   & 8.50   &  699   &   -          &   -          &   -          &   -         \\
 4.15   & 3000   & 8.80   &  699   &   -          &   -          &   -          &   -         \\
 4.15   & 3000   & 9.10   &  699   &   -          &   -          &   -          &   -         \\
 3.85   & 3200   & 8.20   &  393   &   -          &   -          &   -          &   -         \\
 3.85   & 3200   & 8.50   &  393   &   -          &   -          &   -          &   -         \\
 3.85   & 3200   & 8.80   &  393   &   -          &   -          &   -          &   -         \\
 3.85   & 3200   & 9.10   &  393   &   -          &   -          &   -          &   -         \\
 4.00   & 3200   & 8.20   &  524   &   -          &   -          &   -          &   -         \\
 4.00   & 3200   & 8.50   &  524   &   -          &   -          &   -          &   -         \\
 4.00   & 3200   & 8.80   &  524   &   -          &   -          &   -          &   -         \\
 4.00   & 3200   & 9.10   &  524   &   -          &   -          &   -          &   -         \\
 4.15   & 3200   & 8.20   &  699   &   -          &   -          &   -          &   -         \\
 4.15   & 3200   & 8.50   &  699   &   -          &   -          &   -          &   -         \\
 4.15   & 3200   & 8.80   &  699   &   -          &   -          &   -          &   -         \\
 4.15   & 3200   & 9.10   &  699   &   -          &   -          &   -          &   -         \\

\\
  \hline
\\
  \end{tabular}
  \label{models}
  \end{table*}

  \begin{table*}
  \caption{\label{m2_u4} Same as \ref{m2_u2}, but for $\Delta u_{\rm p} = 4.0 \mbox{km s}^{-1}$. 
  }
  \center
  \begin{tabular}{lcccccccccccc}  
  \hline
  \hline

  $\log(L_\star)$ & $T_\mathrm{eff}$  & log(C-O)+12 & $P$ & $\langle\dot{M}\rangle$ & $\langle u_{\rm out} \rangle$ & $\langle {\rm f_c} \rangle$ & 
  $\langle {\rho_{\rm d}/\rho_{\rm g}} \rangle$ \\[1mm]
  [$L_{\odot}$] &  [$\mbox{K}$] & & [days] & [$M_\odot$ yr$^{-1}$] & [km  s$^{-1}$] & & \\

  \hline
\\
 3.85   & 2400   & 8.20   &  393   &   -          &   -          &   -          &   -         \\
 3.85   & 2400   & 8.50   &  393   &   -          &   -          &   -          &   -         \\
 3.85   & 2400   & 8.80   &  393   &   2.09E-06   &   2.87E+01   &   3.68E-01   &   1.99E-03  \\
 3.85   & 2400   & 9.10   &  393   &   2.48E-06   &   5.14E+01   &   5.36E-01   &   5.78E-03  \\
 4.00   & 2400   & 8.20   &  524   &   1.75E-06   &   1.08E+01   &   2.86E-01   &   3.89E-04  \\
 4.00   & 2400   & 8.50   &  524   &   1.68E-06   &   1.13E+01   &   2.93E-01   &   7.94E-04  \\
 4.00   & 2400   & 8.80   &  524   &   4.71E-06   &   3.38E+01   &   4.46E-01   &   2.41E-03  \\
 4.00   & 2400   & 9.10   &  524   &   6.84E-06   &   5.10E+01   &   5.46E-01   &   5.89E-03  \\
 4.15   & 2400   & 8.20   &  699   &   7.02E-06   &   1.91E+01   &   3.73E-01   &   5.07E-04  \\
 4.15   & 2400   & 8.50   &  699   &   5.34E-06   &   1.65E+01   &   3.00E-01   &   8.13E-04  \\
 4.15   & 2400   & 8.80   &  699   &   9.22E-06   &   3.49E+01   &   4.96E-01   &   2.68E-03  \\
 4.15   & 2400   & 9.10   &  699   &   1.68E-05   &   4.85E+01   &   5.85E-01   &   6.31E-03  \\
 3.85   & 2600   & 8.20   &  393   &   -          &   -          &   -          &   -         \\
 3.85   & 2600   & 8.50   &  393   &   -          &   -          &   -          &   -         \\
 3.85   & 2600   & 8.80   &  393   &   -          &   -          &   -          &   -         \\
 3.85   & 2600   & 9.10   &  393   &   9.89E-07   &   5.44E+01   &   4.30E-01   &   4.64E-03  \\
 4.00   & 2600   & 8.20   &  524   &   -          &   -          &   -          &   -         \\
 4.00   & 2600   & 8.50   &  524   &   -          &   -          &   -          &   -         \\
 4.00   & 2600   & 8.80   &  524   &   1.82E-06   &   4.02E+01   &   4.00E-01   &   2.16E-03  \\
 4.00   & 2600   & 9.10   &  524   &   2.44E-06   &   5.70E+01   &   4.46E-01   &   4.81E-03  \\
 4.15   & 2600   & 8.20   &  699   &   -          &   -          &   -          &   -         \\
 4.15   & 2600   & 8.50   &  699   &   -          &   -          &   -          &   -         \\
 4.15   & 2600   & 8.80   &  699   &   4.19E-06   &   3.88E+01   &   4.11E-01   &   2.22E-03  \\
 4.15   & 2600   & 9.10   &  699   &   9.12E-06   &   -          &   -          &   -         \\
 3.85   & 2800   & 8.20   &  393   &   -          &   -          &   -          &   -         \\
 3.85   & 2800   & 8.50   &  393   &   -          &   -          &   -          &   -         \\
 3.85   & 2800   & 8.80   &  393   &   -          &   -          &   -          &   -         \\
 3.85   & 2800   & 9.10   &  393   &   -          &   -          &   -          &   -         \\
 4.00   & 2800   & 8.20   &  524   &   -          &   -          &   -          &   -         \\
 4.00   & 2800   & 8.50   &  524   &   -          &   -          &   -          &   -         \\
 4.00   & 2800   & 8.80   &  524   &   -          &   -          &   -          &   -         \\
 4.00   & 2800   & 9.10   &  524   &   -          &   -          &   -          &   -         \\
 4.15   & 2800   & 8.20   &  699   &   -          &   -          &   -          &   -         \\
 4.15   & 2800   & 8.50   &  699   &   -          &   -          &   -          &   -         \\
 4.15   & 2800   & 8.80   &  699   &   -          &   -          &   -          &   -         \\
 4.15   & 2800   & 9.10   &  699   &   1.39E-06   &   5.90E+01   &   3.20E-01   &   3.45E-03  \\
 3.85   & 3000   & 8.20   &  393   &   -          &   -          &   -          &   -         \\
 3.85   & 3000   & 8.50   &  393   &   -          &   -          &   -          &   -         \\
 3.85   & 3000   & 8.80   &  393   &   -          &   -          &   -          &   -         \\
 3.85   & 3000   & 9.10   &  393   &   -          &   -          &   -          &   -         \\
 4.00   & 3000   & 8.20   &  524   &   -          &   -          &   -          &   -         \\
 4.00   & 3000   & 8.50   &  524   &   -          &   -          &   -          &   -         \\
 4.00   & 3000   & 8.80   &  524   &   -          &   -          &   -          &   -         \\
 4.00   & 3000   & 9.10   &  524   &   -          &   -          &   -          &   -         \\
 4.15   & 3000   & 8.20   &  699   &   -          &   -          &   -          &   -         \\
 4.15   & 3000   & 8.50   &  699   &   -          &   -          &   -          &   -         \\
 4.15   & 3000   & 8.80   &  699   &   -          &   -          &   -          &   -         \\
 4.15   & 3000   & 9.10   &  699   &   -          &   -          &   -          &   -         \\
 3.85   & 3200   & 8.20   &  393   &   -          &   -          &   -          &   -         \\
 3.85   & 3200   & 8.50   &  393   &   -          &   -          &   -          &   -         \\
 3.85   & 3200   & 8.80   &  393   &   -          &   -          &   -          &   -         \\
 3.85   & 3200   & 9.10   &  393   &   -          &   -          &   -          &   -         \\
 4.00   & 3200   & 8.20   &  524   &   -          &   -          &   -          &   -         \\
 4.00   & 3200   & 8.50   &  524   &   -          &   -          &   -          &   -         \\
 4.00   & 3200   & 8.80   &  524   &   -          &   -          &   -          &   -         \\
 4.00   & 3200   & 9.10   &  524   &   -          &   -          &   -          &   -         \\
 4.15   & 3200   & 8.20   &  699   &   -          &   -          &   -          &   -         \\
 4.15   & 3200   & 8.50   &  699   &   -          &   -          &   -          &   -         \\
 4.15   & 3200   & 8.80   &  699   &   -          &   -          &   -          &   -         \\
 4.15   & 3200   & 9.10   &  699   &   -          &   -          &   -          &   -         \\

\\
  \hline
\\
  \end{tabular}
  \label{models}
  \end{table*}

  \begin{table*}
  \caption{\label{m2_u6} Same as \ref{m2_u2}, but for $\Delta u_{\rm p} = 6.0 \mbox{km s}^{-1}$. 
  }
  \center
  \begin{tabular}{lcccccccccccc}  
  \hline
  \hline

  $\log(L_\star)$ & $T_\mathrm{eff}$  & log(C-O)+12 & $P$ & $\langle\dot{M}\rangle$ & $\langle u_{\rm out} \rangle$ & $\langle {\rm f_c} \rangle$ & 
  $\langle {\rho_{\rm d}/\rho_{\rm g}} \rangle$ \\[1mm]
  [$L_{\odot}$] &  [$\mbox{K}$] & & [days] & [$M_\odot$ yr$^{-1}$] & [km  s$^{-1}$] & & \\

  \hline
\\
 3.85   & 2400   & 8.20   &  393   &   1.03E-06   &   6.77E+00   &   3.16E-01   &   4.29E-04  \\
 3.85   & 2400   & 8.50   &  393   &   9.36E-07   &   8.28E+00   &   2.96E-01   &   8.02E-04  \\
 3.85   & 2400   & 8.80   &  393   &   3.99E-06   &   3.00E+01   &   5.80E-01   &   3.14E-03  \\
 3.85   & 2400   & 9.10   &  393   &   5.98E-06   &   4.01E+01   &   6.60E-01   &   7.12E-03  \\
 4.00   & 2400   & 8.20   &  524   &   7.39E-06   &   1.65E+01   &   3.76E-01   &   5.11E-04  \\
 4.00   & 2400   & 8.50   &  524   &   6.32E-06   &   1.55E+01   &   4.00E-01   &   1.08E-03  \\
 4.00   & 2400   & 8.80   &  524   &   9.82E-06   &   2.99E+01   &   5.16E-01   &   2.79E-03  \\
 4.00   & 2400   & 9.10   &  524   &   1.33E-05   &   4.15E+01   &   7.75E-01   &   8.36E-03  \\
 4.15   & 2400   & 8.20   &  699   &   1.27E-06   &   2.35E+01   &   5.14E-01   &   6.98E-04  \\
 4.15   & 2400   & 8.50   &  699   &   1.43E-05   &   1.64E+01   &   5.74E-01   &   1.56E-03  \\
 4.15   & 2400   & 8.80   &  699   &   1.10E-05   &   3.91E+01   &   5.65E-01   &   3.06E-03  \\
 4.15   & 2400   & 9.10   &  699   &   2.53E-05   &   4.68E+01   &   7.74E-01   &   8.35E-03  \\
 3.85   & 2600   & 8.20   &  393   &   -          &   -          &   -          &   -         \\
 3.85   & 2600   & 8.50   &  393   &   -          &   -          &   -          &   -         \\
 3.85   & 2600   & 8.80   &  393   &   1.17E-06   &   3.65E+01   &   4.25E-01   &   2.30E-03  \\
 3.85   & 2600   & 9.10   &  393   &   1.58E-06   &   5.54E+01   &   5.10E-01   &   5.50E-03  \\
 4.00   & 2600   & 8.20   &  524   &   -          &   -          &   -          &   -         \\
 4.00   & 2600   & 8.50   &  524   &   -          &   -          &   -          &   -         \\
 4.00   & 2600   & 8.80   &  524   &   3.08E-06   &   3.95E+01   &   4.97E-01   &   2.69E-03  \\
 4.00   & 2600   & 9.10   &  524   &   4.33E-06   &   4.81E+01   &   4.56E-01   &   4.92E-03  \\
 4.15   & 2600   & 8.20   &  699   &   7.65E-06   &   2.38E+01   &   4.34E-01   &   5.90E-04  \\
 4.15   & 2600   & 8.50   &  699   &   6.65E-06   &   2.12E+01   &   3.61E-01   &   9.78E-04  \\
 4.15   & 2600   & 8.80   &  699   &   8.09E-06   &   3.50E+01   &   4.26E-01   &   2.30E-03  \\
 4.15   & 2600   & 9.10   &  699   &   1.42E-05   &   4.92E+01   &   4.95E-01   &   5.34E-03  \\
 3.85   & 2800   & 8.20   &  393   &   -          &   -          &   -          &   -         \\
 3.85   & 2800   & 8.50   &  393   &   -          &   -          &   -          &   -         \\
 3.85   & 2800   & 8.80   &  393   &   -          &   -          &   -          &   -         \\
 3.85   & 2800   & 9.10   &  393   &        ***   &        ***   &        ***   &        ***  \\
 4.00   & 2800   & 8.20   &  524   &   -          &   -          &   -          &   -         \\
 4.00   & 2800   & 8.50   &  524   &   -          &   -          &   -          &   -         \\
 4.00   & 2800   & 8.80   &  524   &   3.27E-06   &   2.80E+01   &   2.05E-01   &   1.11E-03  \\
 4.00   & 2800   & 9.10   &  524   &   1.77E-06   &   5.57E+01   &   3.88E-01   &   4.19E-03  \\
 4.15   & 2800   & 8.20   &  699   &   -          &   -          &   -          &   -         \\
 4.15   & 2800   & 8.50   &  699   &   -          &   -          &   -          &   -         \\
 4.15   & 2800   & 8.80   &  699   &   3.88E-06   &   4.45E+01   &   4.03E-01   &   2.18E-03  \\
 4.15   & 2800   & 9.10   &  699   &   5.30E-06   &   5.22E+01   &   4.09E-01   &   4.41E-03  \\
 3.85   & 3000   & 8.20   &  393   &   -          &   -          &   -          &   -         \\
 3.85   & 3000   & 8.50   &  393   &   -          &   -          &   -          &   -         \\
 3.85   & 3000   & 8.80   &  393   &   -          &   -          &   -          &   -         \\
 3.85   & 3000   & 9.10   &  393   &   -          &   -          &   -          &   -         \\
 4.00   & 3000   & 8.20   &  524   &   -          &   -          &   -          &   -         \\
 4.00   & 3000   & 8.50   &  524   &   -          &   -          &   -          &   -         \\
 4.00   & 3000   & 8.80   &  524   &   -          &   -          &   -          &   -         \\
 4.00   & 3000   & 9.10   &  524   &   -          &   -          &   -          &   -         \\
 4.15   & 3000   & 8.20   &  699   &   -          &   -          &   -          &   -         \\
 4.15   & 3000   & 8.50   &  699   &   -          &   -          &   -          &   -         \\
 4.15   & 3000   & 8.80   &  699   &   -          &   -          &   -          &   -         \\
 4.15   & 3000   & 9.10   &  699   &   1.76E-06   &   4.80E+01   &   2.87E-01   &   3.10E-03  \\
 3.85   & 3200   & 8.20   &  393   &   -          &   -          &   -          &   -         \\
 3.85   & 3200   & 8.50   &  393   &   -          &   -          &   -          &   -         \\
 3.85   & 3200   & 8.80   &  393   &   -          &   -          &   -          &   -         \\
 3.85   & 3200   & 9.10   &  393   &   -          &   -          &   -          &   -         \\
 4.00   & 3200   & 8.20   &  524   &   -          &   -          &   -          &   -         \\
 4.00   & 3200   & 8.50   &  524   &   -          &   -          &   -          &   -         \\
 4.00   & 3200   & 8.80   &  524   &   -          &   -          &   -          &   -         \\
 4.00   & 3200   & 9.10   &  524   &        ***   &        ***   &        ***   &        ***  \\
 4.15   & 3200   & 8.20   &  699   &   -          &   -          &   -          &   -         \\
 4.15   & 3200   & 8.50   &  699   &   -          &   -          &   -          &   -         \\
 4.15   & 3200   & 8.80   &  699   &   -          &   -          &   -          &   -         \\
 4.15   & 3200   & 9.10   &  699   &        ***   &        ***   &        ***   &        ***  \\

\\
  \hline
\\
  \end{tabular}
  \label{models}
  \end{table*}


\begin{thebibliography}{}

\bibitem[Arndt et al. 1997]{Arndt97}
  Arndt, T.U., Fleischer, A.J., \& Sedlmayr, E. 
  ~1997, \aap, 327, 614

\bibitem[Andersen et al. 2003]{Andersen03}
  Andersen, A.C., H\"ofner, S. \& Gautschy-Liodl, R., 2003, \aap, 400, 981
  
\bibitem[Bergeat \& Chevallier 2005]{Bergeat05}
  Bergeat J. \& Chevallier L., 2005, \aap, 429, 235

\bibitem[Bowen 1988]{Bowen88}
  Bowen, G.H., 1988, \apj,, 329, 299

\bibitem[Bl\"ocker 1995]{Blocker95}
  Bl\"ocker, T., 1995, \aap, 297, 727

\bibitem[Dorfi \& Drury 1987]{Dorfi87}
  Dorfi E. \& Drury L. O'C., 1987, Computer Physics
  Communications 69, 175

\bibitem[Dorfi \& Feuchtinger 1995]{Dorfi95}
  Dorfi E. \& Feuchtinger M., 1995, Computer Physics
  Communications 89, 69

\bibitem[Feast et al 1989]{Feast89}
  Feast M.W., Glass I.S., Whitelock P.A. \& Catchpole R.M. 1989, \mnras 241, 375

\bibitem[Fleischer, Gauger \& Sedlmayr 1992]{Fleischer92}
  Fleischer A.J., Gauger A. \& Sedlmayr E., 1992, \aap, 266, 321

\bibitem[Gail \& Sedlmayr 1986]{Gail86}
  Gail H.-P. \& Sedlmayr E., 1986, \aap, 161, 201

\bibitem[Gail \& Sedlmayr 1988]{Gail88}
  Gail H.-P. \& Sedlmayr E., 1988, \aap, 206, 153

\bibitem[Gauger et al. 1990]{Gauger90}
  Gauger A., Gail H.-P. \& Sedlmayr E., 1990, \aap, 235, 345

\bibitem[Gautschy-Loidl et al. 2004]{Gautschy04}
  Gautschy-Loidl R., H\"ofner S., J\/orgensen U. G. \& Hron J. 2004, \aap, 422, 289

\bibitem[Groenewegen 1993]{Groenewegen93}
  Groenewegen M.A.T., 1993, "On the Evolution and Properties of AGB Stars", PhD Thesis, University of Amsterdam.
  
\bibitem[Groenewegen 1995]{Groenewegen95}
  Groenewegen M.A.T., 1995, ASPC, 83, 141
  
\bibitem[Groenewegen et al. 1998]{Groenewegen98}
  Groenewegen M.A.T., Whitelock P.A., Smith C.H. \& Kerschbaum F., 1998, \mnras, 293, 18

\bibitem[H\"ofner \& Dorfi 1997]{Hofner97}
  H\"ofner S. \& Dorfi E.A., 1997, \aap, 319, 648

\bibitem[H\"ofner et al. 1995]{Hofner95}
  H\"ofner S., Feuchtinger M.U. \& Dorfi E.A., 1995, \aap, 279, 815

\bibitem[H\"ofner et al. 2003]{Hofner03}
  H\"ofner S., Gautschy-Loidl R., Aringer B. \& J\o rgensen U.G.,
  2003, \aap, 399, 589
  
\bibitem[H\"ofner 2008]{Hofner08}
  H\"ofner S., 2008, \aap, 491, L1

\bibitem[Kudritzki \& Reimers 1978]{Kudritzki78}
  Kudritzki R.P. \& Reimers D., 1978, \aap, 70, 229

\bibitem[Lambert et al. 1986]{Lambert86}
  Lambert D.L., Gustafsson B., Eriksson K. \& Hinkle K.H., \apj, 62, 373
  
\bibitem[Lamers \& Cassinelli 1999]{Lamers99}
  Lamers H.J.G.L.M. \& Cassinelli J.P., 1999, "Introduction to Stellar Winds", Cambridge University Press

\bibitem[Mattsson et al. 2007a]{Mattsson07a}
  Mattsson L., H\"ofner S. \& Herwig F., 2007a, \aap, 470, 339
  
\bibitem[Mattsson et al. 2007b]{Mattsson07b}
  Mattsson L., Wahlin R. \& H\"ofner S., 2007b, IAUS, 241, 37

\bibitem[Mattsson et al. 2008]{Mattsson08}
  Mattsson L., Wahlin R., H\"ofner S. \& Eriksson K., 2008, \aap, 484, 5L

\bibitem[Nowotny et al. 2005a]{Nowotny05a}
  Nowotny W., Aringer B., H\"ofner S., Gautschy-Loidl, R. \&  Windsteig W., 2005a, \aap, 437, 273

\bibitem[Nowotny et al. 2005b]{Nowotny05b}
  Nowotny W., Lebzelter T., Hron J. \& H\"ofner S., 2005b, \aap, 437, 285

\bibitem[Ramstedt et al. 2006]{Ramstedt06}
  Ramstedt S., Sch\"oier F.L., Olofsson H., \& Lundgren A.A., 2006, \aap, 454, L103

\bibitem[Reimers 1975]{Reimers75}
  Reimers, D., 1975, Mem. Soc. Roy. Sci. Li\`ege, 6e Ser., 8, 369

\bibitem[Rouleau \& Martin 1991]{Rouleau91}
  Rouleau F. \& Martin P.G., 1991, \apj, 377, 526 

\bibitem[Sandin \& H\"ofner 2003]{Sandin03}
  Sandin, C. \& H\"ofner, S., 2003, \aap, 404, 789

\bibitem[Sandin \& H\"ofner 2004]{Sandin04}
  Sandin, C. \& H\"ofner, S., 2004, \aap, 413, 789

\bibitem[Sch\"oier et al. 2001]{Schoier01}
  Sch\"oier F.L. \& Olofsson H., 2001, \aap, 368, 969

\bibitem[Sch\"oier et al. 2005]{Schoier05}
  Sch\"oier F.L., Lindqvist M. \& Olofsson H., 2005, \aap, 436, 633

\bibitem[Vassiliadis \& Wood 1993]{Vassiliadis93}
  Vassiliadis, E. \& Wood, P.R., 1993, \apj, 413, 614

\bibitem[van Leer 1977]{vanLeer77}
  van Leer B. 1977, J. Comput. Phys., 23, 276

\bibitem[Wachter et al. 2002]{Wachter02}
  Wachter, A., Schr\"oder, K.-P., Winters, J.M., Arndt, T.U. \& Sedlmayr, E., 2002,
  \aap, 384, 452
  
\bibitem[Wachter et al. 2008]{Wachter08}
  Wachter, Winters, J.M., A., Schr\"oder, K.-P., \& Sedlmayr, E., 2008,
  \aap, 486, 497
  
\bibitem[Winters et al. 2000]{Winters00}
  Winters J.M., Le Bertre T., Jeong K.S., Helling C. \& Sedlmayr E., 2000, \aap, 361, 641
  
\bibitem[Wood 1979]{Wood79}
  Wood P., 1979, \apj, 227, 220
  
\end{thebibliography}
\end{document}